\documentclass[%
reprint,
amsmath,amssymb,
aps, prb,
]{revtex4-2}

\usepackage{gensymb}
\usepackage[dvipsnames]{xcolor}
\usepackage[colorlinks]{hyperref}
\hypersetup{
	colorlinks=True,
	linkbordercolor=White,
	linkcolor=BrickRed,
	citecolor=Blue,	 
}

\usepackage{siunitx}
\usepackage{graphicx}
\usepackage{layouts}

\newcommand\LC[1]{\textcolor{black}{#1}}

\usepackage[caption=false]{subfig}

\begin{document}
	\vspace{-0.5cm}
	
	\title{
		From Atoms to Moiré: Combining \textit{Ab-initio} and Machine Learning  to Predict Proximity Effects in van der Waals Heterostructures
	}
	
	%
	
	\author{Lukas Cvitkovich}
	\affiliation{Institute for Theoretical Physics, University of Regensburg, 93040 Regensburg, Germany}
	
	
	\author{Klaus Zollner}
	\affiliation{Institute for Theoretical Physics, University of Regensburg, 93040 Regensburg, Germany}

	\author{Jaroslav Fabian}
	\affiliation{Institute for Theoretical Physics, University of Regensburg, 93040 Regensburg, Germany}

	\newcommand{\CGT}{Cr$_2$Ge$_2$Te$_6$}
	
	\begin{abstract}
		
		We introduce a machine learning framework that efficiently predicts large-scale proximity-induced magnetism in van der Waals heterostructures, overcoming the high computational cost of density functional theory (DFT). We apply it to graphene/\CGT, which exhibits a previously unrecognized dichotomy. Unlike the spin polarization at the Fermi level, which follows the pseudospin, the proximity-induced magnetic moments vary across carbon atoms, defying analytical modeling. To address this, we develop an ensemble-based regression model trained on DFT data and employ local environment descriptors to map the local ($\sim 2$\,nm$^2$) atomic-scale geometry to the carbon magnetic moments. Besides demonstrating locality, the model reveals rich magnetic moiré textures. Crucially, this method can be broadly applied to orbital and spin proximity effects that are highly sensitive to local atomic environments and are beyond analytical description.  
		
	\end{abstract}
	
	\maketitle
	
	
	Integration of twisted 2D materials into van der Waals (vdW) heterostructures allows tailoring of electronic, optical, and magnetic properties~\cite{Geim_vdW_2013, Sierra_vdW_2021, Novoselov_2D_2016, Carr_twistronics_2017, Carr_electronic-structure_2020, Hennighausen_twistronics_2021, Ribeiro-Palau_twistable_2018} for ultrafast and low-power electronic and spintronic~\cite{Zutic_RMP_2004} devices.
	One particularly promising strategy is to harness proximity effects, enabling superconductivity~\cite{Moriya_Superconducting_2020, Han_superconducting_2021} or strong spin-orbit coupling~\cite{Gmitra_2015, Gmitra_2016, Khokhriakov_2018, Zollner_Single_2019, Khokhriakov2020, Karpiak_2020, Zihlmann_2018, Song2018, Safeer2019, Herling2020, Pezo_2022} in materials lacking these features.
	Furthermore, when bringing non-magnetic 2D materials, such as graphene or transition metal dichalcogenides~\cite{Manzeli_2DTMDC_2017}, in contact with a magnetic material, a proximity-induced magnetic exchange interaction can arise~\cite{Yang_Proximity_2013, Hallal_tailoring_2017, Zollner_Proximity_2019, Zollner_Engineering_2022, Chau_CGT/Gr_2022, Pezo_2022, Choi_Emergent_2022, Ghiasi2025, Yang2025, Sahani_giantMR_2025}.
	Once a non-magnetic channel acquires an exchange field via proximity, it can support spin precession~\cite{Tombros_spin_graphene_2007, Raes2016, SaveroTorres_2017}, spin filtering~\cite{Piquemal-Banci2020, Yang_Proximity_2013}, or valley-selective spin dynamics~\cite{Jeong_spin-selective_2024}, enabling spintronic device platforms.
	
	Among 2D materials for proximity-induced phenomena, graphene stands out for its exceptional spin transport ~\cite{Tombros_spin_graphene_2007, Han_spin_transport_2012, Zollner_review_2025, Han_graphene_2014}.
	Magnetism in graphene can be induced via proximity exchange coupling to ferro- or antiferromagnets, ideally preserving its intrinsic transport by avoiding charge transfer and parasitic channels.
	Suitable magnetic materials include semiconductors or insulators such as \CGT~\cite{Zhang_robust_215, Karpiak_2020, Zollner_scattering_2020}, EuO~\cite{Yang_Proximity_2013, Averyanov2018, Pandey_2023, Pandey_2025}, or CrI$_3$~\cite{Zhang_2018, Cardoso_VdW_2018, Seyler2018, Farooq2019, Cardoso_Strong_2023, Jeong_spin-selective_2024, Pandey_2025}, but also ferromagnetic metals such as Co or Ni, separated from graphene by hBN~\cite{Zollner_Theory_2016}.

	\begin{figure*}[htbp]
		\centerline{\includegraphics[width=.9\linewidth]{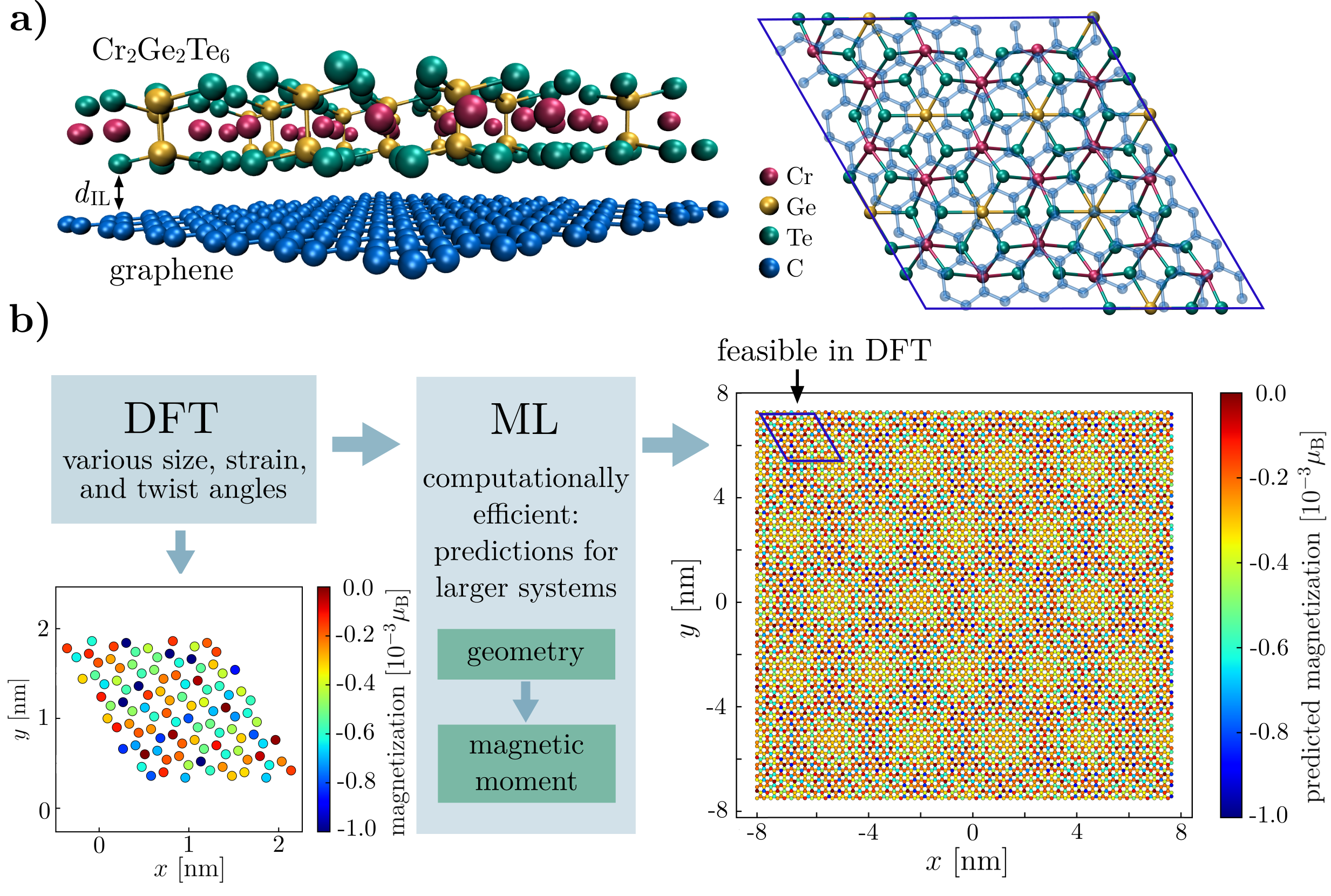}}
		\caption{ML workflow for modeling magnetic proximity effects in graphene/Cr$_2$Ge$_2$Te$_6$.
			\textbf{(a)} 3D and top view of a representative DFT simulation cell. Proximity to the 2D Ising ferromagnet Cr$_2$Ge$_2$Te$_6$ induces a magnetic moment in the carbon atoms.
			\textbf{(b)} We employ DFT to determine the proximitized magnetization of the graphene layer in a set of heterostructures with varying sizes, twist angles $\theta$, and interlayer distances $d_\mathrm{IL}$.
			Together with the structural/geometrical information encoded by the SOAP descriptor, the data set is then used to train an ML regression model to predict the proximitized magnetization in much larger structures, as indicated.
		}
		\label{fig:overview}
	\end{figure*}
	
	Proximity exchange coupling~\cite{Sierra_vdW_2021} has been
	studied theoretically~\cite{Yang_Proximity_2013, Lazic_Effective_2016, Zollner_Theory_2016, Zanolli2016, Zollner_bilayer_2021, Beer_proximity_2024, Zollner_Engineering_2022} and experimentally ~\cite{Beer_proximity_2024, Liu_Electrostatic_2013, Yang_electrostatically_2024}, showing sensitivity
	to interlayer distance~\cite{Yang_Proximity_2013, Zollner_Engineering_2022} and tunability by twisting~\cite{Zollner_Engineering_2022, Carr_twistronics_2017, Carr_electronic-structure_2020, Choi_Emergent_2022} and gating~\cite{Lazic_Effective_2016, Zollner_Theory_2016, Liu_Electrostatic_2013, Yang_electrostatically_2024}. Complementary to scanning tunneling microscopy~\cite{Qiu_visualizing_2021}, density functional theory (DFT) currently provides the only reliable means to investigate proximity effects at the atomic scale~\cite{Yang_Proximity_2013, Lazic_Effective_2016, Zollner_Theory_2016, Zanolli2016, Zollner_bilayer_2021, Zollner_Engineering_2022}, but its high computational cost limits studies to small supercells (a few hundred atoms), making predictions for arbitrary twist angles and large-scale systems infeasible. 
	As a result, phenomena inherently requiring large simulation cells, such as twist-dependent magnetic moiré patterns~\cite{Tong_2019, Carr2020}, as well as spatial variations from local structural inhomogeneities and the role of local stacking configurations (“registries”) in proximity effects \footnote{See Ref.~\cite{Zollner_Engineering_2022}: ``[...] there is a delicate balance in the orbital hybridization [...], which makes the exchange coupling highly sensitive to the atomic registry.'' 
		And in the supplementary material: ``The magnetic moments of the C atoms already give us a first hint that graphene experiences some proximity-induced magnetism from the CGT layer. [...] Here, the atomic registry should play a major role.''}, remain largely unexplored.

	In this work, we reveal pronounced atomic-scale fluctuations in proximity-induced magnetic moments in graphene stacked with a ferromagnetic semiconductor monolayer Cr$_2$Ge$_2$Te$_6$ (CGT). This behavior contrasts sharply with the spin polarization of Dirac electrons, which is locked to the pseudospin, leading to the following classification of proximity effects in graphene: (i) \emph{pseudospin-preserving}, producing a largely uniform and predictable texture captured by a lattice Hamiltonian; and (ii) \emph{pseudospin-breaking}, which are inherently local and defy simple atomistic modeling. Spin polarization at the Fermi level falls into type (i), when the Dirac states are only slightly perturbed~\cite{Zollner_Engineering_2022}. But whenever the 
	hybridization with the proximitizing layer is resonant---as in graphene/hBN/Co~\cite{Zollner_XXX} or, in our case, the deep $p_z$ band states in graphene/CGT affecting the carbon magnetic moments---the proximity effects are of type (ii). 
	
	How can we model non-perturbative, class (ii), proximity effects? Can we, as for class (i), go beyond DFT and develop a methodology to efficiently predict proximitized quantities, such as the magnetization for arbitrary stackings, by mapping the local atomic-scale geometry to the induced response? As we demonstrate here, machine learning (ML) provides a compelling, computationally efficient alternative to expensive DFT simulations. Our premise is simple: assuming that the magnetic moment of each carbon atom is fully determined by its local atomic environment, we train an ML model using atomic geometries as input and DFT-calculated magnetic moments as targets, see Fig.~\ref{fig:overview}. Remarkably, besides accurate predictions, this model also gives a measure of how local the proximity effect is.  
	Below, we apply it to twisted graphene/CGT, which features both proximity types (i) and (ii), and is experimentally relevant for spin transport ~\cite{Chau_CGT/Gr_2022, Sahani_giantMR_2025, Karpiak_2020}.
	We show, for example, that the magnetic moment of a carbon atom depends only on a small surrounding region of $\sim$2 nm$^2$. The ML model predicts large-scale magnetic textures and generalizes to much larger systems, enabling predictions for structures with millions of atoms, arbitrary twist angles, as well as other pseudospin-breaking proximity effects.

	\textit{Crystal structures.}---We consider van der Waals heterostructures composed of graphene and CGT with twist angles  from 0\degree~to 30\degree, sampled in steps of approximately 3\degree, following Ref.~\cite{Zollner_Engineering_2022}. To construct commensurate supercells, as shown in Fig.~\ref{fig:overview}, the monolayers are strained by up to 3\,\%. In addition to the equilibrium interlayer distance of $d_\mathrm{IL} = 3.55$\,\AA, we include configurations with altered separation of $\pm0.1$ and $\pm0.2$\,\AA. The dataset is further diversified by incorporating structures with lateral shifts between the layers.
	For details, see appendix.
	
	\textit{Machine Learning framework.}---We train our ML model on data from the plane-wave DFT code \textit{Quantum Espresso} \cite{Giannozzi_2009}. To ensure that the machine learning model captures only the intrinsic geometry–magnetization relationship (independent of arbitrary rotations or translations) we represent atomic environments using the Smooth Overlap of Atomic Positions (SOAP) descriptor~\cite{Bartok_2013} as implemented in the \textit{DScribe} Python library~\cite{dscribe, dscribe2}. It provides a rotationally and translationally invariant fingerprint of the local atomic structure. SOAP maps the atomic coordinates within the local environment (being specified by a cut-off radius $r_\mathrm{cut}$) onto a single column vector. The underlying machinery calculates overlaps between virtual atomic orbitals placed at the atomic positions of the input structure~\cite{Bartok_2013}.
	The best prediction accuracy is obtained with the highest tested description complexity (for quantitative results see appendix), suggesting that the induced magnetization arises from complex hybridization effects driven by subtle variations in local atomic and chemical environments.
	We note that the choice of descriptor is not unique; more advanced schemes such as the atomic cluster expansion~\cite{Drautz2019}, may match SOAP's accuracy with smaller orbital sets.
	
	While the SOAP vectors serve as input (feature) for our ML model, the output (target) is the induced magnetic moment. The model learns a mapping from SOAP descriptors of the local atomic environment to the DFT-computed magnetization, capturing complex, non-linear dependencies without requiring explicit physical modeling. We train a random forest regression model~\cite{Breiman2001} implemented via the \textit{scikit-learn} Python library~\cite{scikit-learn} on our DFT data. Random Forest regression is an ensemble learning method that predicts a continuous target variable by averaging the outputs of multiple decision trees, each trained on a random subset of the data and features. Other ML algorithms like gradient-boost regression achieve similar prediction accuracy. For a comparison and further technical details, see appendix.
	
	\begin{figure}[htbp]
		\centerline{\includegraphics[width=\linewidth]{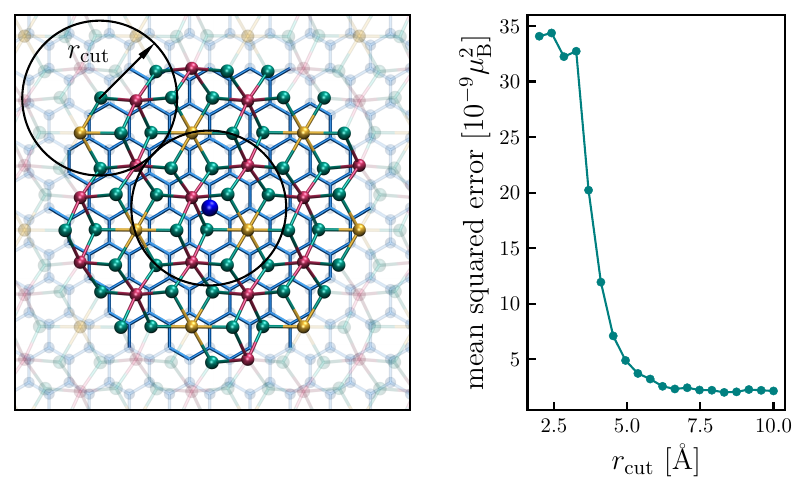}}
		\caption{The accuracy of the ML model is largely determined by the descriptor's cut-off radius $r_\mathrm{cut}$ which defines the local environment. 
			The prediction error saturates at $r_\mathrm{cut}=6$\,\AA.
		}
		\label{fig:locality}
	\end{figure}
	\textit{Locality of proximity-induced magnetization.}---A central assumption of this work, based on observations discussed in the appendix, is that the magnetization of a C atom in the CGT/Gr heterostructure is governed solely by its local atomic environment. To test this, we analyze the dependence of the prediction accuracy on the cut-off radius $r_\mathrm{cut}$ of the SOAP descriptor, which defines the spatial extent of the local environment. Due to the polynomial weighting scheme (see appendix), all atomic orbitals effectively vanish beyond $r_\mathrm{cut}$, meaning that overlap between two orbitals is only captured if the atoms are within a distance of $2 r_\mathrm{cut}$. 
	
	As shown in Fig.~\ref{fig:locality}, the mean squared error (MSE) of the predictions saturates for $r_\mathrm{cut} > 6$\,\AA, indicating that increasing the descriptor range beyond this point provides no additional information. We therefore conclude that the proximitized magnetic moment of each C atom is determined by its local atomic environment within 12\,\AA~containing about 100 CGT atoms.

	\textit{Verification within the DFT data set.}---We assess the accuracy of the ML model using testing data from our DFT dataset. The dataset is randomly split into a training set and a test set for evaluation, see appendix. In total, the dataset comprises 9498 samples, each consisting of a C atom, its local atomic environment, and the corresponding induced magnetic moment. Model performance is quantified using cross-validation on 10\,\% of the data, yielding MSE of $2 \times 10^{-9}\,\mu_\mathrm{B}^2$, corresponding to an average absolute prediction error of approximately $4.5 \times 10^{-5} \mu_\mathrm{B}$. Given average magnetic moments of $0.4\times 10^{-3} \mu_\mathrm{B}$, the deviation between predicted and DFT-computed magnetic moments is around 10\,\%.
	\begin{figure}[htbp]
		\centerline{\includegraphics[width=\linewidth]{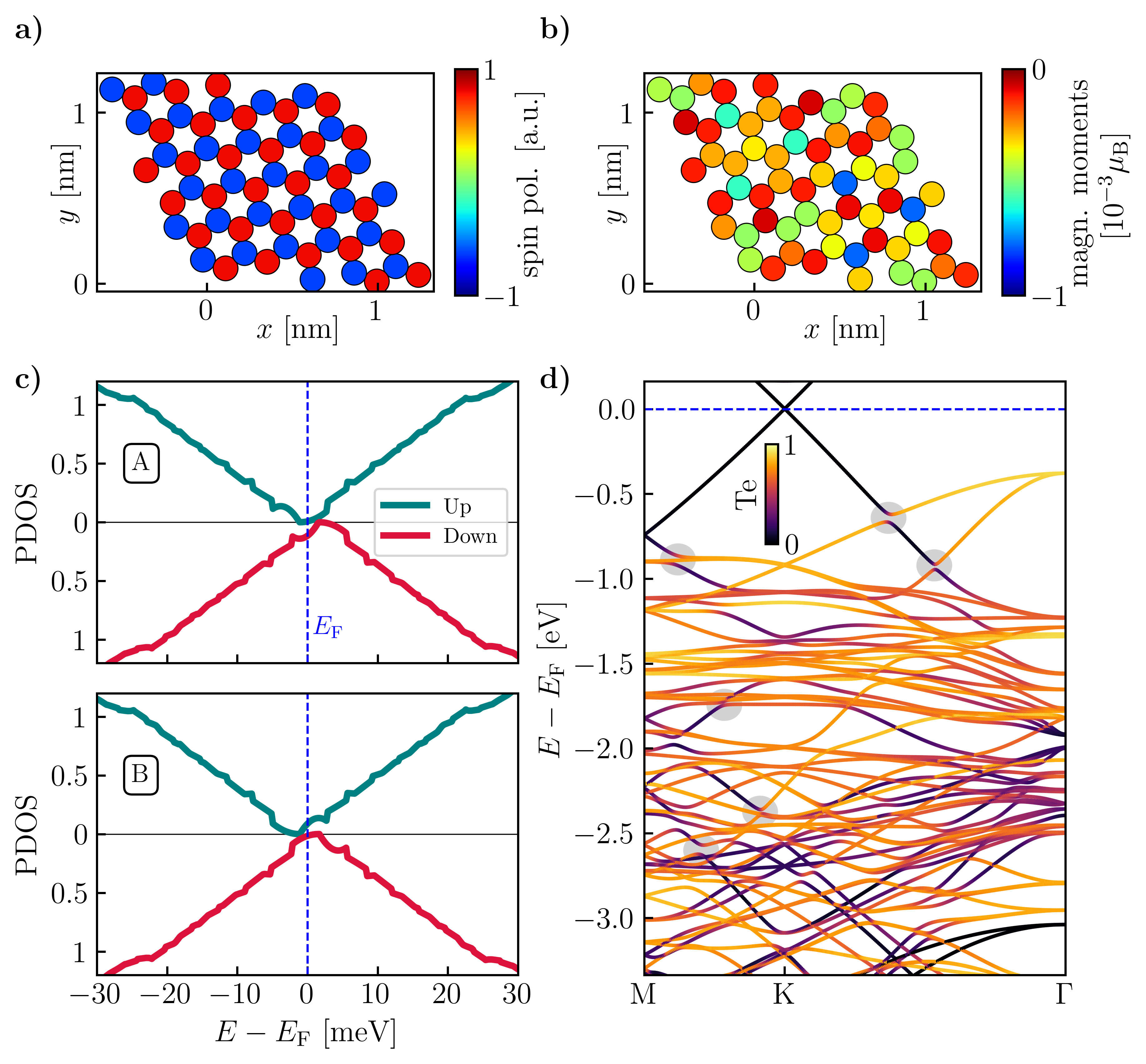}}
		\caption{Two types of proximity effects. Presented are DFT simulations
			for graphene/CGT at twist angle $\theta=8.948\degree$. \textbf{(a)} Sublattice-determined spin polarization $P(E_F)$ of the carbon atoms in the supercell. The spin polarization has only two values, locked to the pseudospin, being of type (i). The pseudospin character of $P(E_F)$ is further reflected in the calculated projected density of states (PDOS) around Fermi energy $E_F$ for the two indicated sublattices, shown in \textbf{(c)}. The curves are not smooth due to computational limitations. \textbf{(b)} Distribution of the carbon magnetic moment. The values are widely spread, showing its type (ii) pseudospin-breaking character.  
			\textbf{(d)} Band structure for spin-down bands with color-coded projections on Te states. Anti-crossings (gray circles) indicate resonant hybridization, affecting the local proximity magnetic moments, between Te valence states and graphene $p_z$ orbitals.
		}
		\label{fig:PDOS}
	\end{figure}
	
	\begin{figure*}[!ht]
		\centerline{\includegraphics[width=\linewidth]{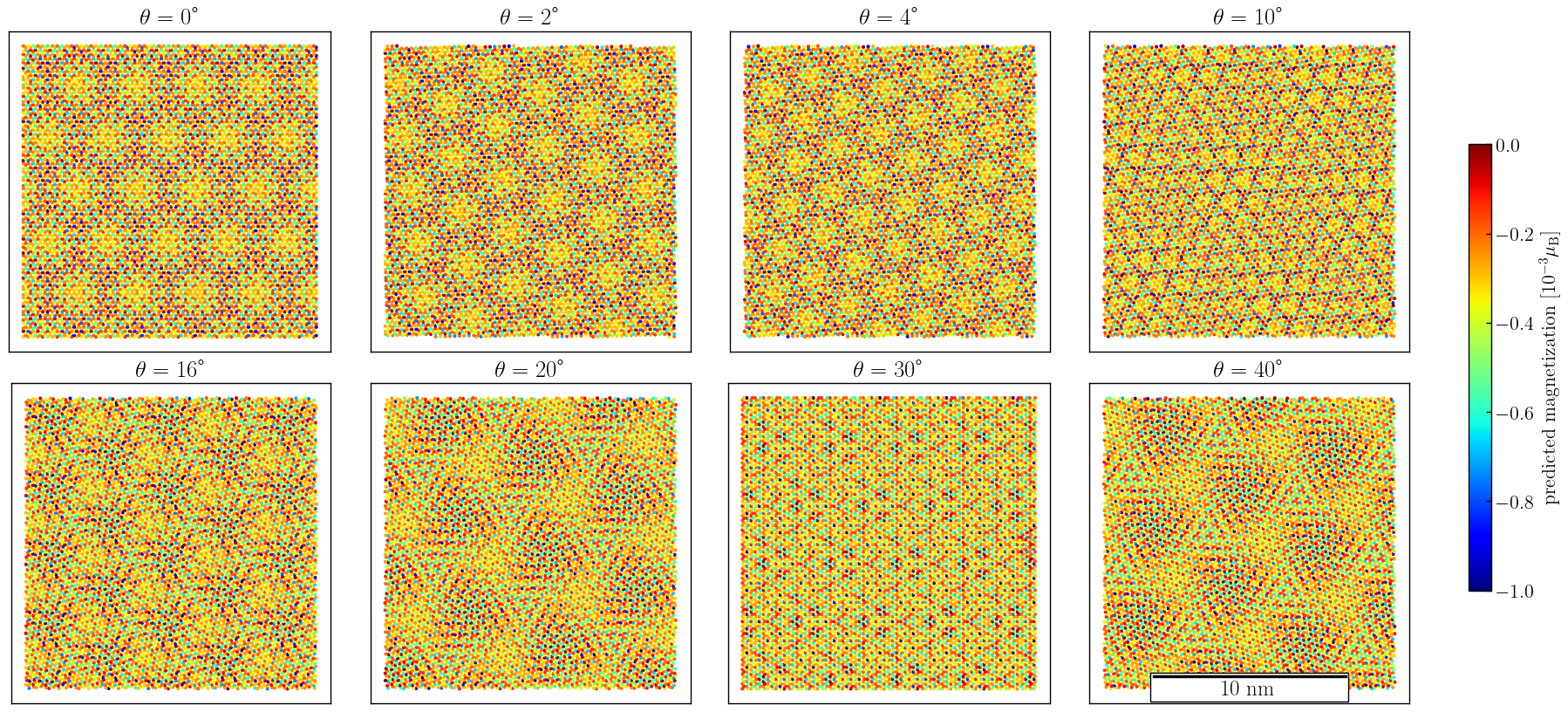}}
		\caption{Magnetic moiré patterns emerging in larger simulation cells for various twist angles $\theta$ as predicted by the ML model. The scale bar in the last panel holds for all panels. Due to the hexagonal shape of both material layers, the pattern repeats with a 60$\degree$ periodicity and is symmetric about $\theta=30\degree$. Thus, $\theta=20\degree$ corresponds to $\theta=40\degree$.
		}
		\label{fig:moire}
	\end{figure*}
	
	\textit{Functional dependency.}---Machine learning proves effective, but is it truly essential? To explore this, we investigated analytical models that relate atomic geometry to induced magnetization.
	A natural guess would be exponentially decaying interactions, modeled as $\sum_j \sum_{i} C_j\exp(-\kappa_j d_i^{(j)})$, where $d_i^{(j)}$ denotes the distance between the central C atom and neighboring atoms $i$ of type $j$, and $\kappa_j$ are decay parameters for each atom type $j=\{$Cr, Te, Ge$\}$. This form, however, reveals no meaningful correlation when all neighbors are considered. A weak correlation emerges only when restricting the model to Te atoms, which lie closest to the graphene layer. 
	Corresponding correlation plots are provided in the appendix.
	On the one hand, the correlation for Te atoms points towards an interaction between the directly adjacent atoms of both monolayers. On the other hand, the weak correlation (Pearson coefficient of 0.3) shows that the proximity-induced magnetization is highly sensitive to microscopic details of the atomic arrangement, which are complex to capture, further justifying the use of ML.
	
	\textit{Fermi level properties.}---An atom’s magnetic moment reflects the net spin polarization accumulated across all occupied electronic states up to the Fermi level. Since spin transport is governed by states at the Fermi level, we next focus on analyzing the spin-resolved properties specifically at this energy. 
	
	In order to resolve exchange splittings on the order of a few meV, sampling of the Brillouin zone (BZ) around the K point, where the graphene Dirac states cross the Fermi energy, requires a dense $k$-point mesh. Uniform $k$-grids in combination with the large simulation cells considered in this work are computationally hardly feasible. We have therefore implemented an adaptive $k$-point mesh, see appendix. This approach allows for a trade-off between sufficient sampling around K, while also delivering accurate states at lower energy and across the full BZ.
	In order to investigate the atom-resolved Fermi level properties, we apply this method to one exemplary structure (102 atoms), using $48\times48$ $k$-points across the BZ while a 15 times denser mesh is used around K.
	The results significantly outperform the uniform QE calculations (see appendix).
	
	To this end, after the electronic structure is obtained from DFT, we perform a projection onto a localized atomic basis and integrate across the BZ using a linear triangular method~\cite{Kurganskii_2D_BZ_integration} to obtain the projected density of states (PDOS) of each C atom, respectively. As shown in Fig.~\ref{fig:PDOS}, only two types of PDOS appear which can readily be associated with the sublattice of the graphene layer. This is a sharp contrast to the wide spread in magnetization among the atoms. Furthermore, while in this example the magnetic moments all align in the same direction, at the considered twist angle $\theta=8.948\degree$, the spin polarization is ferrimagnetic. Validating the simple Hamiltonian from Ref.~\cite{Zollner_Theory_2016, Zollner_Engineering_2022} which treats the sublattice of graphene separately, the PDOS is precisely mirrored about the Dirac point between sublattice A and B. The local atomic environment leaves the spin polarization at the Fermi level unchanged. Figure 3 thus nicely illustrates the dichotomy of proximity effects.

	\textit{Microscopic origin.}---In order to clarify the origins of the distribution of magnetic moments, we consider the band structure in Fig.~\ref{fig:PDOS}d.
	It shows clear anti-crossings (some marked by gray circles) between the graphene $p_z$ bands and -- as apparent upon projection on Te atoms -- Te-dominated bands of CGT.
	Projections on atom types Cr and Ge are provided in the appendix.
	
	For an atom-resolved analysis, we calculate the atom-projected spin polarization $P(E)= \mathrm{PDOS}_\mathrm{up}(E)-\mathrm{PDOS}_\mathrm{dn}(E)$ along the full energy range which gives the total magnetization of each atom by $m(E)=\int_{-\infty}^{E} P(E') dE'$ (see appendix). Close to the Fermi energy, $m(E)$ is virtually the same for each C atom. The difference in magnetization is mainly accumulated at the energy range from -3 to -0.5\,eV, confirming the observations from Fig.~\ref{fig:PDOS}d.
	
	\textit{Large scale results.}---Finally, we apply the ML model to predict magnetic textures of practically relevant system sizes. For this purpose, we combine completely unstrained layers of CGT and graphene and overlay them with varying twist angles $\theta$ considering the equilibrium $d_\mathrm{IL}$ of 3.55\,\AA. Most importantly, we find magnetic moiré patterns arising as a result of periodically recurring local environments of the heterostructures, see Fig.~\ref{fig:moire}. The extent as well as the quantitative fluctuations of these patterns is highly dependent on the twist angle $\theta$. The moiré patterns appear on scales which are far from feasible DFT calculations.
	
	Proximity-induced magnetism in twisted graphene/CGT heterostructures exhibits fluctuations on two distinct length scales. On the atomic scale, neighboring carbon atoms can experience markedly different induced magnetic moments, even within the same local stacking region. This variation arises from subtle changes in their atomic environments which directly affect orbital hybridization. Consequently, such fluctuations cannot be captured by effective moiré potentials that average over several unit cells~\cite{Hu_moire_2021}. On the moiré scale, additional modulation emerges from the global twist-dependent stacking pattern, giving rise to long-range periodicity in the induced magnetization. Capturing both scales simultaneously requires an approach sensitive to local atomic detail while remaining scalable to large systems—precisely the gap addressed by our machine learning framework.

	\textit{Conclusions.}---We have demonstrated that the proximity-induced magnetization in twisted graphene/CGT heterostructures is governed by the local atomic environment. Using a SOAP-based Random Forest model trained on an extensive DFT data set, we achieve accurate predictions of magnetic moments across a diverse set of structures. The model generalizes to large-scale systems and enables rapid exploration of the local magnetic landscape. While simple analytic models based on interatomic distances fail to capture the underlying physics, the machine learning approach successfully encodes complex geometric dependencies. Our results highlight the potential of data-driven methods for investigating proximity-induced properties such as magnetism, charge distribution, and spin–orbit coupling in van der Waals heterostructures, when the local atomic environment governs these effects.
	
	This project has received funding from the Deutsche Forschungsgemeinschaft (DFG, German Research Foundation) 
	SPP 2244 (Project No. 443416183), SFB 1277 (Project. No. 314695032), and the EU 2DSPIN-TECH Program (Graphene Flagship).
	
	\bibliography{my.bib}

\begin{thebibliography}{82}%
\makeatletter
\providecommand \@ifxundefined [1]{%
 \@ifx{#1\undefined}
}%
\providecommand \@ifnum [1]{%
 \ifnum #1\expandafter \@firstoftwo
 \else \expandafter \@secondoftwo
 \fi
}%
\providecommand \@ifx [1]{%
 \ifx #1\expandafter \@firstoftwo
 \else \expandafter \@secondoftwo
 \fi
}%
\providecommand \natexlab [1]{#1}%
\providecommand \enquote  [1]{``#1''}%
\providecommand \bibnamefont  [1]{#1}%
\providecommand \bibfnamefont [1]{#1}%
\providecommand \citenamefont [1]{#1}%
\providecommand \href@noop [0]{\@secondoftwo}%
\providecommand \href [0]{\begingroup \@sanitize@url \@href}%
\providecommand \@href[1]{\@@startlink{#1}\@@href}%
\providecommand \@@href[1]{\endgroup#1\@@endlink}%
\providecommand \@sanitize@url [0]{\catcode `\\12\catcode `\$12\catcode
  `\&12\catcode `\#12\catcode `\^12\catcode `\_12\catcode `\%12\relax}%
\providecommand \@@startlink[1]{}%
\providecommand \@@endlink[0]{}%
\providecommand \url  [0]{\begingroup\@sanitize@url \@url }%
\providecommand \@url [1]{\endgroup\@href {#1}{\urlprefix }}%
\providecommand \urlprefix  [0]{URL }%
\providecommand \Eprint [0]{\href }%
\providecommand \doibase [0]{https://doi.org/}%
\providecommand \selectlanguage [0]{\@gobble}%
\providecommand \bibinfo  [0]{\@secondoftwo}%
\providecommand \bibfield  [0]{\@secondoftwo}%
\providecommand \translation [1]{[#1]}%
\providecommand \BibitemOpen [0]{}%
\providecommand \bibitemStop [0]{}%
\providecommand \bibitemNoStop [0]{.\EOS\space}%
\providecommand \EOS [0]{\spacefactor3000\relax}%
\providecommand \BibitemShut  [1]{\csname bibitem#1\endcsname}%
\let\auto@bib@innerbib\@empty
\bibitem [{\citenamefont {Geim}\ and\ \citenamefont
  {Grigorieva}(2013)}]{Geim_vdW_2013}%
  \BibitemOpen
  \bibfield  {author} {\bibinfo {author} {\bibfnamefont {A.~K.}\ \bibnamefont
  {Geim}}\ and\ \bibinfo {author} {\bibfnamefont {I.~V.}\ \bibnamefont
  {Grigorieva}},\ }\href {https://doi.org/10.1038/nature12385} {\bibfield
  {journal} {\bibinfo  {journal} {Nature}\ }\textbf {\bibinfo {volume} {499}},\
  \bibinfo {pages} {419} (\bibinfo {year} {2013})}\BibitemShut {NoStop}%
\bibitem [{\citenamefont {Sierra}\ \emph {et~al.}(2021)\citenamefont {Sierra},
  \citenamefont {Fabian}, \citenamefont {Kawakami}, \citenamefont {Roche},\
  and\ \citenamefont {Valenzuela}}]{Sierra_vdW_2021}%
  \BibitemOpen
  \bibfield  {author} {\bibinfo {author} {\bibfnamefont {J.~F.}\ \bibnamefont
  {Sierra}}, \bibinfo {author} {\bibfnamefont {J.}~\bibnamefont {Fabian}},
  \bibinfo {author} {\bibfnamefont {R.~K.}\ \bibnamefont {Kawakami}}, \bibinfo
  {author} {\bibfnamefont {S.}~\bibnamefont {Roche}},\ and\ \bibinfo {author}
  {\bibfnamefont {S.~O.}\ \bibnamefont {Valenzuela}},\ }\href
  {https://doi.org/10.1038/s41565-021-00936-x} {\bibfield  {journal} {\bibinfo
  {journal} {Nature Nanotechnology}\ }\textbf {\bibinfo {volume} {16}},\
  \bibinfo {pages} {856} (\bibinfo {year} {2021})}\BibitemShut {NoStop}%
\bibitem [{\citenamefont {Novoselov}\ \emph {et~al.}(2016)\citenamefont
  {Novoselov}, \citenamefont {Mishchenko}, \citenamefont {Carvalho},\ and\
  \citenamefont {Neto}}]{Novoselov_2D_2016}%
  \BibitemOpen
  \bibfield  {author} {\bibinfo {author} {\bibfnamefont {K.~S.}\ \bibnamefont
  {Novoselov}}, \bibinfo {author} {\bibfnamefont {A.}~\bibnamefont
  {Mishchenko}}, \bibinfo {author} {\bibfnamefont {A.}~\bibnamefont
  {Carvalho}},\ and\ \bibinfo {author} {\bibfnamefont {A.~H.~C.}\ \bibnamefont
  {Neto}},\ }\href {https://doi.org/10.1126/science.aac9439} {\bibfield
  {journal} {\bibinfo  {journal} {Science}\ }\textbf {\bibinfo {volume}
  {353}},\ \bibinfo {pages} {aac9439} (\bibinfo {year} {2016})}\BibitemShut
  {NoStop}%
\bibitem [{\citenamefont {Carr}\ \emph {et~al.}(2017)\citenamefont {Carr},
  \citenamefont {Massatt}, \citenamefont {Fang}, \citenamefont {Cazeaux},
  \citenamefont {Luskin},\ and\ \citenamefont
  {Kaxiras}}]{Carr_twistronics_2017}%
  \BibitemOpen
  \bibfield  {author} {\bibinfo {author} {\bibfnamefont {S.}~\bibnamefont
  {Carr}}, \bibinfo {author} {\bibfnamefont {D.}~\bibnamefont {Massatt}},
  \bibinfo {author} {\bibfnamefont {S.}~\bibnamefont {Fang}}, \bibinfo {author}
  {\bibfnamefont {P.}~\bibnamefont {Cazeaux}}, \bibinfo {author} {\bibfnamefont
  {M.}~\bibnamefont {Luskin}},\ and\ \bibinfo {author} {\bibfnamefont
  {E.}~\bibnamefont {Kaxiras}},\ }\href
  {https://doi.org/10.1103/PhysRevB.95.075420} {\bibfield  {journal} {\bibinfo
  {journal} {Phys. Rev. B}\ }\textbf {\bibinfo {volume} {95}},\ \bibinfo
  {pages} {075420} (\bibinfo {year} {2017})}\BibitemShut {NoStop}%
\bibitem [{\citenamefont {Carr}\ \emph
  {et~al.}(2020{\natexlab{a}})\citenamefont {Carr}, \citenamefont {Fang},\ and\
  \citenamefont {Kaxiras}}]{Carr_electronic-structure_2020}%
  \BibitemOpen
  \bibfield  {author} {\bibinfo {author} {\bibfnamefont {S.}~\bibnamefont
  {Carr}}, \bibinfo {author} {\bibfnamefont {S.}~\bibnamefont {Fang}},\ and\
  \bibinfo {author} {\bibfnamefont {E.}~\bibnamefont {Kaxiras}},\ }\href
  {https://doi.org/10.1038/s41578-020-0214-0} {\bibfield  {journal} {\bibinfo
  {journal} {Nature Reviews Materials}\ }\textbf {\bibinfo {volume} {5}},\
  \bibinfo {pages} {748} (\bibinfo {year} {2020}{\natexlab{a}})}\BibitemShut
  {NoStop}%
\bibitem [{\citenamefont {Hennighausen}\ and\ \citenamefont
  {Kar}(2021)}]{Hennighausen_twistronics_2021}%
  \BibitemOpen
  \bibfield  {author} {\bibinfo {author} {\bibfnamefont {Z.}~\bibnamefont
  {Hennighausen}}\ and\ \bibinfo {author} {\bibfnamefont {S.}~\bibnamefont
  {Kar}},\ }\href {https://doi.org/10.1088/2516-1075/abd957} {\bibfield
  {journal} {\bibinfo  {journal} {Electronic Structure}\ }\textbf {\bibinfo
  {volume} {3}},\ \bibinfo {pages} {014004} (\bibinfo {year}
  {2021})}\BibitemShut {NoStop}%
\bibitem [{\citenamefont {Ribeiro-Palau}\ \emph {et~al.}(2018)\citenamefont
  {Ribeiro-Palau}, \citenamefont {Zhang}, \citenamefont {Watanabe},
  \citenamefont {Taniguchi}, \citenamefont {Hone},\ and\ \citenamefont
  {Dean}}]{Ribeiro-Palau_twistable_2018}%
  \BibitemOpen
  \bibfield  {author} {\bibinfo {author} {\bibfnamefont {R.}~\bibnamefont
  {Ribeiro-Palau}}, \bibinfo {author} {\bibfnamefont {C.}~\bibnamefont
  {Zhang}}, \bibinfo {author} {\bibfnamefont {K.}~\bibnamefont {Watanabe}},
  \bibinfo {author} {\bibfnamefont {T.}~\bibnamefont {Taniguchi}}, \bibinfo
  {author} {\bibfnamefont {J.}~\bibnamefont {Hone}},\ and\ \bibinfo {author}
  {\bibfnamefont {C.~R.}\ \bibnamefont {Dean}},\ }\href
  {https://doi.org/10.1126/science.aat6981} {\bibfield  {journal} {\bibinfo
  {journal} {Science}\ }\textbf {\bibinfo {volume} {361}},\ \bibinfo {pages}
  {690} (\bibinfo {year} {2018})}\BibitemShut {NoStop}%
\bibitem [{\citenamefont {\ifmmode \check{Z}\else
  \v{Z}\fi{}uti\ifmmode~\acute{c}\else \'{c}\fi{}}\ \emph
  {et~al.}(2004)\citenamefont {\ifmmode \check{Z}\else
  \v{Z}\fi{}uti\ifmmode~\acute{c}\else \'{c}\fi{}}, \citenamefont {Fabian},\
  and\ \citenamefont {Das~Sarma}}]{Zutic_RMP_2004}%
  \BibitemOpen
  \bibfield  {author} {\bibinfo {author} {\bibfnamefont {I.}~\bibnamefont
  {\ifmmode \check{Z}\else \v{Z}\fi{}uti\ifmmode~\acute{c}\else \'{c}\fi{}}},
  \bibinfo {author} {\bibfnamefont {J.}~\bibnamefont {Fabian}},\ and\ \bibinfo
  {author} {\bibfnamefont {S.}~\bibnamefont {Das~Sarma}},\ }\href
  {https://doi.org/10.1103/RevModPhys.76.323} {\bibfield  {journal} {\bibinfo
  {journal} {Rev. Mod. Phys.}\ }\textbf {\bibinfo {volume} {76}},\ \bibinfo
  {pages} {323} (\bibinfo {year} {2004})}\BibitemShut {NoStop}%
\bibitem [{\citenamefont {Moriya}\ \emph {et~al.}(2020)\citenamefont {Moriya},
  \citenamefont {Yabuki},\ and\ \citenamefont
  {Machida}}]{Moriya_Superconducting_2020}%
  \BibitemOpen
  \bibfield  {author} {\bibinfo {author} {\bibfnamefont {R.}~\bibnamefont
  {Moriya}}, \bibinfo {author} {\bibfnamefont {N.}~\bibnamefont {Yabuki}},\
  and\ \bibinfo {author} {\bibfnamefont {T.}~\bibnamefont {Machida}},\ }\href
  {https://doi.org/10.1103/PhysRevB.101.054503} {\bibfield  {journal} {\bibinfo
   {journal} {Phys. Rev. B}\ }\textbf {\bibinfo {volume} {101}},\ \bibinfo
  {pages} {054503} (\bibinfo {year} {2020})}\BibitemShut {NoStop}%
\bibitem [{\citenamefont {Han}\ \emph {et~al.}(2021)\citenamefont {Han},
  \citenamefont {Ling}, \citenamefont {Liu}, \citenamefont {Li}, \citenamefont
  {Zhang},\ and\ \citenamefont {Wang}}]{Han_superconducting_2021}%
  \BibitemOpen
  \bibfield  {author} {\bibinfo {author} {\bibfnamefont {H.}~\bibnamefont
  {Han}}, \bibinfo {author} {\bibfnamefont {J.}~\bibnamefont {Ling}}, \bibinfo
  {author} {\bibfnamefont {W.}~\bibnamefont {Liu}}, \bibinfo {author}
  {\bibfnamefont {H.}~\bibnamefont {Li}}, \bibinfo {author} {\bibfnamefont
  {C.}~\bibnamefont {Zhang}},\ and\ \bibinfo {author} {\bibfnamefont
  {J.}~\bibnamefont {Wang}},\ }\href {https://doi.org/10.1063/5.0051968}
  {\bibfield  {journal} {\bibinfo  {journal} {Applied Physics Letters}\
  }\textbf {\bibinfo {volume} {118}},\ \bibinfo {pages} {253101} (\bibinfo
  {year} {2021})}\BibitemShut {NoStop}%
\bibitem [{\citenamefont {Gmitra}\ and\ \citenamefont
  {Fabian}(2015)}]{Gmitra_2015}%
  \BibitemOpen
  \bibfield  {author} {\bibinfo {author} {\bibfnamefont {M.}~\bibnamefont
  {Gmitra}}\ and\ \bibinfo {author} {\bibfnamefont {J.}~\bibnamefont
  {Fabian}},\ }\href {https://doi.org/10.1103/PhysRevB.92.155403} {\bibfield
  {journal} {\bibinfo  {journal} {Phys. Rev. B}\ }\textbf {\bibinfo {volume}
  {92}},\ \bibinfo {pages} {155403} (\bibinfo {year} {2015})}\BibitemShut
  {NoStop}%
\bibitem [{\citenamefont {Gmitra}\ \emph {et~al.}(2016)\citenamefont {Gmitra},
  \citenamefont {Kochan}, \citenamefont {H\"ogl},\ and\ \citenamefont
  {Fabian}}]{Gmitra_2016}%
  \BibitemOpen
  \bibfield  {author} {\bibinfo {author} {\bibfnamefont {M.}~\bibnamefont
  {Gmitra}}, \bibinfo {author} {\bibfnamefont {D.}~\bibnamefont {Kochan}},
  \bibinfo {author} {\bibfnamefont {P.}~\bibnamefont {H\"ogl}},\ and\ \bibinfo
  {author} {\bibfnamefont {J.}~\bibnamefont {Fabian}},\ }\href
  {https://doi.org/10.1103/PhysRevB.93.155104} {\bibfield  {journal} {\bibinfo
  {journal} {Phys. Rev. B}\ }\textbf {\bibinfo {volume} {93}},\ \bibinfo
  {pages} {155104} (\bibinfo {year} {2016})}\BibitemShut {NoStop}%
\bibitem [{\citenamefont {Khokhriakov}\ \emph {et~al.}(2018)\citenamefont
  {Khokhriakov}, \citenamefont {Cummings}, \citenamefont {Song}, \citenamefont
  {Vila}, \citenamefont {Karpiak}, \citenamefont {Dankert}, \citenamefont
  {Roche},\ and\ \citenamefont {Dash}}]{Khokhriakov_2018}%
  \BibitemOpen
  \bibfield  {author} {\bibinfo {author} {\bibfnamefont {D.}~\bibnamefont
  {Khokhriakov}}, \bibinfo {author} {\bibfnamefont {A.~W.}\ \bibnamefont
  {Cummings}}, \bibinfo {author} {\bibfnamefont {K.}~\bibnamefont {Song}},
  \bibinfo {author} {\bibfnamefont {M.}~\bibnamefont {Vila}}, \bibinfo {author}
  {\bibfnamefont {B.}~\bibnamefont {Karpiak}}, \bibinfo {author} {\bibfnamefont
  {A.}~\bibnamefont {Dankert}}, \bibinfo {author} {\bibfnamefont
  {S.}~\bibnamefont {Roche}},\ and\ \bibinfo {author} {\bibfnamefont {S.~P.}\
  \bibnamefont {Dash}},\ }\href {https://doi.org/10.1126/sciadv.aat9349}
  {\bibfield  {journal} {\bibinfo  {journal} {Science Advances}\ }\textbf
  {\bibinfo {volume} {4}},\ \bibinfo {pages} {eaat9349} (\bibinfo {year}
  {2018})}\BibitemShut {NoStop}%
\bibitem [{\citenamefont {Zollner}\ and\ \citenamefont
  {Fabian}(2019)}]{Zollner_Single_2019}%
  \BibitemOpen
  \bibfield  {author} {\bibinfo {author} {\bibfnamefont {K.}~\bibnamefont
  {Zollner}}\ and\ \bibinfo {author} {\bibfnamefont {J.}~\bibnamefont
  {Fabian}},\ }\href {https://doi.org/10.1103/PhysRevB.100.165141} {\bibfield
  {journal} {\bibinfo  {journal} {Phys. Rev. B}\ }\textbf {\bibinfo {volume}
  {100}},\ \bibinfo {pages} {165141} (\bibinfo {year} {2019})}\BibitemShut
  {NoStop}%
\bibitem [{\citenamefont {Khokhriakov}\ \emph {et~al.}(2020)\citenamefont
  {Khokhriakov}, \citenamefont {Hoque}, \citenamefont {Karpiak},\ and\
  \citenamefont {Dash}}]{Khokhriakov2020}%
  \BibitemOpen
  \bibfield  {author} {\bibinfo {author} {\bibfnamefont {D.}~\bibnamefont
  {Khokhriakov}}, \bibinfo {author} {\bibfnamefont {A.~M.}\ \bibnamefont
  {Hoque}}, \bibinfo {author} {\bibfnamefont {B.}~\bibnamefont {Karpiak}},\
  and\ \bibinfo {author} {\bibfnamefont {S.~P.}\ \bibnamefont {Dash}},\ }\href
  {https://doi.org/10.1038/s41467-020-17481-1} {\bibfield  {journal} {\bibinfo
  {journal} {Nature Communications}\ }\textbf {\bibinfo {volume} {11}},\
  \bibinfo {pages} {3657} (\bibinfo {year} {2020})}\BibitemShut {NoStop}%
\bibitem [{\citenamefont {Karpiak}\ \emph {et~al.}(2019)\citenamefont
  {Karpiak}, \citenamefont {Cummings}, \citenamefont {Zollner}, \citenamefont
  {Vila}, \citenamefont {Khokhriakov}, \citenamefont {Hoque}, \citenamefont
  {Dankert}, \citenamefont {Svedlindh}, \citenamefont {Fabian}, \citenamefont
  {Roche},\ and\ \citenamefont {Dash}}]{Karpiak_2020}%
  \BibitemOpen
  \bibfield  {author} {\bibinfo {author} {\bibfnamefont {B.}~\bibnamefont
  {Karpiak}}, \bibinfo {author} {\bibfnamefont {A.~W.}\ \bibnamefont
  {Cummings}}, \bibinfo {author} {\bibfnamefont {K.}~\bibnamefont {Zollner}},
  \bibinfo {author} {\bibfnamefont {M.}~\bibnamefont {Vila}}, \bibinfo {author}
  {\bibfnamefont {D.}~\bibnamefont {Khokhriakov}}, \bibinfo {author}
  {\bibfnamefont {A.~M.}\ \bibnamefont {Hoque}}, \bibinfo {author}
  {\bibfnamefont {A.}~\bibnamefont {Dankert}}, \bibinfo {author} {\bibfnamefont
  {P.}~\bibnamefont {Svedlindh}}, \bibinfo {author} {\bibfnamefont
  {J.}~\bibnamefont {Fabian}}, \bibinfo {author} {\bibfnamefont
  {S.}~\bibnamefont {Roche}},\ and\ \bibinfo {author} {\bibfnamefont {S.~P.}\
  \bibnamefont {Dash}},\ }\href {https://doi.org/10.1088/2053-1583/ab5915}
  {\bibfield  {journal} {\bibinfo  {journal} {2D Materials}\ }\textbf {\bibinfo
  {volume} {7}},\ \bibinfo {pages} {015026} (\bibinfo {year}
  {2019})}\BibitemShut {NoStop}%
\bibitem [{\citenamefont {Zihlmann}\ \emph {et~al.}(2018)\citenamefont
  {Zihlmann}, \citenamefont {Cummings}, \citenamefont {Garcia}, \citenamefont
  {Kedves}, \citenamefont {Watanabe}, \citenamefont {Taniguchi}, \citenamefont
  {Sch\"onenberger},\ and\ \citenamefont {Makk}}]{Zihlmann_2018}%
  \BibitemOpen
  \bibfield  {author} {\bibinfo {author} {\bibfnamefont {S.}~\bibnamefont
  {Zihlmann}}, \bibinfo {author} {\bibfnamefont {A.~W.}\ \bibnamefont
  {Cummings}}, \bibinfo {author} {\bibfnamefont {J.~H.}\ \bibnamefont
  {Garcia}}, \bibinfo {author} {\bibfnamefont {M.}~\bibnamefont {Kedves}},
  \bibinfo {author} {\bibfnamefont {K.}~\bibnamefont {Watanabe}}, \bibinfo
  {author} {\bibfnamefont {T.}~\bibnamefont {Taniguchi}}, \bibinfo {author}
  {\bibfnamefont {C.}~\bibnamefont {Sch\"onenberger}},\ and\ \bibinfo {author}
  {\bibfnamefont {P.}~\bibnamefont {Makk}},\ }\href
  {https://doi.org/10.1103/PhysRevB.97.075434} {\bibfield  {journal} {\bibinfo
  {journal} {Phys. Rev. B}\ }\textbf {\bibinfo {volume} {97}},\ \bibinfo
  {pages} {075434} (\bibinfo {year} {2018})}\BibitemShut {NoStop}%
\bibitem [{\citenamefont {Song}\ \emph {et~al.}(2018)\citenamefont {Song},
  \citenamefont {Soriano}, \citenamefont {Cummings}, \citenamefont {Robles},
  \citenamefont {Ordej{\'o}n},\ and\ \citenamefont {Roche}}]{Song2018}%
  \BibitemOpen
  \bibfield  {author} {\bibinfo {author} {\bibfnamefont {K.}~\bibnamefont
  {Song}}, \bibinfo {author} {\bibfnamefont {D.}~\bibnamefont {Soriano}},
  \bibinfo {author} {\bibfnamefont {A.~W.}\ \bibnamefont {Cummings}}, \bibinfo
  {author} {\bibfnamefont {R.}~\bibnamefont {Robles}}, \bibinfo {author}
  {\bibfnamefont {P.}~\bibnamefont {Ordej{\'o}n}},\ and\ \bibinfo {author}
  {\bibfnamefont {S.}~\bibnamefont {Roche}},\ }\href
  {https://doi.org/10.1021/acs.nanolett.7b05482} {\bibfield  {journal}
  {\bibinfo  {journal} {Nano Letters}\ }\textbf {\bibinfo {volume} {18}},\
  \bibinfo {pages} {2033} (\bibinfo {year} {2018})}\BibitemShut {NoStop}%
\bibitem [{\citenamefont {Safeer}\ \emph {et~al.}(2019)\citenamefont {Safeer},
  \citenamefont {Ingla-Ayn{\'e}s}, \citenamefont {Herling}, \citenamefont
  {Garcia}, \citenamefont {Vila}, \citenamefont {Ontoso}, \citenamefont
  {Calvo}, \citenamefont {Roche}, \citenamefont {Hueso},\ and\ \citenamefont
  {Casanova}}]{Safeer2019}%
  \BibitemOpen
  \bibfield  {author} {\bibinfo {author} {\bibfnamefont {C.~K.}\ \bibnamefont
  {Safeer}}, \bibinfo {author} {\bibfnamefont {J.}~\bibnamefont
  {Ingla-Ayn{\'e}s}}, \bibinfo {author} {\bibfnamefont {F.}~\bibnamefont
  {Herling}}, \bibinfo {author} {\bibfnamefont {J.~H.}\ \bibnamefont {Garcia}},
  \bibinfo {author} {\bibfnamefont {M.}~\bibnamefont {Vila}}, \bibinfo {author}
  {\bibfnamefont {N.}~\bibnamefont {Ontoso}}, \bibinfo {author} {\bibfnamefont
  {M.~R.}\ \bibnamefont {Calvo}}, \bibinfo {author} {\bibfnamefont
  {S.}~\bibnamefont {Roche}}, \bibinfo {author} {\bibfnamefont {L.~E.}\
  \bibnamefont {Hueso}},\ and\ \bibinfo {author} {\bibfnamefont
  {F.}~\bibnamefont {Casanova}},\ }\href
  {https://doi.org/10.1021/acs.nanolett.8b04368} {\bibfield  {journal}
  {\bibinfo  {journal} {Nano Letters}\ }\textbf {\bibinfo {volume} {19}},\
  \bibinfo {pages} {1074} (\bibinfo {year} {2019})}\BibitemShut {NoStop}%
\bibitem [{\citenamefont {Herling}\ \emph {et~al.}(2020)\citenamefont
  {Herling}, \citenamefont {Safeer}, \citenamefont {Ingla-Aynés},
  \citenamefont {Ontoso}, \citenamefont {Hueso},\ and\ \citenamefont
  {Casanova}}]{Herling2020}%
  \BibitemOpen
  \bibfield  {author} {\bibinfo {author} {\bibfnamefont {F.}~\bibnamefont
  {Herling}}, \bibinfo {author} {\bibfnamefont {C.~K.}\ \bibnamefont {Safeer}},
  \bibinfo {author} {\bibfnamefont {J.}~\bibnamefont {Ingla-Aynés}}, \bibinfo
  {author} {\bibfnamefont {N.}~\bibnamefont {Ontoso}}, \bibinfo {author}
  {\bibfnamefont {L.~E.}\ \bibnamefont {Hueso}},\ and\ \bibinfo {author}
  {\bibfnamefont {F.}~\bibnamefont {Casanova}},\ }\href
  {https://doi.org/10.1063/5.0006101} {\bibfield  {journal} {\bibinfo
  {journal} {APL Materials}\ }\textbf {\bibinfo {volume} {8}},\ \bibinfo
  {pages} {071103} (\bibinfo {year} {2020})}\BibitemShut {NoStop}%
\bibitem [{\citenamefont {Pezo}\ \emph {et~al.}(2021)\citenamefont {Pezo},
  \citenamefont {Zanolli}, \citenamefont {Wittemeier}, \citenamefont
  {Ordejón}, \citenamefont {Fazzio}, \citenamefont {Roche},\ and\
  \citenamefont {Garcia}}]{Pezo_2022}%
  \BibitemOpen
  \bibfield  {author} {\bibinfo {author} {\bibfnamefont {A.}~\bibnamefont
  {Pezo}}, \bibinfo {author} {\bibfnamefont {Z.}~\bibnamefont {Zanolli}},
  \bibinfo {author} {\bibfnamefont {N.}~\bibnamefont {Wittemeier}}, \bibinfo
  {author} {\bibfnamefont {P.}~\bibnamefont {Ordejón}}, \bibinfo {author}
  {\bibfnamefont {A.}~\bibnamefont {Fazzio}}, \bibinfo {author} {\bibfnamefont
  {S.}~\bibnamefont {Roche}},\ and\ \bibinfo {author} {\bibfnamefont {J.~H.}\
  \bibnamefont {Garcia}},\ }\href {https://doi.org/10.1088/2053-1583/ac3378}
  {\bibfield  {journal} {\bibinfo  {journal} {2D Materials}\ }\textbf {\bibinfo
  {volume} {9}},\ \bibinfo {pages} {015008} (\bibinfo {year}
  {2021})}\BibitemShut {NoStop}%
\bibitem [{\citenamefont {Manzeli}\ \emph {et~al.}(2017)\citenamefont
  {Manzeli}, \citenamefont {Ovchinnikov}, \citenamefont {Pasquier},
  \citenamefont {Yazyev},\ and\ \citenamefont {Kis}}]{Manzeli_2DTMDC_2017}%
  \BibitemOpen
  \bibfield  {author} {\bibinfo {author} {\bibfnamefont {S.}~\bibnamefont
  {Manzeli}}, \bibinfo {author} {\bibfnamefont {D.}~\bibnamefont
  {Ovchinnikov}}, \bibinfo {author} {\bibfnamefont {D.}~\bibnamefont
  {Pasquier}}, \bibinfo {author} {\bibfnamefont {O.~V.}\ \bibnamefont
  {Yazyev}},\ and\ \bibinfo {author} {\bibfnamefont {A.}~\bibnamefont {Kis}},\
  }\href {https://doi.org/10.1038/natrevmats.2017.33} {\bibfield  {journal}
  {\bibinfo  {journal} {Nature Reviews Materials}\ }\textbf {\bibinfo {volume}
  {2}},\ \bibinfo {pages} {17033} (\bibinfo {year} {2017})}\BibitemShut
  {NoStop}%
\bibitem [{\citenamefont {Yang}\ \emph {et~al.}(2013)\citenamefont {Yang},
  \citenamefont {Hallal}, \citenamefont {Terrade}, \citenamefont {Waintal},
  \citenamefont {Roche},\ and\ \citenamefont {Chshiev}}]{Yang_Proximity_2013}%
  \BibitemOpen
  \bibfield  {author} {\bibinfo {author} {\bibfnamefont {H.~X.}\ \bibnamefont
  {Yang}}, \bibinfo {author} {\bibfnamefont {A.}~\bibnamefont {Hallal}},
  \bibinfo {author} {\bibfnamefont {D.}~\bibnamefont {Terrade}}, \bibinfo
  {author} {\bibfnamefont {X.}~\bibnamefont {Waintal}}, \bibinfo {author}
  {\bibfnamefont {S.}~\bibnamefont {Roche}},\ and\ \bibinfo {author}
  {\bibfnamefont {M.}~\bibnamefont {Chshiev}},\ }\href
  {https://doi.org/10.1103/PhysRevLett.110.046603} {\bibfield  {journal}
  {\bibinfo  {journal} {Phys. Rev. Lett.}\ }\textbf {\bibinfo {volume} {110}},\
  \bibinfo {pages} {046603} (\bibinfo {year} {2013})}\BibitemShut {NoStop}%
\bibitem [{\citenamefont {Hallal}\ \emph {et~al.}(2017)\citenamefont {Hallal},
  \citenamefont {Ibrahim}, \citenamefont {Yang}, \citenamefont {Roche},\ and\
  \citenamefont {Chshiev}}]{Hallal_tailoring_2017}%
  \BibitemOpen
  \bibfield  {author} {\bibinfo {author} {\bibfnamefont {A.}~\bibnamefont
  {Hallal}}, \bibinfo {author} {\bibfnamefont {F.}~\bibnamefont {Ibrahim}},
  \bibinfo {author} {\bibfnamefont {H.}~\bibnamefont {Yang}}, \bibinfo {author}
  {\bibfnamefont {S.}~\bibnamefont {Roche}},\ and\ \bibinfo {author}
  {\bibfnamefont {M.}~\bibnamefont {Chshiev}},\ }\href
  {https://doi.org/10.1088/2053-1583/aa6663} {\bibfield  {journal} {\bibinfo
  {journal} {2D Materials}\ }\textbf {\bibinfo {volume} {4}},\ \bibinfo {pages}
  {025074} (\bibinfo {year} {2017})}\BibitemShut {NoStop}%
\bibitem [{\citenamefont {Zollner}\ \emph {et~al.}(2019)\citenamefont
  {Zollner}, \citenamefont {Faria~Junior},\ and\ \citenamefont
  {Fabian}}]{Zollner_Proximity_2019}%
  \BibitemOpen
  \bibfield  {author} {\bibinfo {author} {\bibfnamefont {K.}~\bibnamefont
  {Zollner}}, \bibinfo {author} {\bibfnamefont {P.~E.}\ \bibnamefont
  {Faria~Junior}},\ and\ \bibinfo {author} {\bibfnamefont {J.}~\bibnamefont
  {Fabian}},\ }\href {https://doi.org/10.1103/PhysRevB.100.085128} {\bibfield
  {journal} {\bibinfo  {journal} {Phys. Rev. B}\ }\textbf {\bibinfo {volume}
  {100}},\ \bibinfo {pages} {085128} (\bibinfo {year} {2019})}\BibitemShut
  {NoStop}%
\bibitem [{\citenamefont {Zollner}\ and\ \citenamefont
  {Fabian}(2022)}]{Zollner_Engineering_2022}%
  \BibitemOpen
  \bibfield  {author} {\bibinfo {author} {\bibfnamefont {K.}~\bibnamefont
  {Zollner}}\ and\ \bibinfo {author} {\bibfnamefont {J.}~\bibnamefont
  {Fabian}},\ }\href {https://doi.org/10.1103/PhysRevLett.128.106401}
  {\bibfield  {journal} {\bibinfo  {journal} {Phys. Rev. Lett.}\ }\textbf
  {\bibinfo {volume} {128}},\ \bibinfo {pages} {106401} (\bibinfo {year}
  {2022})}\BibitemShut {NoStop}%
\bibitem [{\citenamefont {Chau}\ \emph {et~al.}(2022)\citenamefont {Chau},
  \citenamefont {Hong}, \citenamefont {Kang},\ and\ \citenamefont
  {Suh}}]{Chau_CGT/Gr_2022}%
  \BibitemOpen
  \bibfield  {author} {\bibinfo {author} {\bibfnamefont {T.~K.}\ \bibnamefont
  {Chau}}, \bibinfo {author} {\bibfnamefont {S.~J.}\ \bibnamefont {Hong}},
  \bibinfo {author} {\bibfnamefont {H.}~\bibnamefont {Kang}},\ and\ \bibinfo
  {author} {\bibfnamefont {D.}~\bibnamefont {Suh}},\ }\href
  {https://doi.org/10.1038/s41535-022-00435-9} {\bibfield  {journal} {\bibinfo
  {journal} {npj Quantum Materials}\ }\textbf {\bibinfo {volume} {7}},\
  \bibinfo {pages} {27} (\bibinfo {year} {2022})}\BibitemShut {NoStop}%
\bibitem [{\citenamefont {Choi}\ \emph {et~al.}(2022)\citenamefont {Choi},
  \citenamefont {Sim}, \citenamefont {Burch},\ and\ \citenamefont
  {Lee}}]{Choi_Emergent_2022}%
  \BibitemOpen
  \bibfield  {author} {\bibinfo {author} {\bibfnamefont {E.-M.}\ \bibnamefont
  {Choi}}, \bibinfo {author} {\bibfnamefont {K.~I.}\ \bibnamefont {Sim}},
  \bibinfo {author} {\bibfnamefont {K.~S.}\ \bibnamefont {Burch}},\ and\
  \bibinfo {author} {\bibfnamefont {Y.~H.}\ \bibnamefont {Lee}},\ }\href
  {https://doi.org/https://doi.org/10.1002/advs.202200186} {\bibfield
  {journal} {\bibinfo  {journal} {Advanced Science}\ }\textbf {\bibinfo
  {volume} {9}},\ \bibinfo {pages} {2200186} (\bibinfo {year}
  {2022})}\BibitemShut {NoStop}%
\bibitem [{\citenamefont {Ghiasi}\ \emph {et~al.}(2025)\citenamefont {Ghiasi},
  \citenamefont {Petrosyan}, \citenamefont {Ingla-Ayn{\'e}s}, \citenamefont
  {Bras}, \citenamefont {Watanabe}, \citenamefont {Taniguchi}, \citenamefont
  {Ma{\~{n}}as-Valero}, \citenamefont {Coronado}, \citenamefont {Zollner},
  \citenamefont {Fabian}, \citenamefont {Kim},\ and\ \citenamefont {van~der
  Zant}}]{Ghiasi2025}%
  \BibitemOpen
  \bibfield  {author} {\bibinfo {author} {\bibfnamefont {T.~S.}\ \bibnamefont
  {Ghiasi}}, \bibinfo {author} {\bibfnamefont {D.}~\bibnamefont {Petrosyan}},
  \bibinfo {author} {\bibfnamefont {J.}~\bibnamefont {Ingla-Ayn{\'e}s}},
  \bibinfo {author} {\bibfnamefont {T.}~\bibnamefont {Bras}}, \bibinfo {author}
  {\bibfnamefont {K.}~\bibnamefont {Watanabe}}, \bibinfo {author}
  {\bibfnamefont {T.}~\bibnamefont {Taniguchi}}, \bibinfo {author}
  {\bibfnamefont {S.}~\bibnamefont {Ma{\~{n}}as-Valero}}, \bibinfo {author}
  {\bibfnamefont {E.}~\bibnamefont {Coronado}}, \bibinfo {author}
  {\bibfnamefont {K.}~\bibnamefont {Zollner}}, \bibinfo {author} {\bibfnamefont
  {J.}~\bibnamefont {Fabian}}, \bibinfo {author} {\bibfnamefont
  {P.}~\bibnamefont {Kim}},\ and\ \bibinfo {author} {\bibfnamefont {H.~S.~J.}\
  \bibnamefont {van~der Zant}},\ }\href
  {https://doi.org/10.1038/s41467-025-60377-1} {\bibfield  {journal} {\bibinfo
  {journal} {Nature Communications}\ }\textbf {\bibinfo {volume} {16}},\
  \bibinfo {pages} {5336} (\bibinfo {year} {2025})}\BibitemShut {NoStop}%
\bibitem [{\citenamefont {Yang}\ \emph {et~al.}(2025)\citenamefont {Yang},
  \citenamefont {Gobbi}, \citenamefont {Herling}, \citenamefont {Pham},
  \citenamefont {Calavalle}, \citenamefont {Mart{\'i}n-Garc{\'i}a},
  \citenamefont {Fert}, \citenamefont {Hueso},\ and\ \citenamefont
  {Casanova}}]{Yang2025}%
  \BibitemOpen
  \bibfield  {author} {\bibinfo {author} {\bibfnamefont {H.}~\bibnamefont
  {Yang}}, \bibinfo {author} {\bibfnamefont {M.}~\bibnamefont {Gobbi}},
  \bibinfo {author} {\bibfnamefont {F.}~\bibnamefont {Herling}}, \bibinfo
  {author} {\bibfnamefont {V.~T.}\ \bibnamefont {Pham}}, \bibinfo {author}
  {\bibfnamefont {F.}~\bibnamefont {Calavalle}}, \bibinfo {author}
  {\bibfnamefont {B.}~\bibnamefont {Mart{\'i}n-Garc{\'i}a}}, \bibinfo {author}
  {\bibfnamefont {A.}~\bibnamefont {Fert}}, \bibinfo {author} {\bibfnamefont
  {L.~E.}\ \bibnamefont {Hueso}},\ and\ \bibinfo {author} {\bibfnamefont
  {F.}~\bibnamefont {Casanova}},\ }\href
  {https://doi.org/10.1038/s41928-024-01267-0} {\bibfield  {journal} {\bibinfo
  {journal} {Nature Electronics}\ }\textbf {\bibinfo {volume} {8}},\ \bibinfo
  {pages} {15} (\bibinfo {year} {2025})}\BibitemShut {NoStop}%
\bibitem [{\citenamefont {Sahani}\ \emph {et~al.}(2025)\citenamefont {Sahani},
  \citenamefont {Das}, \citenamefont {Watanabe}, \citenamefont {Taniguchi},
  \citenamefont {Agarwal},\ and\ \citenamefont {Bid}}]{Sahani_giantMR_2025}%
  \BibitemOpen
  \bibfield  {author} {\bibinfo {author} {\bibfnamefont {D.}~\bibnamefont
  {Sahani}}, \bibinfo {author} {\bibfnamefont {S.}~\bibnamefont {Das}},
  \bibinfo {author} {\bibfnamefont {K.}~\bibnamefont {Watanabe}}, \bibinfo
  {author} {\bibfnamefont {T.}~\bibnamefont {Taniguchi}}, \bibinfo {author}
  {\bibfnamefont {A.}~\bibnamefont {Agarwal}},\ and\ \bibinfo {author}
  {\bibfnamefont {A.}~\bibnamefont {Bid}},\ }\href
  {https://doi.org/10.1103/PhysRevLett.134.106301} {\bibfield  {journal}
  {\bibinfo  {journal} {Phys. Rev. Lett.}\ }\textbf {\bibinfo {volume} {134}},\
  \bibinfo {pages} {106301} (\bibinfo {year} {2025})}\BibitemShut {NoStop}%
\bibitem [{\citenamefont {Tombros}\ \emph {et~al.}(2007)\citenamefont
  {Tombros}, \citenamefont {Jozsa}, \citenamefont {Popinciuc}, \citenamefont
  {Jonkman},\ and\ \citenamefont {van Wees}}]{Tombros_spin_graphene_2007}%
  \BibitemOpen
  \bibfield  {author} {\bibinfo {author} {\bibfnamefont {N.}~\bibnamefont
  {Tombros}}, \bibinfo {author} {\bibfnamefont {C.}~\bibnamefont {Jozsa}},
  \bibinfo {author} {\bibfnamefont {M.}~\bibnamefont {Popinciuc}}, \bibinfo
  {author} {\bibfnamefont {H.~T.}\ \bibnamefont {Jonkman}},\ and\ \bibinfo
  {author} {\bibfnamefont {B.~J.}\ \bibnamefont {van Wees}},\ }\href
  {https://doi.org/10.1038/nature06037} {\bibfield  {journal} {\bibinfo
  {journal} {Nature}\ }\textbf {\bibinfo {volume} {448}},\ \bibinfo {pages}
  {571} (\bibinfo {year} {2007})}\BibitemShut {NoStop}%
\bibitem [{\citenamefont {Raes}\ \emph {et~al.}(2016)\citenamefont {Raes},
  \citenamefont {Scheerder}, \citenamefont {Costache}, \citenamefont {Bonell},
  \citenamefont {Sierra}, \citenamefont {Cuppens}, \citenamefont {Van~de
  Vondel},\ and\ \citenamefont {Valenzuela}}]{Raes2016}%
  \BibitemOpen
  \bibfield  {author} {\bibinfo {author} {\bibfnamefont {B.}~\bibnamefont
  {Raes}}, \bibinfo {author} {\bibfnamefont {J.~E.}\ \bibnamefont {Scheerder}},
  \bibinfo {author} {\bibfnamefont {M.~V.}\ \bibnamefont {Costache}}, \bibinfo
  {author} {\bibfnamefont {F.}~\bibnamefont {Bonell}}, \bibinfo {author}
  {\bibfnamefont {J.~F.}\ \bibnamefont {Sierra}}, \bibinfo {author}
  {\bibfnamefont {J.}~\bibnamefont {Cuppens}}, \bibinfo {author} {\bibfnamefont
  {J.}~\bibnamefont {Van~de Vondel}},\ and\ \bibinfo {author} {\bibfnamefont
  {S.~O.}\ \bibnamefont {Valenzuela}},\ }\href
  {https://doi.org/10.1038/ncomms11444} {\bibfield  {journal} {\bibinfo
  {journal} {Nature Communications}\ }\textbf {\bibinfo {volume} {7}},\
  \bibinfo {pages} {11444} (\bibinfo {year} {2016})}\BibitemShut {NoStop}%
\bibitem [{\citenamefont {Savero~Torres}\ \emph {et~al.}(2017)\citenamefont
  {Savero~Torres}, \citenamefont {Sierra}, \citenamefont {Benítez},
  \citenamefont {Bonell}, \citenamefont {Costache},\ and\ \citenamefont
  {Valenzuela}}]{SaveroTorres_2017}%
  \BibitemOpen
  \bibfield  {author} {\bibinfo {author} {\bibfnamefont {W.}~\bibnamefont
  {Savero~Torres}}, \bibinfo {author} {\bibfnamefont {J.~F.}\ \bibnamefont
  {Sierra}}, \bibinfo {author} {\bibfnamefont {L.~A.}\ \bibnamefont
  {Benítez}}, \bibinfo {author} {\bibfnamefont {F.}~\bibnamefont {Bonell}},
  \bibinfo {author} {\bibfnamefont {M.~V.}\ \bibnamefont {Costache}},\ and\
  \bibinfo {author} {\bibfnamefont {S.~O.}\ \bibnamefont {Valenzuela}},\ }\href
  {https://doi.org/10.1088/2053-1583/aa8823} {\bibfield  {journal} {\bibinfo
  {journal} {2D Materials}\ }\textbf {\bibinfo {volume} {4}},\ \bibinfo {pages}
  {041008} (\bibinfo {year} {2017})}\BibitemShut {NoStop}%
\bibitem [{\citenamefont {Piquemal-Banci}\ \emph {et~al.}(2020)\citenamefont
  {Piquemal-Banci}, \citenamefont {Galceran}, \citenamefont {Dubois},
  \citenamefont {Zatko}, \citenamefont {Galbiati}, \citenamefont {Godel},
  \citenamefont {Martin}, \citenamefont {Weatherup}, \citenamefont {Petroff},
  \citenamefont {Fert}, \citenamefont {Charlier}, \citenamefont {Robertson},
  \citenamefont {Hofmann}, \citenamefont {Dlubak},\ and\ \citenamefont
  {Seneor}}]{Piquemal-Banci2020}%
  \BibitemOpen
  \bibfield  {author} {\bibinfo {author} {\bibfnamefont {M.}~\bibnamefont
  {Piquemal-Banci}}, \bibinfo {author} {\bibfnamefont {R.}~\bibnamefont
  {Galceran}}, \bibinfo {author} {\bibfnamefont {S.~M.-M.}\ \bibnamefont
  {Dubois}}, \bibinfo {author} {\bibfnamefont {V.}~\bibnamefont {Zatko}},
  \bibinfo {author} {\bibfnamefont {M.}~\bibnamefont {Galbiati}}, \bibinfo
  {author} {\bibfnamefont {F.}~\bibnamefont {Godel}}, \bibinfo {author}
  {\bibfnamefont {M.-B.}\ \bibnamefont {Martin}}, \bibinfo {author}
  {\bibfnamefont {R.~S.}\ \bibnamefont {Weatherup}}, \bibinfo {author}
  {\bibfnamefont {F.}~\bibnamefont {Petroff}}, \bibinfo {author} {\bibfnamefont
  {A.}~\bibnamefont {Fert}}, \bibinfo {author} {\bibfnamefont {J.-C.}\
  \bibnamefont {Charlier}}, \bibinfo {author} {\bibfnamefont {J.}~\bibnamefont
  {Robertson}}, \bibinfo {author} {\bibfnamefont {S.}~\bibnamefont {Hofmann}},
  \bibinfo {author} {\bibfnamefont {B.}~\bibnamefont {Dlubak}},\ and\ \bibinfo
  {author} {\bibfnamefont {P.}~\bibnamefont {Seneor}},\ }\href
  {https://doi.org/10.1038/s41467-020-19420-6} {\bibfield  {journal} {\bibinfo
  {journal} {Nature Communications}\ }\textbf {\bibinfo {volume} {11}},\
  \bibinfo {pages} {5670} (\bibinfo {year} {2020})}\BibitemShut {NoStop}%
\bibitem [{\citenamefont {Jeong}\ \emph {et~al.}(2024)\citenamefont {Jeong},
  \citenamefont {Kiem}, \citenamefont {Guo}, \citenamefont {Duan},
  \citenamefont {Watanabe}, \citenamefont {Taniguchi}, \citenamefont {Liu},
  \citenamefont {Han}, \citenamefont {Zheng},\ and\ \citenamefont
  {Yang}}]{Jeong_spin-selective_2024}%
  \BibitemOpen
  \bibfield  {author} {\bibinfo {author} {\bibfnamefont {J.}~\bibnamefont
  {Jeong}}, \bibinfo {author} {\bibfnamefont {D.~H.}\ \bibnamefont {Kiem}},
  \bibinfo {author} {\bibfnamefont {D.}~\bibnamefont {Guo}}, \bibinfo {author}
  {\bibfnamefont {R.}~\bibnamefont {Duan}}, \bibinfo {author} {\bibfnamefont
  {K.}~\bibnamefont {Watanabe}}, \bibinfo {author} {\bibfnamefont
  {T.}~\bibnamefont {Taniguchi}}, \bibinfo {author} {\bibfnamefont
  {Z.}~\bibnamefont {Liu}}, \bibinfo {author} {\bibfnamefont {M.~J.}\
  \bibnamefont {Han}}, \bibinfo {author} {\bibfnamefont {S.}~\bibnamefont
  {Zheng}},\ and\ \bibinfo {author} {\bibfnamefont {H.}~\bibnamefont {Yang}},\
  }\href {https://doi.org/https://doi.org/10.1002/adma.202310291} {\bibfield
  {journal} {\bibinfo  {journal} {Advanced Materials}\ }\textbf {\bibinfo
  {volume} {36}},\ \bibinfo {pages} {2310291} (\bibinfo {year}
  {2024})}\BibitemShut {NoStop}%
\bibitem [{\citenamefont {Han}\ \emph {et~al.}(2012)\citenamefont {Han},
  \citenamefont {McCreary}, \citenamefont {Pi}, \citenamefont {Wang},
  \citenamefont {Li}, \citenamefont {Wen}, \citenamefont {Chen},\ and\
  \citenamefont {Kawakami}}]{Han_spin_transport_2012}%
  \BibitemOpen
  \bibfield  {author} {\bibinfo {author} {\bibfnamefont {W.}~\bibnamefont
  {Han}}, \bibinfo {author} {\bibfnamefont {K.}~\bibnamefont {McCreary}},
  \bibinfo {author} {\bibfnamefont {K.}~\bibnamefont {Pi}}, \bibinfo {author}
  {\bibfnamefont {W.}~\bibnamefont {Wang}}, \bibinfo {author} {\bibfnamefont
  {Y.}~\bibnamefont {Li}}, \bibinfo {author} {\bibfnamefont {H.}~\bibnamefont
  {Wen}}, \bibinfo {author} {\bibfnamefont {J.}~\bibnamefont {Chen}},\ and\
  \bibinfo {author} {\bibfnamefont {R.}~\bibnamefont {Kawakami}},\ }\href
  {https://doi.org/https://doi.org/10.1016/j.jmmm.2011.08.001} {\bibfield
  {journal} {\bibinfo  {journal} {Journal of Magnetism and Magnetic Materials}\
  }\textbf {\bibinfo {volume} {324}},\ \bibinfo {pages} {369} (\bibinfo {year}
  {2012})}\BibitemShut {NoStop}%
\bibitem [{\citenamefont {Zollner}\ \emph
  {et~al.}(2025{\natexlab{a}})\citenamefont {Zollner}, \citenamefont {Kurpas},
  \citenamefont {Gmitra},\ and\ \citenamefont {Fabian}}]{Zollner_review_2025}%
  \BibitemOpen
  \bibfield  {author} {\bibinfo {author} {\bibfnamefont {K.}~\bibnamefont
  {Zollner}}, \bibinfo {author} {\bibfnamefont {M.}~\bibnamefont {Kurpas}},
  \bibinfo {author} {\bibfnamefont {M.}~\bibnamefont {Gmitra}},\ and\ \bibinfo
  {author} {\bibfnamefont {J.}~\bibnamefont {Fabian}},\ }\href
  {https://doi.org/10.1038/s42254-025-00818-4} {\bibfield  {journal} {\bibinfo
  {journal} {Nature Reviews Physics}\ }\textbf {\bibinfo {volume} {7}},\
  \bibinfo {pages} {255} (\bibinfo {year} {2025}{\natexlab{a}})}\BibitemShut
  {NoStop}%
\bibitem [{\citenamefont {Han}\ \emph {et~al.}(2014)\citenamefont {Han},
  \citenamefont {Kawakami}, \citenamefont {Gmitra},\ and\ \citenamefont
  {Fabian}}]{Han_graphene_2014}%
  \BibitemOpen
  \bibfield  {author} {\bibinfo {author} {\bibfnamefont {W.}~\bibnamefont
  {Han}}, \bibinfo {author} {\bibfnamefont {R.~K.}\ \bibnamefont {Kawakami}},
  \bibinfo {author} {\bibfnamefont {M.}~\bibnamefont {Gmitra}},\ and\ \bibinfo
  {author} {\bibfnamefont {J.}~\bibnamefont {Fabian}},\ }\href
  {https://doi.org/10.1038/nnano.2014.214} {\bibfield  {journal} {\bibinfo
  {journal} {Nature Nanotechnology}\ }\textbf {\bibinfo {volume} {9}},\
  \bibinfo {pages} {794} (\bibinfo {year} {2014})}\BibitemShut {NoStop}%
\bibitem [{\citenamefont {Zhang}\ \emph {et~al.}(2015)\citenamefont {Zhang},
  \citenamefont {Zhao}, \citenamefont {Yao},\ and\ \citenamefont
  {Yang}}]{Zhang_robust_215}%
  \BibitemOpen
  \bibfield  {author} {\bibinfo {author} {\bibfnamefont {J.}~\bibnamefont
  {Zhang}}, \bibinfo {author} {\bibfnamefont {B.}~\bibnamefont {Zhao}},
  \bibinfo {author} {\bibfnamefont {Y.}~\bibnamefont {Yao}},\ and\ \bibinfo
  {author} {\bibfnamefont {Z.}~\bibnamefont {Yang}},\ }\href
  {https://doi.org/10.1103/PhysRevB.92.165418} {\bibfield  {journal} {\bibinfo
  {journal} {Phys. Rev. B}\ }\textbf {\bibinfo {volume} {92}},\ \bibinfo
  {pages} {165418} (\bibinfo {year} {2015})}\BibitemShut {NoStop}%
\bibitem [{\citenamefont {Zollner}\ \emph {et~al.}(2020)\citenamefont
  {Zollner}, \citenamefont {Petrovi\ifmmode~\acute{c}\else \'{c}\fi{}},
  \citenamefont {Dolui}, \citenamefont {Plech\'a\ifmmode~\check{c}\else
  \v{c}\fi{}}, \citenamefont {Nikoli\ifmmode~\acute{c}\else \'{c}\fi{}},\ and\
  \citenamefont {Fabian}}]{Zollner_scattering_2020}%
  \BibitemOpen
  \bibfield  {author} {\bibinfo {author} {\bibfnamefont {K.}~\bibnamefont
  {Zollner}}, \bibinfo {author} {\bibfnamefont {M.~D.}\ \bibnamefont
  {Petrovi\ifmmode~\acute{c}\else \'{c}\fi{}}}, \bibinfo {author}
  {\bibfnamefont {K.}~\bibnamefont {Dolui}}, \bibinfo {author} {\bibfnamefont
  {P.}~\bibnamefont {Plech\'a\ifmmode~\check{c}\else \v{c}\fi{}}}, \bibinfo
  {author} {\bibfnamefont {B.~K.}\ \bibnamefont {Nikoli\ifmmode~\acute{c}\else
  \'{c}\fi{}}},\ and\ \bibinfo {author} {\bibfnamefont {J.}~\bibnamefont
  {Fabian}},\ }\href {https://doi.org/10.1103/PhysRevResearch.2.043057}
  {\bibfield  {journal} {\bibinfo  {journal} {Phys. Rev. Res.}\ }\textbf
  {\bibinfo {volume} {2}},\ \bibinfo {pages} {043057} (\bibinfo {year}
  {2020})}\BibitemShut {NoStop}%
\bibitem [{\citenamefont {Averyanov}\ \emph {et~al.}(2018)\citenamefont
  {Averyanov}, \citenamefont {Sokolov}, \citenamefont {Tokmachev},
  \citenamefont {Parfenov}, \citenamefont {Karateev}, \citenamefont
  {Taldenkov},\ and\ \citenamefont {Storchak}}]{Averyanov2018}%
  \BibitemOpen
  \bibfield  {author} {\bibinfo {author} {\bibfnamefont {D.~V.}\ \bibnamefont
  {Averyanov}}, \bibinfo {author} {\bibfnamefont {I.~S.}\ \bibnamefont
  {Sokolov}}, \bibinfo {author} {\bibfnamefont {A.~M.}\ \bibnamefont
  {Tokmachev}}, \bibinfo {author} {\bibfnamefont {O.~E.}\ \bibnamefont
  {Parfenov}}, \bibinfo {author} {\bibfnamefont {I.~A.}\ \bibnamefont
  {Karateev}}, \bibinfo {author} {\bibfnamefont {A.~N.}\ \bibnamefont
  {Taldenkov}},\ and\ \bibinfo {author} {\bibfnamefont {V.~G.}\ \bibnamefont
  {Storchak}},\ }\href {https://doi.org/10.1021/acsami.8b04289} {\bibfield
  {journal} {\bibinfo  {journal} {ACS Applied Materials {\&} Interfaces}\
  }\textbf {\bibinfo {volume} {10}},\ \bibinfo {pages} {20767} (\bibinfo {year}
  {2018})}\BibitemShut {NoStop}%
\bibitem [{\citenamefont {Pandey}\ \emph {et~al.}(2023)\citenamefont {Pandey},
  \citenamefont {Hettler}, \citenamefont {Arenal}, \citenamefont {Bouillet},
  \citenamefont {Moghe}, \citenamefont {Berciaud}, \citenamefont {Robert},
  \citenamefont {Dayen},\ and\ \citenamefont {Halley}}]{Pandey_2023}%
  \BibitemOpen
  \bibfield  {author} {\bibinfo {author} {\bibfnamefont {S.}~\bibnamefont
  {Pandey}}, \bibinfo {author} {\bibfnamefont {S.}~\bibnamefont {Hettler}},
  \bibinfo {author} {\bibfnamefont {R.}~\bibnamefont {Arenal}}, \bibinfo
  {author} {\bibfnamefont {C.}~\bibnamefont {Bouillet}}, \bibinfo {author}
  {\bibfnamefont {A.~R.}\ \bibnamefont {Moghe}}, \bibinfo {author}
  {\bibfnamefont {S.}~\bibnamefont {Berciaud}}, \bibinfo {author}
  {\bibfnamefont {J.}~\bibnamefont {Robert}}, \bibinfo {author} {\bibfnamefont
  {J.-F. m.~c.}\ \bibnamefont {Dayen}},\ and\ \bibinfo {author} {\bibfnamefont
  {D.}~\bibnamefont {Halley}},\ }\href
  {https://doi.org/10.1103/PhysRevB.108.144423} {\bibfield  {journal} {\bibinfo
   {journal} {Phys. Rev. B}\ }\textbf {\bibinfo {volume} {108}},\ \bibinfo
  {pages} {144423} (\bibinfo {year} {2023})}\BibitemShut {NoStop}%
\bibitem [{\citenamefont {Pandey}\ \emph {et~al.}(2025)\citenamefont {Pandey},
  \citenamefont {Pin}, \citenamefont {Hettler}, \citenamefont {Arenal},
  \citenamefont {Bouillet}, \citenamefont {Maroutian}, \citenamefont {Robert},
  \citenamefont {Gobaut}, \citenamefont {Kundys}, \citenamefont {Dayen},\ and\
  \citenamefont {Halley}}]{Pandey_2025}%
  \BibitemOpen
  \bibfield  {author} {\bibinfo {author} {\bibfnamefont {S.}~\bibnamefont
  {Pandey}}, \bibinfo {author} {\bibfnamefont {T.}~\bibnamefont {Pin}},
  \bibinfo {author} {\bibfnamefont {S.}~\bibnamefont {Hettler}}, \bibinfo
  {author} {\bibfnamefont {R.}~\bibnamefont {Arenal}}, \bibinfo {author}
  {\bibfnamefont {C.}~\bibnamefont {Bouillet}}, \bibinfo {author}
  {\bibfnamefont {T.}~\bibnamefont {Maroutian}}, \bibinfo {author}
  {\bibfnamefont {J.}~\bibnamefont {Robert}}, \bibinfo {author} {\bibfnamefont
  {B.}~\bibnamefont {Gobaut}}, \bibinfo {author} {\bibfnamefont
  {B.}~\bibnamefont {Kundys}}, \bibinfo {author} {\bibfnamefont {J.-F.}\
  \bibnamefont {Dayen}},\ and\ \bibinfo {author} {\bibfnamefont
  {D.}~\bibnamefont {Halley}},\ }\href
  {https://doi.org/https://doi.org/10.1002/adma.202417669} {\bibfield
  {journal} {\bibinfo  {journal} {Advanced Materials}\ }\textbf {\bibinfo
  {volume} {37}},\ \bibinfo {pages} {2417669} (\bibinfo {year}
  {2025})}\BibitemShut {NoStop}%
\bibitem [{\citenamefont {Zhang}\ \emph {et~al.}(2018)\citenamefont {Zhang},
  \citenamefont {Zhao}, \citenamefont {Zhou}, \citenamefont {Xue},
  \citenamefont {Ma},\ and\ \citenamefont {Yang}}]{Zhang_2018}%
  \BibitemOpen
  \bibfield  {author} {\bibinfo {author} {\bibfnamefont {J.}~\bibnamefont
  {Zhang}}, \bibinfo {author} {\bibfnamefont {B.}~\bibnamefont {Zhao}},
  \bibinfo {author} {\bibfnamefont {T.}~\bibnamefont {Zhou}}, \bibinfo {author}
  {\bibfnamefont {Y.}~\bibnamefont {Xue}}, \bibinfo {author} {\bibfnamefont
  {C.}~\bibnamefont {Ma}},\ and\ \bibinfo {author} {\bibfnamefont
  {Z.}~\bibnamefont {Yang}},\ }\href
  {https://doi.org/10.1103/PhysRevB.97.085401} {\bibfield  {journal} {\bibinfo
  {journal} {Phys. Rev. B}\ }\textbf {\bibinfo {volume} {97}},\ \bibinfo
  {pages} {085401} (\bibinfo {year} {2018})}\BibitemShut {NoStop}%
\bibitem [{\citenamefont {Cardoso}\ \emph {et~al.}(2018)\citenamefont
  {Cardoso}, \citenamefont {Soriano}, \citenamefont
  {Garc\'{\i}a-Mart\'{\i}nez},\ and\ \citenamefont
  {Fern\'andez-Rossier}}]{Cardoso_VdW_2018}%
  \BibitemOpen
  \bibfield  {author} {\bibinfo {author} {\bibfnamefont {C.}~\bibnamefont
  {Cardoso}}, \bibinfo {author} {\bibfnamefont {D.}~\bibnamefont {Soriano}},
  \bibinfo {author} {\bibfnamefont {N.~A.}\ \bibnamefont
  {Garc\'{\i}a-Mart\'{\i}nez}},\ and\ \bibinfo {author} {\bibfnamefont
  {J.}~\bibnamefont {Fern\'andez-Rossier}},\ }\href
  {https://doi.org/10.1103/PhysRevLett.121.067701} {\bibfield  {journal}
  {\bibinfo  {journal} {Phys. Rev. Lett.}\ }\textbf {\bibinfo {volume} {121}},\
  \bibinfo {pages} {067701} (\bibinfo {year} {2018})}\BibitemShut {NoStop}%
\bibitem [{\citenamefont {Seyler}\ \emph {et~al.}(2018)\citenamefont {Seyler},
  \citenamefont {Zhong}, \citenamefont {Huang}, \citenamefont {Linpeng},
  \citenamefont {Wilson}, \citenamefont {Taniguchi}, \citenamefont {Watanabe},
  \citenamefont {Yao}, \citenamefont {Xiao}, \citenamefont {McGuire},
  \citenamefont {Fu},\ and\ \citenamefont {Xu}}]{Seyler2018}%
  \BibitemOpen
  \bibfield  {author} {\bibinfo {author} {\bibfnamefont {K.~L.}\ \bibnamefont
  {Seyler}}, \bibinfo {author} {\bibfnamefont {D.}~\bibnamefont {Zhong}},
  \bibinfo {author} {\bibfnamefont {B.}~\bibnamefont {Huang}}, \bibinfo
  {author} {\bibfnamefont {X.}~\bibnamefont {Linpeng}}, \bibinfo {author}
  {\bibfnamefont {N.~P.}\ \bibnamefont {Wilson}}, \bibinfo {author}
  {\bibfnamefont {T.}~\bibnamefont {Taniguchi}}, \bibinfo {author}
  {\bibfnamefont {K.}~\bibnamefont {Watanabe}}, \bibinfo {author}
  {\bibfnamefont {W.}~\bibnamefont {Yao}}, \bibinfo {author} {\bibfnamefont
  {D.}~\bibnamefont {Xiao}}, \bibinfo {author} {\bibfnamefont {M.~A.}\
  \bibnamefont {McGuire}}, \bibinfo {author} {\bibfnamefont {K.-M.~C.}\
  \bibnamefont {Fu}},\ and\ \bibinfo {author} {\bibfnamefont {X.}~\bibnamefont
  {Xu}},\ }\href {https://doi.org/10.1021/acs.nanolett.8b01105} {\bibfield
  {journal} {\bibinfo  {journal} {Nano Letters}\ }\textbf {\bibinfo {volume}
  {18}},\ \bibinfo {pages} {3823} (\bibinfo {year} {2018})}\BibitemShut
  {NoStop}%
\bibitem [{\citenamefont {Farooq}\ and\ \citenamefont
  {Hong}(2019)}]{Farooq2019}%
  \BibitemOpen
  \bibfield  {author} {\bibinfo {author} {\bibfnamefont {M.~U.}\ \bibnamefont
  {Farooq}}\ and\ \bibinfo {author} {\bibfnamefont {J.}~\bibnamefont {Hong}},\
  }\href {https://doi.org/10.1038/s41699-019-0086-6} {\bibfield  {journal}
  {\bibinfo  {journal} {npj 2D Materials and Applications}\ }\textbf {\bibinfo
  {volume} {3}},\ \bibinfo {pages} {3} (\bibinfo {year} {2019})}\BibitemShut
  {NoStop}%
\bibitem [{\citenamefont {Cardoso}\ \emph {et~al.}(2023)\citenamefont
  {Cardoso}, \citenamefont {Costa}, \citenamefont {MacDonald},\ and\
  \citenamefont {Fern\'andez-Rossier}}]{Cardoso_Strong_2023}%
  \BibitemOpen
  \bibfield  {author} {\bibinfo {author} {\bibfnamefont {C.}~\bibnamefont
  {Cardoso}}, \bibinfo {author} {\bibfnamefont {A.~T.}\ \bibnamefont {Costa}},
  \bibinfo {author} {\bibfnamefont {A.~H.}\ \bibnamefont {MacDonald}},\ and\
  \bibinfo {author} {\bibfnamefont {J.}~\bibnamefont {Fern\'andez-Rossier}},\
  }\href {https://doi.org/10.1103/PhysRevB.108.184423} {\bibfield  {journal}
  {\bibinfo  {journal} {Phys. Rev. B}\ }\textbf {\bibinfo {volume} {108}},\
  \bibinfo {pages} {184423} (\bibinfo {year} {2023})}\BibitemShut {NoStop}%
\bibitem [{\citenamefont {Zollner}\ \emph {et~al.}(2016)\citenamefont
  {Zollner}, \citenamefont {Gmitra}, \citenamefont {Frank},\ and\ \citenamefont
  {Fabian}}]{Zollner_Theory_2016}%
  \BibitemOpen
  \bibfield  {author} {\bibinfo {author} {\bibfnamefont {K.}~\bibnamefont
  {Zollner}}, \bibinfo {author} {\bibfnamefont {M.}~\bibnamefont {Gmitra}},
  \bibinfo {author} {\bibfnamefont {T.}~\bibnamefont {Frank}},\ and\ \bibinfo
  {author} {\bibfnamefont {J.}~\bibnamefont {Fabian}},\ }\href
  {https://doi.org/10.1103/PhysRevB.94.155441} {\bibfield  {journal} {\bibinfo
  {journal} {Phys. Rev. B}\ }\textbf {\bibinfo {volume} {94}},\ \bibinfo
  {pages} {155441} (\bibinfo {year} {2016})}\BibitemShut {NoStop}%
\bibitem [{\citenamefont {Lazi\ifmmode~\acute{c}\else \'{c}\fi{}}\ \emph
  {et~al.}(2016)\citenamefont {Lazi\ifmmode~\acute{c}\else \'{c}\fi{}},
  \citenamefont {Belashchenko},\ and\ \citenamefont {\ifmmode \check{Z}\else
  \v{Z}\fi{}uti\ifmmode~\acute{c}\else \'{c}\fi{}}}]{Lazic_Effective_2016}%
  \BibitemOpen
  \bibfield  {author} {\bibinfo {author} {\bibfnamefont {P.}~\bibnamefont
  {Lazi\ifmmode~\acute{c}\else \'{c}\fi{}}}, \bibinfo {author} {\bibfnamefont
  {K.~D.}\ \bibnamefont {Belashchenko}},\ and\ \bibinfo {author} {\bibfnamefont
  {I.}~\bibnamefont {\ifmmode \check{Z}\else
  \v{Z}\fi{}uti\ifmmode~\acute{c}\else \'{c}\fi{}}},\ }\href
  {https://doi.org/10.1103/PhysRevB.93.241401} {\bibfield  {journal} {\bibinfo
  {journal} {Phys. Rev. B}\ }\textbf {\bibinfo {volume} {93}},\ \bibinfo
  {pages} {241401} (\bibinfo {year} {2016})}\BibitemShut {NoStop}%
\bibitem [{\citenamefont {Zanolli}(2016)}]{Zanolli2016}%
  \BibitemOpen
  \bibfield  {author} {\bibinfo {author} {\bibfnamefont {Z.}~\bibnamefont
  {Zanolli}},\ }\href {https://doi.org/10.1038/srep31346} {\bibfield  {journal}
  {\bibinfo  {journal} {Scientific Reports}\ }\textbf {\bibinfo {volume} {6}},\
  \bibinfo {pages} {31346} (\bibinfo {year} {2016})}\BibitemShut {NoStop}%
\bibitem [{\citenamefont {Zollner}\ and\ \citenamefont
  {Fabian}(2021)}]{Zollner_bilayer_2021}%
  \BibitemOpen
  \bibfield  {author} {\bibinfo {author} {\bibfnamefont {K.}~\bibnamefont
  {Zollner}}\ and\ \bibinfo {author} {\bibfnamefont {J.}~\bibnamefont
  {Fabian}},\ }\href {https://doi.org/10.1103/PhysRevB.104.075126} {\bibfield
  {journal} {\bibinfo  {journal} {Phys. Rev. B}\ }\textbf {\bibinfo {volume}
  {104}},\ \bibinfo {pages} {075126} (\bibinfo {year} {2021})}\BibitemShut
  {NoStop}%
\bibitem [{\citenamefont {Beer}\ \emph {et~al.}(2024)\citenamefont {Beer},
  \citenamefont {Zollner}, \citenamefont {Serati~de Brito}, \citenamefont
  {Faria~Junior}, \citenamefont {Parzefall}, \citenamefont {Ghiasi},
  \citenamefont {Ingla-Ayn{\'e}s}, \citenamefont {Ma{\~{n}}as-Valero},
  \citenamefont {Boix-Constant}, \citenamefont {Watanabe}, \citenamefont
  {Taniguchi}, \citenamefont {Fabian}, \citenamefont {van~der Zant},
  \citenamefont {Galv{\~a}o~Gobato},\ and\ \citenamefont
  {Sch{\"u}ller}}]{Beer_proximity_2024}%
  \BibitemOpen
  \bibfield  {author} {\bibinfo {author} {\bibfnamefont {A.}~\bibnamefont
  {Beer}}, \bibinfo {author} {\bibfnamefont {K.}~\bibnamefont {Zollner}},
  \bibinfo {author} {\bibfnamefont {C.}~\bibnamefont {Serati~de Brito}},
  \bibinfo {author} {\bibfnamefont {P.~E.}\ \bibnamefont {Faria~Junior}},
  \bibinfo {author} {\bibfnamefont {P.}~\bibnamefont {Parzefall}}, \bibinfo
  {author} {\bibfnamefont {T.~S.}\ \bibnamefont {Ghiasi}}, \bibinfo {author}
  {\bibfnamefont {J.}~\bibnamefont {Ingla-Ayn{\'e}s}}, \bibinfo {author}
  {\bibfnamefont {S.}~\bibnamefont {Ma{\~{n}}as-Valero}}, \bibinfo {author}
  {\bibfnamefont {C.}~\bibnamefont {Boix-Constant}}, \bibinfo {author}
  {\bibfnamefont {K.}~\bibnamefont {Watanabe}}, \bibinfo {author}
  {\bibfnamefont {T.}~\bibnamefont {Taniguchi}}, \bibinfo {author}
  {\bibfnamefont {J.}~\bibnamefont {Fabian}}, \bibinfo {author} {\bibfnamefont
  {H.~S.~J.}\ \bibnamefont {van~der Zant}}, \bibinfo {author} {\bibfnamefont
  {Y.}~\bibnamefont {Galv{\~a}o~Gobato}},\ and\ \bibinfo {author}
  {\bibfnamefont {C.}~\bibnamefont {Sch{\"u}ller}},\ }\href
  {https://doi.org/10.1021/acsnano.4c07336} {\bibfield  {journal} {\bibinfo
  {journal} {ACS Nano}\ }\textbf {\bibinfo {volume} {18}},\ \bibinfo {pages}
  {31044} (\bibinfo {year} {2024})}\BibitemShut {NoStop}%
\bibitem [{\citenamefont {Liu}\ \emph {et~al.}(2013)\citenamefont {Liu},
  \citenamefont {Prestigiacomo},\ and\ \citenamefont
  {Adams}}]{Liu_Electrostatic_2013}%
  \BibitemOpen
  \bibfield  {author} {\bibinfo {author} {\bibfnamefont {T.~J.}\ \bibnamefont
  {Liu}}, \bibinfo {author} {\bibfnamefont {J.~C.}\ \bibnamefont
  {Prestigiacomo}},\ and\ \bibinfo {author} {\bibfnamefont {P.~W.}\
  \bibnamefont {Adams}},\ }\href
  {https://doi.org/10.1103/PhysRevLett.111.027207} {\bibfield  {journal}
  {\bibinfo  {journal} {Phys. Rev. Lett.}\ }\textbf {\bibinfo {volume} {111}},\
  \bibinfo {pages} {027207} (\bibinfo {year} {2013})}\BibitemShut {NoStop}%
\bibitem [{\citenamefont {Yang}\ \emph {et~al.}(2024)\citenamefont {Yang},
  \citenamefont {Bhujel}, \citenamefont {Chica}, \citenamefont {Telford},
  \citenamefont {Roy}, \citenamefont {Ibrahim}, \citenamefont {Chshiev},
  \citenamefont {Cosset-Ch{\'e}neau},\ and\ \citenamefont
  {Wees}}]{Yang_electrostatically_2024}%
  \BibitemOpen
  \bibfield  {author} {\bibinfo {author} {\bibfnamefont {B.}~\bibnamefont
  {Yang}}, \bibinfo {author} {\bibfnamefont {B.}~\bibnamefont {Bhujel}},
  \bibinfo {author} {\bibfnamefont {D.~G.}\ \bibnamefont {Chica}}, \bibinfo
  {author} {\bibfnamefont {E.~J.}\ \bibnamefont {Telford}}, \bibinfo {author}
  {\bibfnamefont {X.}~\bibnamefont {Roy}}, \bibinfo {author} {\bibfnamefont
  {F.}~\bibnamefont {Ibrahim}}, \bibinfo {author} {\bibfnamefont
  {M.}~\bibnamefont {Chshiev}}, \bibinfo {author} {\bibfnamefont
  {M.}~\bibnamefont {Cosset-Ch{\'e}neau}},\ and\ \bibinfo {author}
  {\bibfnamefont {B.~J.~v.}\ \bibnamefont {Wees}},\ }\href
  {https://doi.org/10.1038/s41467-024-48809-w} {\bibfield  {journal} {\bibinfo
  {journal} {Nature Communications}\ }\textbf {\bibinfo {volume} {15}},\
  \bibinfo {pages} {4459} (\bibinfo {year} {2024})}\BibitemShut {NoStop}%
\bibitem [{\citenamefont {Qiu}\ \emph {et~al.}(2021)\citenamefont {Qiu},
  \citenamefont {Holwill}, \citenamefont {Olsen}, \citenamefont {Lyu},
  \citenamefont {Li}, \citenamefont {Fang}, \citenamefont {Yang}, \citenamefont
  {Kashchenko}, \citenamefont {Novoselov},\ and\ \citenamefont
  {Lu}}]{Qiu_visualizing_2021}%
  \BibitemOpen
  \bibfield  {author} {\bibinfo {author} {\bibfnamefont {Z.}~\bibnamefont
  {Qiu}}, \bibinfo {author} {\bibfnamefont {M.}~\bibnamefont {Holwill}},
  \bibinfo {author} {\bibfnamefont {T.}~\bibnamefont {Olsen}}, \bibinfo
  {author} {\bibfnamefont {P.}~\bibnamefont {Lyu}}, \bibinfo {author}
  {\bibfnamefont {J.}~\bibnamefont {Li}}, \bibinfo {author} {\bibfnamefont
  {H.}~\bibnamefont {Fang}}, \bibinfo {author} {\bibfnamefont {H.}~\bibnamefont
  {Yang}}, \bibinfo {author} {\bibfnamefont {M.}~\bibnamefont {Kashchenko}},
  \bibinfo {author} {\bibfnamefont {K.~S.}\ \bibnamefont {Novoselov}},\ and\
  \bibinfo {author} {\bibfnamefont {J.}~\bibnamefont {Lu}},\ }\href
  {https://doi.org/10.1038/s41467-020-20376-w} {\bibfield  {journal} {\bibinfo
  {journal} {Nature Communications}\ }\textbf {\bibinfo {volume} {12}},\
  \bibinfo {pages} {70} (\bibinfo {year} {2021})}\BibitemShut {NoStop}%
\bibitem [{\citenamefont {Tong}\ \emph {et~al.}(2019)\citenamefont {Tong},
  \citenamefont {Chen},\ and\ \citenamefont {Yao}}]{Tong_2019}%
  \BibitemOpen
  \bibfield  {author} {\bibinfo {author} {\bibfnamefont {Q.}~\bibnamefont
  {Tong}}, \bibinfo {author} {\bibfnamefont {M.}~\bibnamefont {Chen}},\ and\
  \bibinfo {author} {\bibfnamefont {W.}~\bibnamefont {Yao}},\ }\href
  {https://doi.org/10.1103/PhysRevApplied.12.024031} {\bibfield  {journal}
  {\bibinfo  {journal} {Phys. Rev. Appl.}\ }\textbf {\bibinfo {volume} {12}},\
  \bibinfo {pages} {024031} (\bibinfo {year} {2019})}\BibitemShut {NoStop}%
\bibitem [{\citenamefont {Carr}\ \emph
  {et~al.}(2020{\natexlab{b}})\citenamefont {Carr}, \citenamefont {Fang},\ and\
  \citenamefont {Kaxiras}}]{Carr2020}%
  \BibitemOpen
  \bibfield  {author} {\bibinfo {author} {\bibfnamefont {S.}~\bibnamefont
  {Carr}}, \bibinfo {author} {\bibfnamefont {S.}~\bibnamefont {Fang}},\ and\
  \bibinfo {author} {\bibfnamefont {E.}~\bibnamefont {Kaxiras}},\ }\href
  {https://doi.org/10.1038/s41578-020-0214-0} {\bibfield  {journal} {\bibinfo
  {journal} {Nature Reviews Materials}\ }\textbf {\bibinfo {volume} {5}},\
  \bibinfo {pages} {748} (\bibinfo {year} {2020}{\natexlab{b}})}\BibitemShut
  {NoStop}%
\bibitem [{Note1()}]{Note1}%
  \BibitemOpen
  \bibinfo {note} {See Ref.~\cite {Zollner_Engineering_2022}: ``[...] there is
  a delicate balance in the orbital hybridization [...], which makes the
  exchange coupling highly sensitive to the atomic registry.'' And in the
  supplementary material: ``The magnetic moments of the C atoms already give us
  a first hint that graphene experiences some proximity-induced magnetism from
  the CGT layer. [...] Here, the atomic registry should play a major
  role.''}\BibitemShut {NoStop}%
\bibitem [{\citenamefont {Zollner}\ \emph
  {et~al.}(2025{\natexlab{b}})\citenamefont {Zollner}, \citenamefont
  {Cvitkovich}, \citenamefont {Silvioli}, \citenamefont {Stier},\ and\
  \citenamefont {Fabian}}]{Zollner_XXX}%
  \BibitemOpen
  \bibfield  {author} {\bibinfo {author} {\bibfnamefont {K.}~\bibnamefont
  {Zollner}}, \bibinfo {author} {\bibfnamefont {L.}~\bibnamefont {Cvitkovich}},
  \bibinfo {author} {\bibfnamefont {R.}~\bibnamefont {Silvioli}}, \bibinfo
  {author} {\bibfnamefont {A.}~\bibnamefont {Stier}},\ and\ \bibinfo {author}
  {\bibfnamefont {J.}~\bibnamefont {Fabian}},\ }\href@noop {} {\bibinfo {title}
  {Local variations of proximity exchange in {C}obalt/h{BN}/{G}raphene spin
  injection geometries}} (\bibinfo {year} {2025}{\natexlab{b}}),\ \bibinfo
  {note} {to be published}\BibitemShut {NoStop}%
\bibitem [{\citenamefont {Giannozzi}\ \emph {et~al.}(2009)\citenamefont
  {Giannozzi}, \citenamefont {Baroni}, \citenamefont {Bonini}, \citenamefont
  {Calandra}, \citenamefont {Car}, \citenamefont {Cavazzoni}, \citenamefont
  {Ceresoli}, \citenamefont {Chiarotti}, \citenamefont {Cococcioni},
  \citenamefont {Dabo}, \citenamefont {Dal~Corso}, \citenamefont
  {de~Gironcoli}, \citenamefont {Fabris}, \citenamefont {Fratesi},
  \citenamefont {Gebauer}, \citenamefont {Gerstmann}, \citenamefont
  {Gougoussis}, \citenamefont {Kokalj}, \citenamefont {Lazzeri}, \citenamefont
  {Martin-Samos}, \citenamefont {Marzari}, \citenamefont {Mauri}, \citenamefont
  {Mazzarello}, \citenamefont {Paolini}, \citenamefont {Pasquarello},
  \citenamefont {Paulatto}, \citenamefont {Sbraccia}, \citenamefont {Scandolo},
  \citenamefont {Sclauzero}, \citenamefont {Seitsonen}, \citenamefont
  {Smogunov}, \citenamefont {Umari},\ and\ \citenamefont
  {Wentzcovitch}}]{Giannozzi_2009}%
  \BibitemOpen
  \bibfield  {author} {\bibinfo {author} {\bibfnamefont {P.}~\bibnamefont
  {Giannozzi}}, \bibinfo {author} {\bibfnamefont {S.}~\bibnamefont {Baroni}},
  \bibinfo {author} {\bibfnamefont {N.}~\bibnamefont {Bonini}}, \bibinfo
  {author} {\bibfnamefont {M.}~\bibnamefont {Calandra}}, \bibinfo {author}
  {\bibfnamefont {R.}~\bibnamefont {Car}}, \bibinfo {author} {\bibfnamefont
  {C.}~\bibnamefont {Cavazzoni}}, \bibinfo {author} {\bibfnamefont
  {D.}~\bibnamefont {Ceresoli}}, \bibinfo {author} {\bibfnamefont {G.~L.}\
  \bibnamefont {Chiarotti}}, \bibinfo {author} {\bibfnamefont {M.}~\bibnamefont
  {Cococcioni}}, \bibinfo {author} {\bibfnamefont {I.}~\bibnamefont {Dabo}},
  \bibinfo {author} {\bibfnamefont {A.}~\bibnamefont {Dal~Corso}}, \bibinfo
  {author} {\bibfnamefont {S.}~\bibnamefont {de~Gironcoli}}, \bibinfo {author}
  {\bibfnamefont {S.}~\bibnamefont {Fabris}}, \bibinfo {author} {\bibfnamefont
  {G.}~\bibnamefont {Fratesi}}, \bibinfo {author} {\bibfnamefont
  {R.}~\bibnamefont {Gebauer}}, \bibinfo {author} {\bibfnamefont
  {U.}~\bibnamefont {Gerstmann}}, \bibinfo {author} {\bibfnamefont
  {C.}~\bibnamefont {Gougoussis}}, \bibinfo {author} {\bibfnamefont
  {A.}~\bibnamefont {Kokalj}}, \bibinfo {author} {\bibfnamefont
  {M.}~\bibnamefont {Lazzeri}}, \bibinfo {author} {\bibfnamefont
  {L.}~\bibnamefont {Martin-Samos}}, \bibinfo {author} {\bibfnamefont
  {N.}~\bibnamefont {Marzari}}, \bibinfo {author} {\bibfnamefont
  {F.}~\bibnamefont {Mauri}}, \bibinfo {author} {\bibfnamefont
  {R.}~\bibnamefont {Mazzarello}}, \bibinfo {author} {\bibfnamefont
  {S.}~\bibnamefont {Paolini}}, \bibinfo {author} {\bibfnamefont
  {A.}~\bibnamefont {Pasquarello}}, \bibinfo {author} {\bibfnamefont
  {L.}~\bibnamefont {Paulatto}}, \bibinfo {author} {\bibfnamefont
  {C.}~\bibnamefont {Sbraccia}}, \bibinfo {author} {\bibfnamefont
  {S.}~\bibnamefont {Scandolo}}, \bibinfo {author} {\bibfnamefont
  {G.}~\bibnamefont {Sclauzero}}, \bibinfo {author} {\bibfnamefont {A.~P.}\
  \bibnamefont {Seitsonen}}, \bibinfo {author} {\bibfnamefont {A.}~\bibnamefont
  {Smogunov}}, \bibinfo {author} {\bibfnamefont {P.}~\bibnamefont {Umari}},\
  and\ \bibinfo {author} {\bibfnamefont {R.~M.}\ \bibnamefont {Wentzcovitch}},\
  }\href {https://doi.org/10.1088/0953-8984/21/39/395502} {\bibfield  {journal}
  {\bibinfo  {journal} {Journal of Physics: Condensed Matter}\ }\textbf
  {\bibinfo {volume} {21}},\ \bibinfo {pages} {395502} (\bibinfo {year}
  {2009})}\BibitemShut {NoStop}%
\bibitem [{\citenamefont {Bart\'ok}\ \emph {et~al.}(2013)\citenamefont
  {Bart\'ok}, \citenamefont {Kondor},\ and\ \citenamefont
  {Cs\'anyi}}]{Bartok_2013}%
  \BibitemOpen
  \bibfield  {author} {\bibinfo {author} {\bibfnamefont {A.~P.}\ \bibnamefont
  {Bart\'ok}}, \bibinfo {author} {\bibfnamefont {R.}~\bibnamefont {Kondor}},\
  and\ \bibinfo {author} {\bibfnamefont {G.}~\bibnamefont {Cs\'anyi}},\ }\href
  {https://doi.org/10.1103/PhysRevB.87.184115} {\bibfield  {journal} {\bibinfo
  {journal} {Phys. Rev. B}\ }\textbf {\bibinfo {volume} {87}},\ \bibinfo
  {pages} {184115} (\bibinfo {year} {2013})}\BibitemShut {NoStop}%
\bibitem [{\citenamefont {Himanen}\ \emph {et~al.}(2020)\citenamefont
  {Himanen}, \citenamefont {J{\"a}ger}, \citenamefont {Morooka}, \citenamefont
  {Federici~Canova}, \citenamefont {Ranawat}, \citenamefont {Gao},
  \citenamefont {Rinke},\ and\ \citenamefont {Foster}}]{dscribe}%
  \BibitemOpen
  \bibfield  {author} {\bibinfo {author} {\bibfnamefont {L.}~\bibnamefont
  {Himanen}}, \bibinfo {author} {\bibfnamefont {M.~O.~J.}\ \bibnamefont
  {J{\"a}ger}}, \bibinfo {author} {\bibfnamefont {E.~V.}\ \bibnamefont
  {Morooka}}, \bibinfo {author} {\bibfnamefont {F.}~\bibnamefont
  {Federici~Canova}}, \bibinfo {author} {\bibfnamefont {Y.~S.}\ \bibnamefont
  {Ranawat}}, \bibinfo {author} {\bibfnamefont {D.~Z.}\ \bibnamefont {Gao}},
  \bibinfo {author} {\bibfnamefont {P.}~\bibnamefont {Rinke}},\ and\ \bibinfo
  {author} {\bibfnamefont {A.~S.}\ \bibnamefont {Foster}},\ }\href
  {https://doi.org/10.1016/j.cpc.2019.106949} {\bibfield  {journal} {\bibinfo
  {journal} {Computer Physics Communications}\ }\textbf {\bibinfo {volume}
  {247}},\ \bibinfo {pages} {106949} (\bibinfo {year} {2020})}\BibitemShut
  {NoStop}%
\bibitem [{\citenamefont {Laakso}\ \emph {et~al.}(2023)\citenamefont {Laakso},
  \citenamefont {Himanen}, \citenamefont {Homm}, \citenamefont {Morooka},
  \citenamefont {J{\"a}ger}, \citenamefont {Todorovi{\'c}},\ and\ \citenamefont
  {Rinke}}]{dscribe2}%
  \BibitemOpen
  \bibfield  {author} {\bibinfo {author} {\bibfnamefont {J.}~\bibnamefont
  {Laakso}}, \bibinfo {author} {\bibfnamefont {L.}~\bibnamefont {Himanen}},
  \bibinfo {author} {\bibfnamefont {H.}~\bibnamefont {Homm}}, \bibinfo {author}
  {\bibfnamefont {E.~V.}\ \bibnamefont {Morooka}}, \bibinfo {author}
  {\bibfnamefont {M.~O.}\ \bibnamefont {J{\"a}ger}}, \bibinfo {author}
  {\bibfnamefont {M.}~\bibnamefont {Todorovi{\'c}}},\ and\ \bibinfo {author}
  {\bibfnamefont {P.}~\bibnamefont {Rinke}},\ }\href@noop {} {\bibfield
  {journal} {\bibinfo  {journal} {The Journal of Chemical Physics}\ }\textbf
  {\bibinfo {volume} {158}} (\bibinfo {year} {2023})}\BibitemShut {NoStop}%
\bibitem [{\citenamefont {Drautz}(2019)}]{Drautz2019}%
  \BibitemOpen
  \bibfield  {author} {\bibinfo {author} {\bibfnamefont {R.}~\bibnamefont
  {Drautz}},\ }\href {https://doi.org/10.1103/PhysRevB.99.014104} {\bibfield
  {journal} {\bibinfo  {journal} {Phys. Rev. B}\ }\textbf {\bibinfo {volume}
  {99}},\ \bibinfo {pages} {014104} (\bibinfo {year} {2019})}\BibitemShut
  {NoStop}%
\bibitem [{\citenamefont {Breiman}(2001)}]{Breiman2001}%
  \BibitemOpen
  \bibfield  {author} {\bibinfo {author} {\bibfnamefont {L.}~\bibnamefont
  {Breiman}},\ }\href {https://doi.org/10.1023/A:1010933404324} {\bibfield
  {journal} {\bibinfo  {journal} {Machine Learning}\ }\textbf {\bibinfo
  {volume} {45}},\ \bibinfo {pages} {5} (\bibinfo {year} {2001})}\BibitemShut
  {NoStop}%
\bibitem [{\citenamefont {Pedregosa}\ \emph {et~al.}(2011)\citenamefont
  {Pedregosa}, \citenamefont {Varoquaux}, \citenamefont {Gramfort},
  \citenamefont {Michel}, \citenamefont {Thirion}, \citenamefont {Grisel},
  \citenamefont {Blondel}, \citenamefont {Prettenhofer}, \citenamefont {Weiss},
  \citenamefont {Dubourg}, \citenamefont {Vanderplas}, \citenamefont {Passos},
  \citenamefont {Cournapeau}, \citenamefont {Brucher}, \citenamefont {Perrot},\
  and\ \citenamefont {Duchesnay}}]{scikit-learn}%
  \BibitemOpen
  \bibfield  {author} {\bibinfo {author} {\bibfnamefont {F.}~\bibnamefont
  {Pedregosa}}, \bibinfo {author} {\bibfnamefont {G.}~\bibnamefont
  {Varoquaux}}, \bibinfo {author} {\bibfnamefont {A.}~\bibnamefont {Gramfort}},
  \bibinfo {author} {\bibfnamefont {V.}~\bibnamefont {Michel}}, \bibinfo
  {author} {\bibfnamefont {B.}~\bibnamefont {Thirion}}, \bibinfo {author}
  {\bibfnamefont {O.}~\bibnamefont {Grisel}}, \bibinfo {author} {\bibfnamefont
  {M.}~\bibnamefont {Blondel}}, \bibinfo {author} {\bibfnamefont
  {P.}~\bibnamefont {Prettenhofer}}, \bibinfo {author} {\bibfnamefont
  {R.}~\bibnamefont {Weiss}}, \bibinfo {author} {\bibfnamefont
  {V.}~\bibnamefont {Dubourg}}, \bibinfo {author} {\bibfnamefont
  {J.}~\bibnamefont {Vanderplas}}, \bibinfo {author} {\bibfnamefont
  {A.}~\bibnamefont {Passos}}, \bibinfo {author} {\bibfnamefont
  {D.}~\bibnamefont {Cournapeau}}, \bibinfo {author} {\bibfnamefont
  {M.}~\bibnamefont {Brucher}}, \bibinfo {author} {\bibfnamefont
  {M.}~\bibnamefont {Perrot}},\ and\ \bibinfo {author} {\bibfnamefont
  {E.}~\bibnamefont {Duchesnay}},\ }\href@noop {} {\bibfield  {journal}
  {\bibinfo  {journal} {Journal of Machine Learning Research}\ }\textbf
  {\bibinfo {volume} {12}},\ \bibinfo {pages} {2825} (\bibinfo {year}
  {2011})}\BibitemShut {NoStop}%
\bibitem [{\citenamefont {Kurganskii}\ \emph {et~al.}(1985)\citenamefont
  {Kurganskii}, \citenamefont {Dubrovskii},\ and\ \citenamefont
  {Domashevskaya}}]{Kurganskii_2D_BZ_integration}%
  \BibitemOpen
  \bibfield  {author} {\bibinfo {author} {\bibfnamefont {S.~I.}\ \bibnamefont
  {Kurganskii}}, \bibinfo {author} {\bibfnamefont {O.~I.}\ \bibnamefont
  {Dubrovskii}},\ and\ \bibinfo {author} {\bibfnamefont {E.~P.}\ \bibnamefont
  {Domashevskaya}},\ }\href
  {https://doi.org/https://doi.org/10.1002/pssb.2221290129} {\bibfield
  {journal} {\bibinfo  {journal} {Physica Status Solidi (B)}\ }\textbf
  {\bibinfo {volume} {129}},\ \bibinfo {pages} {293} (\bibinfo {year}
  {1985})}\BibitemShut {NoStop}%
\bibitem [{\citenamefont {Hu}\ and\ \citenamefont
  {MacDonald}(2021)}]{Hu_moire_2021}%
  \BibitemOpen
  \bibfield  {author} {\bibinfo {author} {\bibfnamefont {N.~C.}\ \bibnamefont
  {Hu}}\ and\ \bibinfo {author} {\bibfnamefont {A.~H.}\ \bibnamefont
  {MacDonald}},\ }\href {https://doi.org/10.1103/PhysRevB.104.214403}
  {\bibfield  {journal} {\bibinfo  {journal} {Phys. Rev. B}\ }\textbf {\bibinfo
  {volume} {104}},\ \bibinfo {pages} {214403} (\bibinfo {year}
  {2021})}\BibitemShut {NoStop}%
\bibitem [{\citenamefont {{Dal Corso}}(2014)}]{DalCorso_2014}%
  \BibitemOpen
  \bibfield  {author} {\bibinfo {author} {\bibfnamefont {A.}~\bibnamefont {{Dal
  Corso}}},\ }\href
  {https://doi.org/https://doi.org/10.1016/j.commatsci.2014.07.043} {\bibfield
  {journal} {\bibinfo  {journal} {Computational Materials Science}\ }\textbf
  {\bibinfo {volume} {95}},\ \bibinfo {pages} {337} (\bibinfo {year}
  {2014})}\BibitemShut {NoStop}%
\bibitem [{\citenamefont {Perdew}\ \emph {et~al.}(1996)\citenamefont {Perdew},
  \citenamefont {Burke},\ and\ \citenamefont {Ernzerhof}}]{PBE}%
  \BibitemOpen
  \bibfield  {author} {\bibinfo {author} {\bibfnamefont {J.~P.}\ \bibnamefont
  {Perdew}}, \bibinfo {author} {\bibfnamefont {K.}~\bibnamefont {Burke}},\ and\
  \bibinfo {author} {\bibfnamefont {M.}~\bibnamefont {Ernzerhof}},\ }\href
  {https://doi.org/10.1103/PhysRevLett.77.3865} {\bibfield  {journal} {\bibinfo
   {journal} {Phys. Rev. Lett.}\ }\textbf {\bibinfo {volume} {77}},\ \bibinfo
  {pages} {3865} (\bibinfo {year} {1996})}\BibitemShut {NoStop}%
\bibitem [{\citenamefont {Gong}\ \emph {et~al.}(2017)\citenamefont {Gong},
  \citenamefont {Li}, \citenamefont {Li}, \citenamefont {Ji}, \citenamefont
  {Stern}, \citenamefont {Xia}, \citenamefont {Cao}, \citenamefont {Bao},
  \citenamefont {Wang}, \citenamefont {Wang}, \citenamefont {Qiu},
  \citenamefont {Cava}, \citenamefont {Louie}, \citenamefont {Xia},\ and\
  \citenamefont {Zhang}}]{Gong2017}%
  \BibitemOpen
  \bibfield  {author} {\bibinfo {author} {\bibfnamefont {C.}~\bibnamefont
  {Gong}}, \bibinfo {author} {\bibfnamefont {L.}~\bibnamefont {Li}}, \bibinfo
  {author} {\bibfnamefont {Z.}~\bibnamefont {Li}}, \bibinfo {author}
  {\bibfnamefont {H.}~\bibnamefont {Ji}}, \bibinfo {author} {\bibfnamefont
  {A.}~\bibnamefont {Stern}}, \bibinfo {author} {\bibfnamefont
  {Y.}~\bibnamefont {Xia}}, \bibinfo {author} {\bibfnamefont {T.}~\bibnamefont
  {Cao}}, \bibinfo {author} {\bibfnamefont {W.}~\bibnamefont {Bao}}, \bibinfo
  {author} {\bibfnamefont {C.}~\bibnamefont {Wang}}, \bibinfo {author}
  {\bibfnamefont {Y.}~\bibnamefont {Wang}}, \bibinfo {author} {\bibfnamefont
  {Z.~Q.}\ \bibnamefont {Qiu}}, \bibinfo {author} {\bibfnamefont {R.~J.}\
  \bibnamefont {Cava}}, \bibinfo {author} {\bibfnamefont {S.~G.}\ \bibnamefont
  {Louie}}, \bibinfo {author} {\bibfnamefont {J.}~\bibnamefont {Xia}},\ and\
  \bibinfo {author} {\bibfnamefont {X.}~\bibnamefont {Zhang}},\ }\href
  {https://doi.org/10.1038/nature22060} {\bibfield  {journal} {\bibinfo
  {journal} {Nature}\ }\textbf {\bibinfo {volume} {546}},\ \bibinfo {pages}
  {265} (\bibinfo {year} {2017})}\BibitemShut {NoStop}%
\bibitem [{\citenamefont {Milardovich}\ \emph {et~al.}(2023)\citenamefont
  {Milardovich}, \citenamefont {Wilhelmer}, \citenamefont {Waldhoer},
  \citenamefont {Cvitkovich}, \citenamefont {Sivaraman},\ and\ \citenamefont
  {Grasser}}]{Milardovich_ML_2023}%
  \BibitemOpen
  \bibfield  {author} {\bibinfo {author} {\bibfnamefont {D.}~\bibnamefont
  {Milardovich}}, \bibinfo {author} {\bibfnamefont {C.}~\bibnamefont
  {Wilhelmer}}, \bibinfo {author} {\bibfnamefont {D.}~\bibnamefont {Waldhoer}},
  \bibinfo {author} {\bibfnamefont {L.}~\bibnamefont {Cvitkovich}}, \bibinfo
  {author} {\bibfnamefont {G.}~\bibnamefont {Sivaraman}},\ and\ \bibinfo
  {author} {\bibfnamefont {T.}~\bibnamefont {Grasser}},\ }\href
  {https://doi.org/10.1063/5.0146753} {\bibfield  {journal} {\bibinfo
  {journal} {The Journal of Chemical Physics}\ }\textbf {\bibinfo {volume}
  {158}},\ \bibinfo {pages} {194802} (\bibinfo {year} {2023})}\BibitemShut
  {NoStop}%
\bibitem [{\citenamefont {Cvitkovich}\ \emph {et~al.}(2024)\citenamefont
  {Cvitkovich}, \citenamefont {Fehringer}, \citenamefont {Wilhelmer},
  \citenamefont {Milardovich}, \citenamefont {Waldhör},\ and\ \citenamefont
  {Grasser}}]{Cvitkovich_ML_2024}%
  \BibitemOpen
  \bibfield  {author} {\bibinfo {author} {\bibfnamefont {L.}~\bibnamefont
  {Cvitkovich}}, \bibinfo {author} {\bibfnamefont {F.}~\bibnamefont
  {Fehringer}}, \bibinfo {author} {\bibfnamefont {C.}~\bibnamefont
  {Wilhelmer}}, \bibinfo {author} {\bibfnamefont {D.}~\bibnamefont
  {Milardovich}}, \bibinfo {author} {\bibfnamefont {D.}~\bibnamefont
  {Waldhör}},\ and\ \bibinfo {author} {\bibfnamefont {T.}~\bibnamefont
  {Grasser}},\ }\href {https://doi.org/10.1063/5.0220091} {\bibfield  {journal}
  {\bibinfo  {journal} {The Journal of Chemical Physics}\ }\textbf {\bibinfo
  {volume} {161}},\ \bibinfo {pages} {144706} (\bibinfo {year}
  {2024})}\BibitemShut {NoStop}%
\bibitem [{\citenamefont {Caro}(2019)}]{Caro_2019}%
  \BibitemOpen
  \bibfield  {author} {\bibinfo {author} {\bibfnamefont {M.~A.}\ \bibnamefont
  {Caro}},\ }\href {https://doi.org/10.1103/PhysRevB.100.024112} {\bibfield
  {journal} {\bibinfo  {journal} {Phys. Rev. B}\ }\textbf {\bibinfo {volume}
  {100}},\ \bibinfo {pages} {024112} (\bibinfo {year} {2019})}\BibitemShut
  {NoStop}%
\bibitem [{\citenamefont {Rasmussen}\ and\ \citenamefont
  {Williams}(2006)}]{GPR2006}%
  \BibitemOpen
  \bibfield  {author} {\bibinfo {author} {\bibfnamefont {C.~E.}\ \bibnamefont
  {Rasmussen}}\ and\ \bibinfo {author} {\bibfnamefont {C.~K.~I.}\ \bibnamefont
  {Williams}},\ }\href@noop {} {\emph {\bibinfo {title} {Gaussian Processes for
  Machine Learning}}}\ (\bibinfo  {publisher} {MIT Press},\ \bibinfo {year}
  {2006})\BibitemShut {NoStop}%
\bibitem [{\citenamefont {Friedman}(2001)}]{gradient_boosting}%
  \BibitemOpen
  \bibfield  {author} {\bibinfo {author} {\bibfnamefont {J.~H.}\ \bibnamefont
  {Friedman}},\ }\href {https://doi.org/10.1214/aos/1013203451} {\bibfield
  {journal} {\bibinfo  {journal} {The Annals of Statistics}\ }\textbf {\bibinfo
  {volume} {29}},\ \bibinfo {pages} {1189 } (\bibinfo {year}
  {2001})}\BibitemShut {NoStop}%
\bibitem [{\citenamefont {Chen}\ and\ \citenamefont
  {Guestrin}(2016)}]{XGBoost}%
  \BibitemOpen
  \bibfield  {author} {\bibinfo {author} {\bibfnamefont {T.}~\bibnamefont
  {Chen}}\ and\ \bibinfo {author} {\bibfnamefont {C.}~\bibnamefont
  {Guestrin}},\ }in\ \href {https://doi.org/10.1145/2939672.2939785} {\emph
  {\bibinfo {booktitle} {Proceedings of the 22nd ACM SIGKDD International
  Conference on Knowledge Discovery and Data Mining}}},\ \bibinfo {series and
  number} {KDD '16}\ (\bibinfo  {publisher} {Association for Computing
  Machinery},\ \bibinfo {address} {New York, NY, USA},\ \bibinfo {year}
  {2016})\ p.\ \bibinfo {pages} {785–794}\BibitemShut {NoStop}%
\bibitem [{\citenamefont {Cvitkovich}()}]{Link_2025}%
  \BibitemOpen
  \bibfield  {author} {\bibinfo {author} {\bibfnamefont {L.}~\bibnamefont
  {Cvitkovich}},\ }\href@noop {} {\bibinfo {title} {Link to git repository}},\
  \bibinfo {note} {to be published after acceptance}\BibitemShut {NoStop}%
\bibitem [{\citenamefont {Cranmer}\ \emph {et~al.}(2020)\citenamefont
  {Cranmer}, \citenamefont {Sanchez-Gonzalez}, \citenamefont {Battaglia},
  \citenamefont {Xu}, \citenamefont {Cranmer}, \citenamefont {Spergel},\ and\
  \citenamefont {Ho}}]{cranmerDiscovering2020}%
  \BibitemOpen
  \bibfield  {author} {\bibinfo {author} {\bibfnamefont {M.}~\bibnamefont
  {Cranmer}}, \bibinfo {author} {\bibfnamefont {A.}~\bibnamefont
  {Sanchez-Gonzalez}}, \bibinfo {author} {\bibfnamefont {P.}~\bibnamefont
  {Battaglia}}, \bibinfo {author} {\bibfnamefont {R.}~\bibnamefont {Xu}},
  \bibinfo {author} {\bibfnamefont {K.}~\bibnamefont {Cranmer}}, \bibinfo
  {author} {\bibfnamefont {D.}~\bibnamefont {Spergel}},\ and\ \bibinfo {author}
  {\bibfnamefont {S.}~\bibnamefont {Ho}},\ }\href@noop {} {\bibfield  {journal}
  {\bibinfo  {journal} {NeurIPS 2020}\ } (\bibinfo {year} {2020})}\BibitemShut
  {NoStop}%
\bibitem [{\citenamefont
  {Cranmer}(2023)}]{cranmerInterpretableMachineLearning2023}%
  \BibitemOpen
  \bibfield  {author} {\bibinfo {author} {\bibfnamefont {M.}~\bibnamefont
  {Cranmer}},\ }\href {https://doi.org/10.48550/arXiv.2305.01582} {\bibinfo
  {title} {Interpretable {Machine} {Learning} for {Science} with {PySR} and
  {SymbolicRegression}.jl}} (\bibinfo {year} {2023}),\ \bibinfo {note}
  {arXiv:2305.01582 [astro-ph, physics:physics]}\BibitemShut {NoStop}%
\end{thebibliography}%
	

	\clearpage
	\onecolumngrid
	
	\appendix
	\section{Training Dataset and DFT setup}
	\label{app:training}
	
	An overview of the complete dataset used for training of the ML model is given in Tab.~\ref{tab:training}. All structures are relaxed to consider correct interlayer distances $d_\mathrm{IL}$ and  rippling of the graphene layer ($<1$\,pm)~\cite{Zollner_Engineering_2022}. We consider heterostructures with twist angles $\theta$ ranging from 0 to 30$\degree$ in steps of 3$\degree$. One structure (at $\theta \approx18\degree$) was removed due to unrealistic strains and band offsets in the commensurate supercell. 
	Lateral shifts are applied in two ways. The dataset is based on commensurate supercells with a three-fold rotation symmetry. The rotation symmetry is handy in terms of computational costs, however, reduces the information contained in such a calculation. For ML, a more diverse dataset is required (which is not clear a priori and only realized after obtaining large errors in the predictions using the too small data sets). We have thus extended our dataset by three measures. First, by translation of the graphene layer, retaining the three-fold rotation symmetry. Second, by rigid, lateral, and randomized shifts which break the remaining symmetries (apart from time-reversal symmetry). Finally, we consider additional interlayer distances where the whole graphene layer is moved toward (away) from the CGT layer by 0.1\,\AA. In total, 75 different structures with 9498 C atoms are contained in the data set.
	
	The calculations were carried out using plane-wave DFT as implemented in the code \textit{Quantum ESPRESSO}. According to the system size, the $k$-point mesh has been chosen as listed in Tab.~\ref{tab:training}. Data for the ML training set -- in contrast to the data for the detailed analysis of the magnetization mechanism -- is calculated based on standard uniform meshes.
	Furthermore, we used PAW pseudopotentials with scalar relativistic non-linear core correction~\cite{DalCorso_2014} in combination with the PBE functional~\cite{PBE}. Correlation effects of Cr 3d electrons are considered by a Hubbard correction of $U=1.0$\,eV~. \LC{In Ref.~\cite{Gong2017}, it has been shown that the system becomes in-plane anisotropic for $U<0.2$\,eV, and that the interlayer coupling becomes antiferromagnetic for $U>1.7$\,eV, contradicting the experiment. Therefore, the $U$ values within the range of 0.2-1.7\,eV are argued to be a reasonable choice. Our Hubbard U is in the middle of this range.}

	\begin{table}[ht!]
		\caption{CGT/Gr structures in the training dataset of the ML model. In the first two columns, the number of (carbon) atoms, are given. The third row lists the number of considered structures. This number results from the changes (relaxation, lateral shifts, interlayer displacements) that were additionally applied to each structure and added to the dataset. The twist angle $\theta$ denotes the rotational angle between both layers. 
			Lateral shifts are applied in two ways, symmetry-retaining and symmetry-breaking shifts. The number of such operations applied to each structure are given in column 5 and 6. Next, the considered vertical shifts $\delta_\mathrm{IL}$ from the equilibrium $d_\mathrm{IL}=3.55$\,\AA, and finally, the number of used kpoints $n_k$ are indicated.
		}
		\centering
		\renewcommand{\arraystretch}{1.5}
		\begin{tabular}{|c|c|c|c|c|c|c|c|}
			\hline
			\# atoms & \# C atoms & \# structures  & \hspace{0.5cm}$\theta$ [\textdegree]\hspace{0.5cm} & sym. shift & sym. broken & $\delta_\mathrm{IL}$[\AA]&\hspace{0.2cm} $n_k$ \hspace{0.4cm}\\ \hline
			80       & 50         & 4     &      30.00   &  1  & 0 &  $\pm0.1$ & 42 \\ \hline
			102      & 62         & 14    &      8.94    &  2  & 5 &  $\pm0.1, \pm0.2$ &  42  \\ \hline
			174      & 104        & 14    &      26.99   &  2  & 5 &  $\pm0.1, \pm0.2$ &  18  \\ \hline
			218      & 128        & 1     &      0.00    &  0  & 0 &   & 18  \\ \hline
			224      & 134        & 9     &      12.21   &  1  & 1 &  $\pm0.1, \pm0.2$  & 12  \\ \hline
			236      & 146        & 9     &      5.81    &  1  & 1 &  $\pm0.1, \pm0.2$  & 12 \\ \hline
			242      & 152        & 5     &      23.41   &  1  & 1 &  $\pm0.1, \pm0.2$  & 12 \\ \hline
			302      & 182        & 10    &      3.00    &  2  & 2 &  $\pm0.1, \pm0.2$  &  9 \\ \hline
			314      & 194        & 9     &      14.7    &  2  & 1 &  $\pm0.1, \pm0.2$  &  9 \\ \hline 
		\end{tabular}
		\label{tab:training}
	\end{table}
	
	\section{Descriptors for atomic environment}
	\label{app:descriptor}
	Suitable descriptors which transform a configuration of atoms into an abstract mathematical object are motivated by the key requirement of any machine learning model, namely the availability of objectively evaluated features of the data. 
	
\begin{figure}[h]
		\centerline{\includegraphics[width=.4\linewidth]{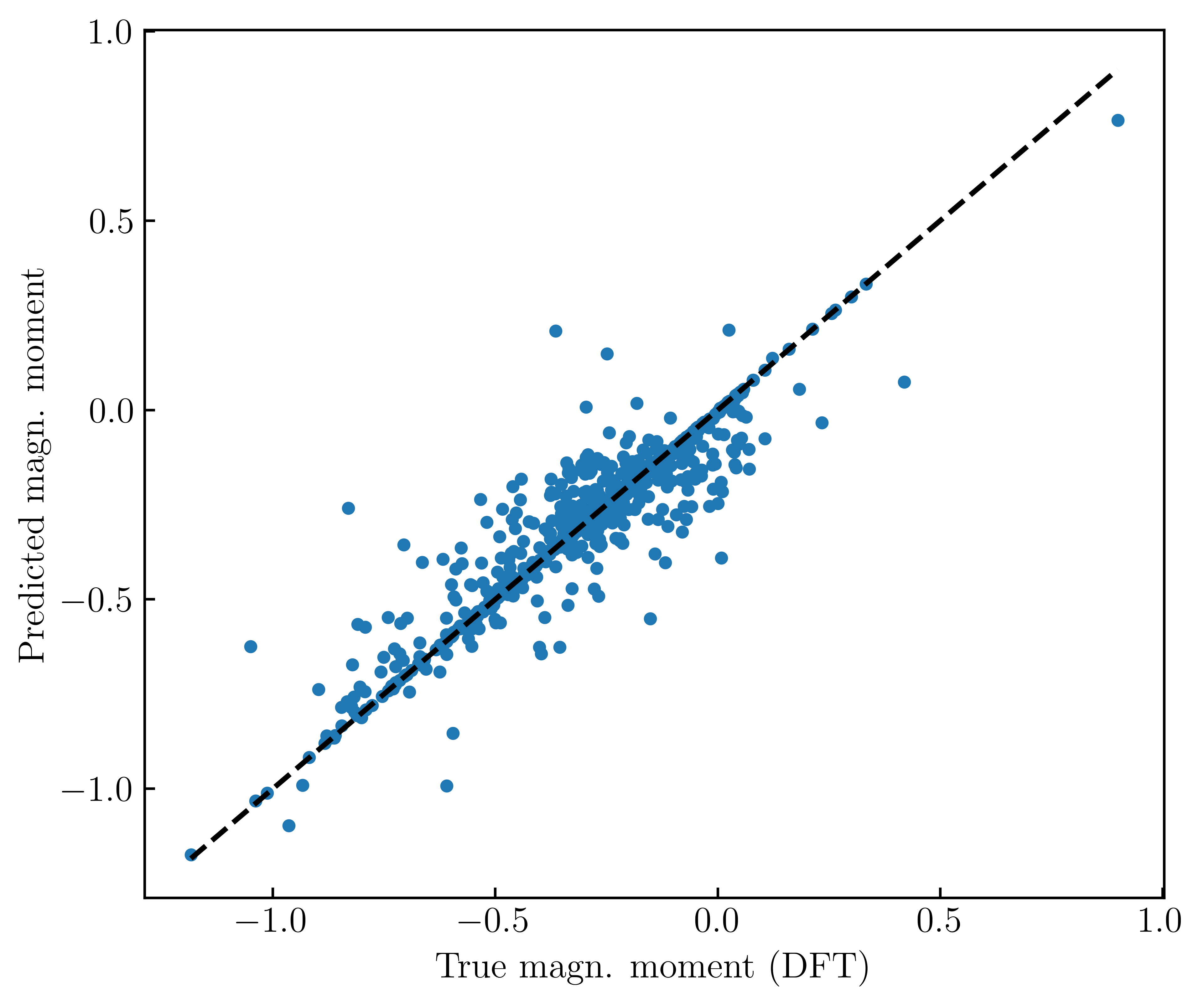}}
		\caption{Correlation between DFT-calculated and ML-predicted values when using our home-made descriptor (see text).
		}
		\label{fig:homemade_desc}
	\end{figure}

	We start with a simple home-made descriptor. This descriptor takes the three spherical coordinates $r, \Theta, \phi$ and the magnetic moment of the respective atom type (intended to mirror the type of atom as well as it's capability to proximitize the C atom in the graphene layer) from its nearest neighbors within a cut-off radius of 16\,\AA. The correlation between DFT and ML predictions is shown in Fig.~\ref{fig:homemade_desc}. Although the prediction works well in some cases, delivering seemingly satisfying results, the mean squared error is $7.5\times 10^{-9}\mu_\mathrm{B}^2$, resulting in an prediction error of $8.6\times 10^{-5}\mu_\mathrm{B}$ -- two times higher than the error obtained with SOAP-based predictions. The improved predictions when using the more sophisticated SOAP approach is another evidence for the complexity of the hybridization process.
	
\begin{figure}[h]
		\centerline{\includegraphics[width=.5\linewidth]{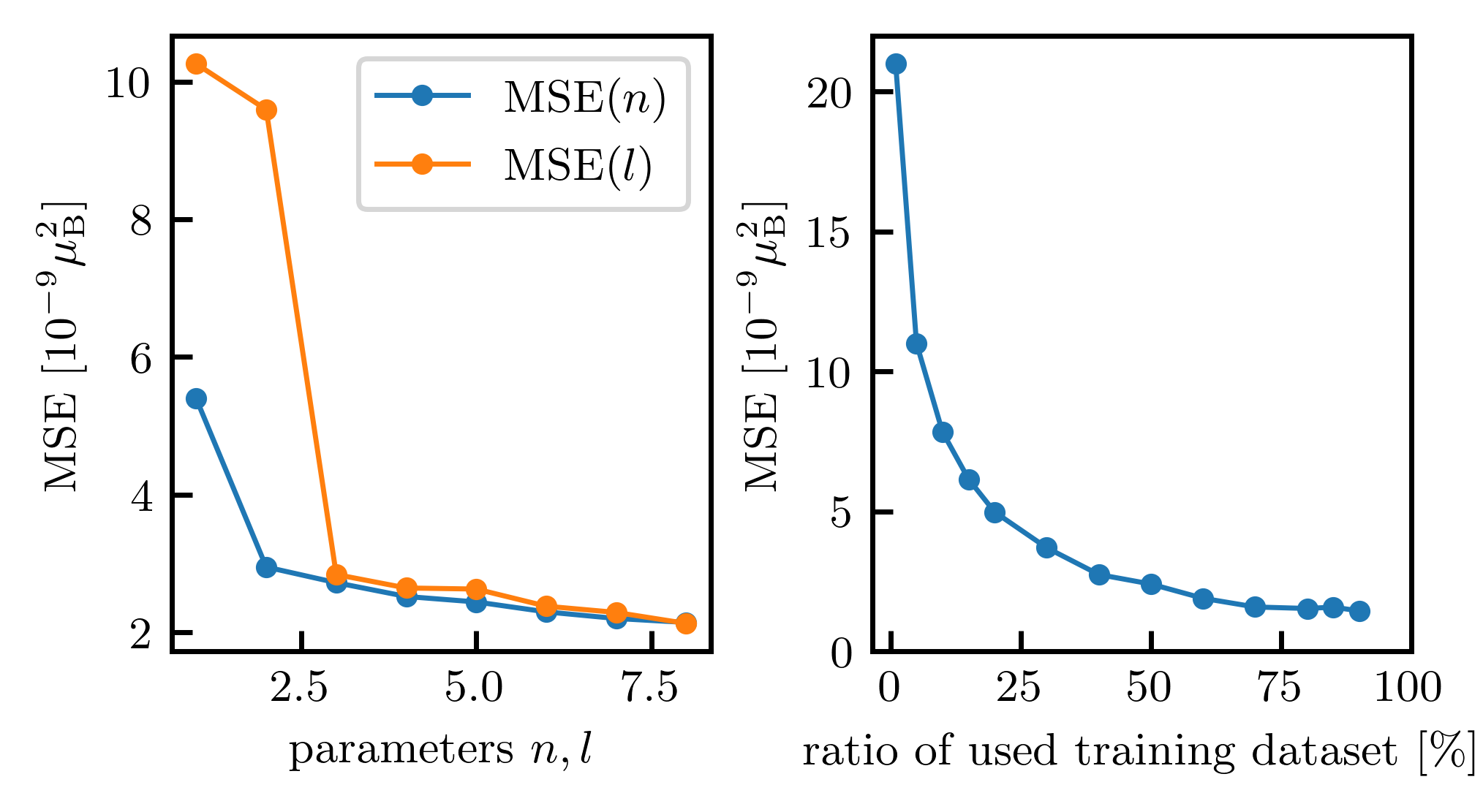}}
		\caption{MSE as a function of the SOAP basis set size and used training set size.
		}
		\label{fig:SOAP_basis}
	\end{figure}

	Due to the above mentioned limitation, we have moved to the Smooth Overlap of Atomic Positions (SOAP) descriptor~\cite{Bartok_2013} as implemented in the \textit{DScribe} Python library~\cite{dscribe, dscribe2}. The SOAP descriptor is design to represent atomic structures and has been found useful for interatomic force fields~\cite{Milardovich_ML_2023, Cvitkovich_ML_2024}. Effectively, it maps an input atomic structure to a column vector.
	The entries of this vector are determined by the overlap of hydrogen-like orbitals which are placed on each atom within a distance of $2 r_{cut}$~\cite{Bartok_2013}.
	Its length depends on the specified parameters (determining e.g. the size of the orbital basis sets). In general, more precise descriptions can be expected for larger orbital basis sets. 
	The descriptor uses a basis of radial functions and spherical harmonics, controlled by parameters $n$ and $l$, respectively. These define the resolution of radial and angular features in the local environment. As shown in Fig.~\ref{fig:SOAP_basis}, a minimum of $n=2$ and $l=3$ (i.e., including up to $f$-like angular components) is necessary to qualitatively capture the induced magnetization. The lowest MSE is achieved with the highest tested complexity, $n=l=8$, showing that the induced magnetization is a result of complex hybridization rooted in small differences between local atomic and chemical environments.
	We find increased accuracy in the ML predictions when weighting the atomic density by a polynomial function $[1+2(r/r_\mathrm{cut})^3-3(r/r_\mathrm{cut})]^m$ as suggested in Ref.~\cite{Caro_2019}. This polynomial goes exactly to zero at $r=r_\mathrm{cut}$ if $m>0$.
	For our predictions, we choose $m=1$, use a cutoff radius of $r_\mathrm{cut}=8$\,\AA, and a Gaussian smearing width of 0.5\,\AA. 
	
	\LC{We note that the optimal orbital basis found above, characterized by $n=l=8$, lacks explicit chemical meaning and primarily reflects the internal workings of SOAP, whose description is limited to 3-body interactions. Other descriptors can reach comparable accuracy using smaller orbital sets. The atomic cluster expansion (ASE) from Ref.~\cite{Drautz2019} builds a systematic expansion over multi-body terms. As a result, a smaller orbital basis can suffice to reach similar prediction accuracy. However, as SOAP provides highly satisfactory accuracy, we leave further optimization of the ML framework for future studies.}
	
	\section{Machine Learning Model}
	\label{app:ML}
	In the input data for training the ML model, each datapoint, i.e. each C atom in the DFT dataset, gets a SOAP vector assigned by considering the orbital overlaps in its local vicinity. The target is the magnetic moment from DFT. As a matter of fact, the ML model has the same purpose as a DFT calculation: To determine electronic and spin properties from a given atomic configuration. Its efficiency, however, exceeds DFT by several orders of magnitude.
	
	The forest regression model~\cite{Breiman2001} was trained with 400 decision trees, using a square root strategy for feature selection at each split, a maximum tree depth of 30, and no bootstrapping. 
	
\begin{figure}
		\centering
		\includegraphics[width=0.75\linewidth]{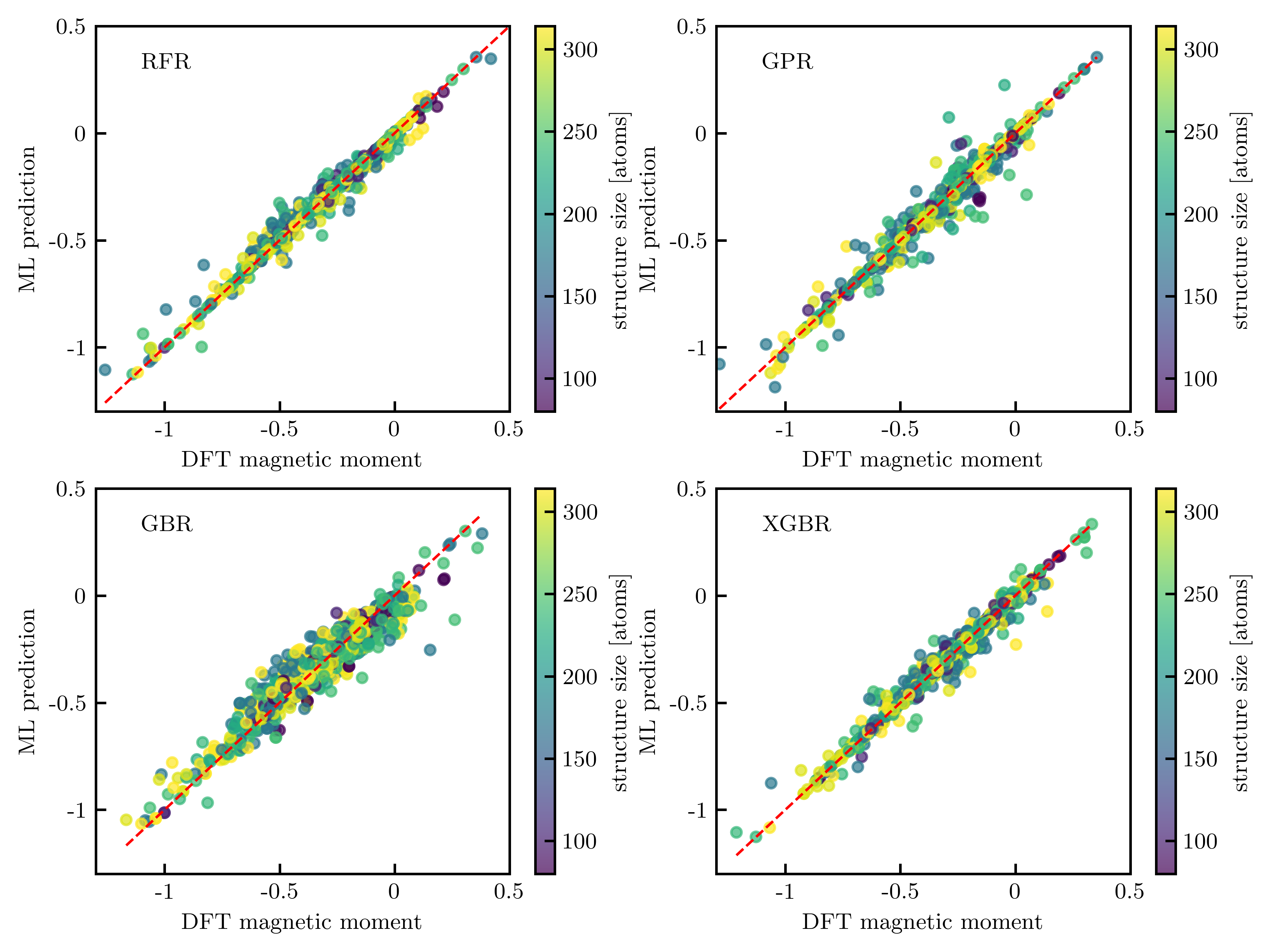}
		\caption{Comparison of four ML models trained on data encoded by the SOAP descriptor.}
		\label{fig:ML_comparison}
	\end{figure}

	\LC{Next, we compare the performance of the random forest regression model to other types of ML algorithms. To this end, we compare the previously employed random forest regressor (RFR) to three other ML algorithms. 
		Two of those, the Gaussian process regressor (GPR)~\cite{GPR2006} and the gradient boosting regressor (GBR)~\cite{gradient_boosting} are taken from the same \textit{SKlearn} Python package. The last one, XGBR is also a gradient boosting regressor but taken from the package \textit{xgboost}~\cite{XGBoost} and is expected to train faster and to achieve slightly better accuracy. For this comparison, we renounce to tune the hyper-parameters of each model separately, but simply match the parameters of each model to the RFR reference. As shown in Fig.~\ref{fig:ML_comparison}, all models exhibit comparable performance. Notably, the XGBR model achieves the lowest error, improving upon the RFR baseline by approximately 20\,\%. While the predictive power of our approach primarily arises from the accurate representation of the atomic environment, the specific choice of machine learning algorithm can further enhance the overall prediction accuracy.}
	
	The ML model is computationally efficient: training with optimal SOAP parameters completes in under 10 minutes on a single Intel Core Ultra 7 165U core. Large-scale predictions are similarly fast; for example, a heterostructure containing 106{,}000 atoms as used for plots in Fig.~4 (main text) is evaluated in 950 seconds. The primary computational bottleneck remains the generation of the DFT training data, which required about 1.5 million core hours. 
	
	The fully trained ML model is available in an open-source repository\cite{Link_2025} as a \textit{pickle} object, ready for direct use in predicting magnetization in arbitrary graphene/CGT heterostructures. A Python example script is provided for reference. The full training dataset, including all atomic structures and corresponding DFT-computed magnetic moments, are also publicly accessible there.
	
	\section{Impact of local atomic environment}
	~\label{app:local_environment}
	This section is devoted to investigate the induced magnetic moments with respect to small changes of the environment, considering only DFT calculations. We demonstrate how sensitive the proximity-induced magnetization is with respect to these changes and specifically discuss the origin of the (anti-)parallel alignment between proximity-induced moments and CGT magnetization.
	
	In all our DFT calculations, we find spreads of the magnetic moments. Although, in 96\,\% of the cases, the magnetization is anti-parallel to the magnetization of the CGT layer (indicated by negative values in our dataset), in some cases, the magnetic moments are aligned with the CGT magnetic moment. In both single-layer CGT and our heterobilayer, Cr atoms possess a large magnetic moment of about 3.5\,$\mu_\mathrm{B}$, driving the overall out-of-plane magnetization. The Cr atoms are in the center of the CGT layer. Te atoms, on the other hand, are at the outermost positions and exhibit a magnetic moment of -0.2$\mu_\mathrm{B}$. The magnetic moments of directly neighboring Te and C atoms are usually aligned in parallel. 
	
\begin{figure}[h]
		\centerline{\includegraphics[width=.9\linewidth]{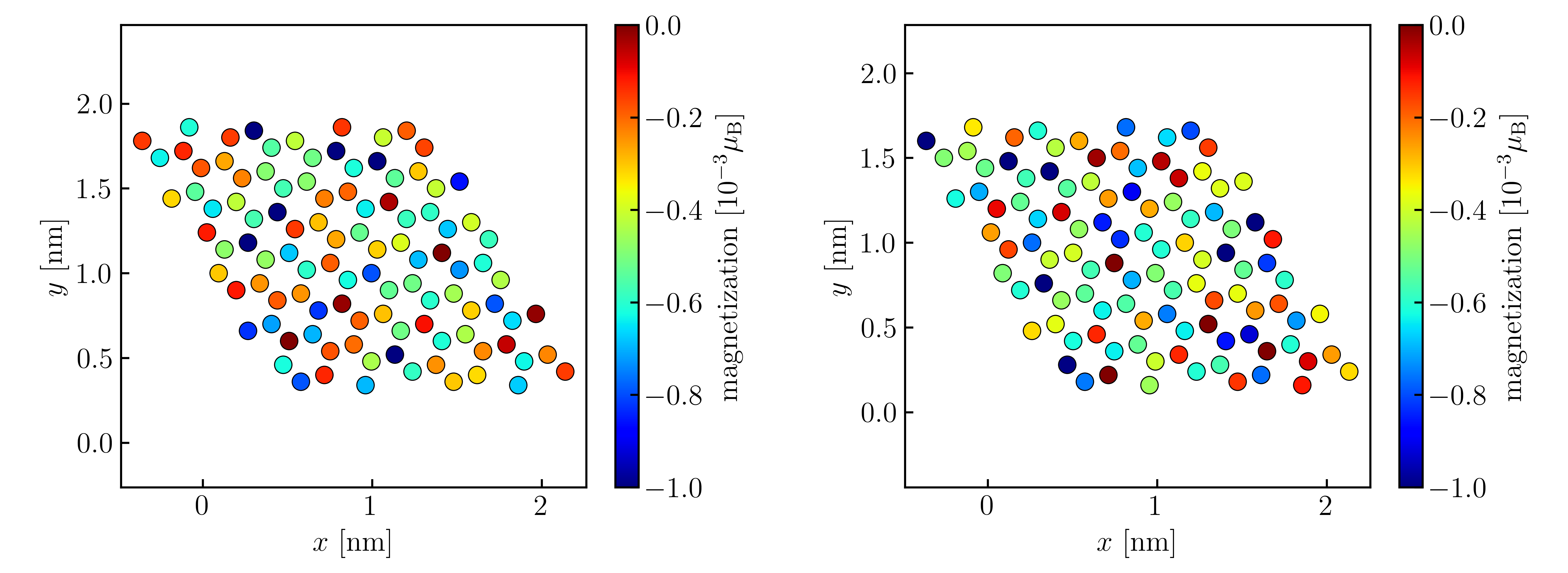}}
		\caption{
			Induced magnetic moment when the lattice is shifted by 0.1\,\AA~with all other properties like the twist angle held constant. Even such small changes in the atomic environment influence the effective proximitized interactions.
		}
		\label{fig:lateral_shift}
	\end{figure}

	First, in Fig.~\ref{fig:lateral_shift} we show the impact of a small lateral shift of the graphene layer with respect to the CGT using one exemplary structure (174 atoms in Tab.~\ref{tab:training}). Despite the small shift of 0.2\,\AA~ -- 3\,\% of the CGT unit cell -- the resulting magnetization differs strongly from the initial pattern. Surprisingly, even the relative differences between neighboring atoms are lost.
	
\begin{figure}[h]
		\centerline{\includegraphics[width=.9\linewidth]{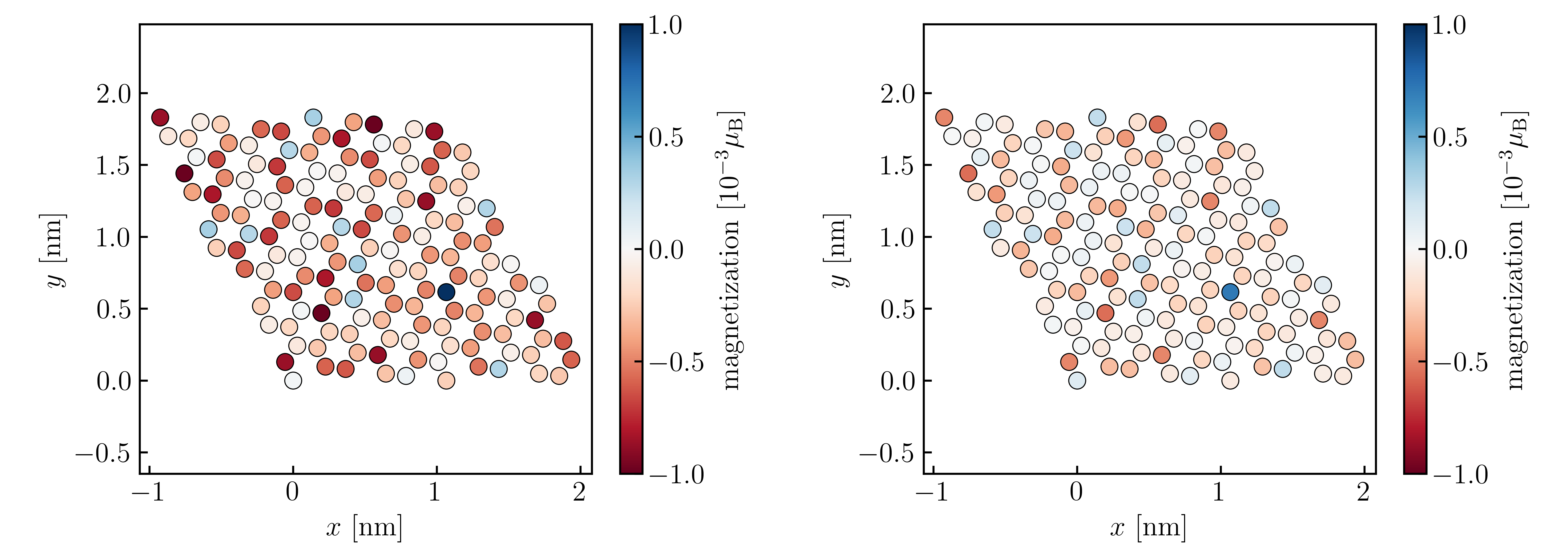}}
		\caption{Induced magnetic moment with the interlayer distance decreased (left) or increased (right) with respect by 0.1\,\AA~with respect to the equilibrium position. While the relative magnetization pattern between the atoms remains, the induced magnetic moments are in general strongly diminished in the second case.}
		\label{fig:perp_shift}
	\end{figure}

	Next, we present the impact of varying interlayer distance. We consider the structure with 242 atoms shifted by -0.1\,\AA~and +0.1\,\AA~from the equilibrium position, the latter being one particular example showing many positive moments at the C atoms.
	In Fig.~\ref{fig:perp_shift}, a different color map is chosen to better distinguish between positive and negative values. In contrast to the lateral shift in Fig.~\ref{fig:lateral_shift}, the magnetization pattern persists between both cases. The lateral position is decisive for the relative magnetization among the atoms. It is quite clear, however, that the proximity exchange depends very strongly on the interlayer distance. Moving the graphene layer away from the CGT results in drastically reduced magnetic moments. As a result, these structures are also more likely to have C atoms with positive magnetic moments, i.e. in parallel with the CGT magnetization. This can be interpreted as a decay of the proximity interaction with Te atoms while the (previously covered) interaction with Cr atoms (or the whole CGT layer) becomes dominant.
	
\begin{figure}[h]
		\centerline{\includegraphics[width=\linewidth]{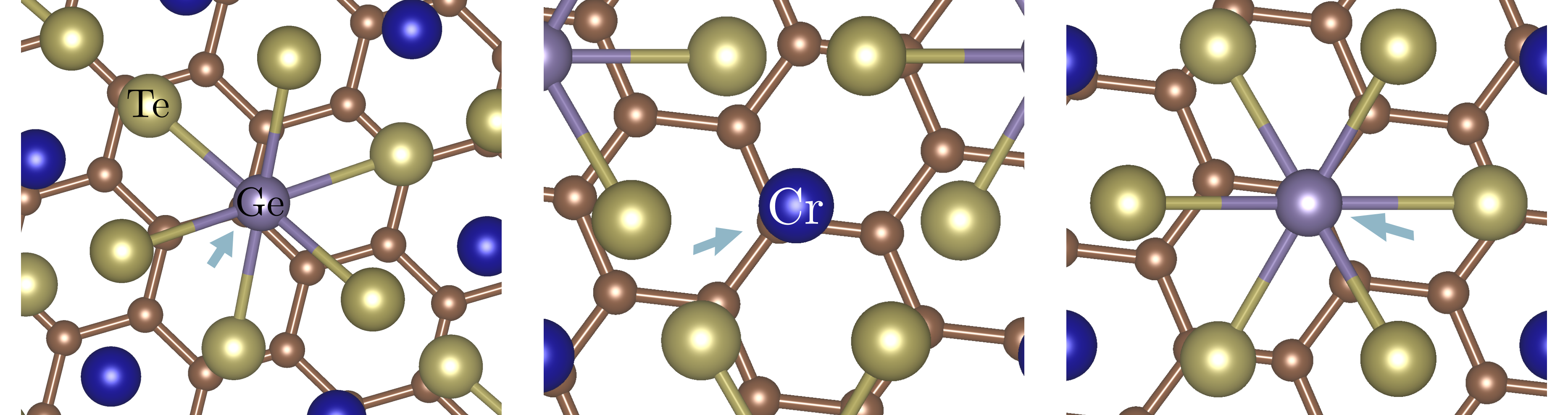}}
		\caption{Atomic registries resulting in a parallel alignment of the magnetization of the considered C atom (indicated by an arrow) and the CGT layer. The prototypical configuration are C atoms above Cr or Ge atoms. Their neighboring atoms face the ``opposite'' atomic environment, with mostly Te atoms as their neighbors. Therefore, their magnetization is often flipped.}
		\label{fig:atomic_registries}
	\end{figure}

	Finally, we discuss typical atomic environments (``registries'') associated with the rare finding of positive magnetic moments appearing even in structures with equilibrium $d_\mathrm{IL}$ or below (so strong interactions can be expected). Three representative cases are presented in Fig.~\ref{fig:atomic_registries}. It is obvious that these cases fulfill two conditions. First, the positively polarized C atom is maximally distanced to all negatively polarized Te atoms. Second, the C atom overlaps with another atom which also carries a positive magnetic moment, so Cr or Ge (which has a magnetic moment of 0.6\,$\mu_\mathrm{B}$).
	Nevertheless, these prerequisites are not more than a rule of thumb. We have also observed configurations with C atoms overlapping with both Cr and Ge atoms where the induced magnetic moment is negative. Furthermore, significant interactions with Cr or Ge are not expected from the band structure in Sec.~\ref{app:hybridization} and because they are not directly neighboring the C atoms. 
	We conclude that the exact moment is set by other details of the local environment, whose complexity justify and even necessitate the application of machine learning techniques.
	
	Given the quite different lattice parameters of the two materials, directly neighboring atoms face a completely different atomic configuration above the CGT layer. Therefore, the induced magnetization changes on a very short scale, the scale of a bond length, as can be seen for example in the magnetization maps of Fig.~\ref{fig:lateral_shift}.
	
	\section{Verification of the ML model within the DFT data set}
	\label{app:verification}
\begin{figure}[htbp]
		\centerline{\includegraphics[width=0.5\linewidth]{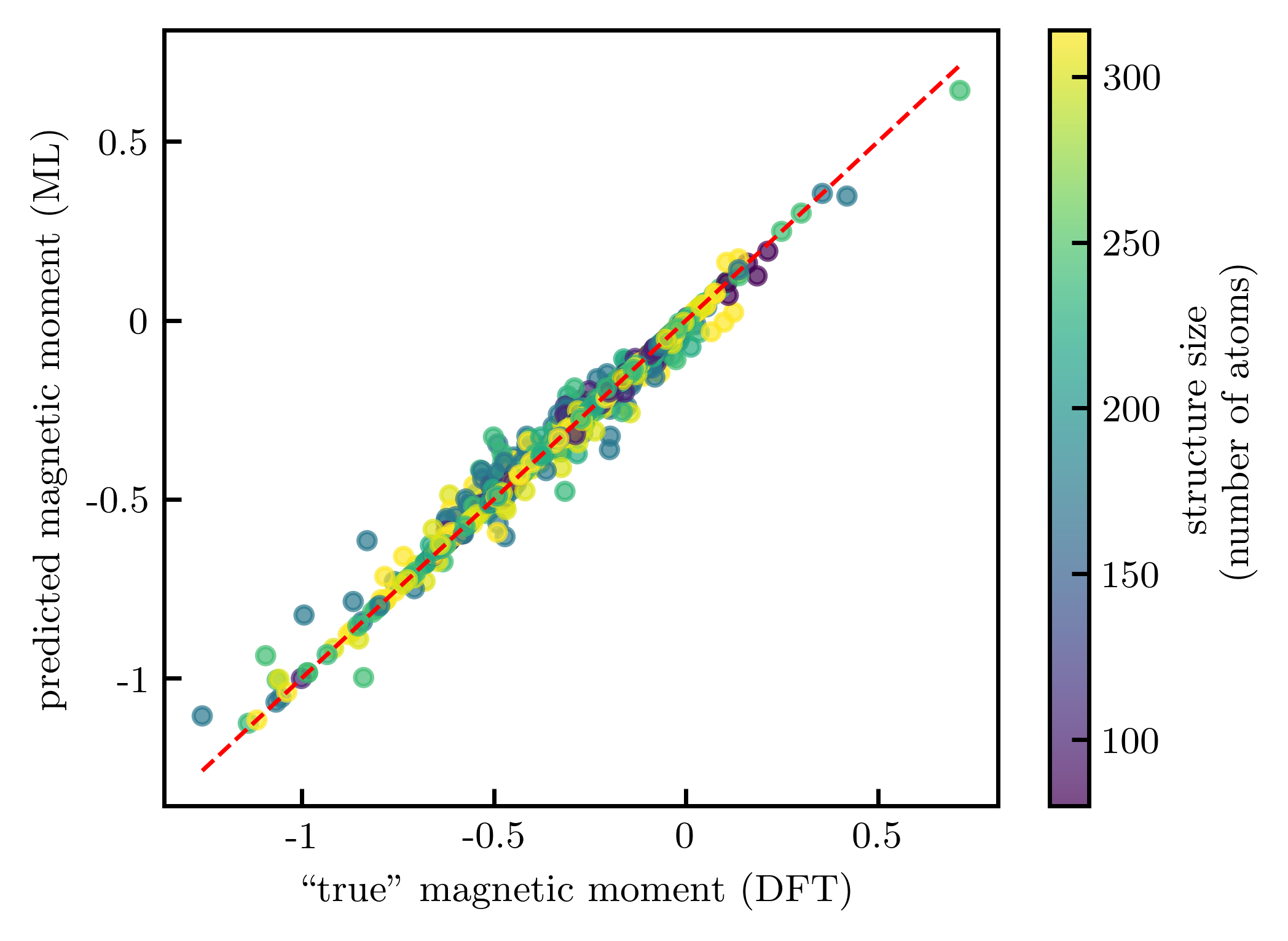}}
		\caption{Verification of the ML model within the DFT dataset (data scaled for visualization). 90\% of the data has been used for training, the remaining 10\% serve as the testing data.
		}
		\label{fig:verification}
	\end{figure}

	As already described in the main text, we use our DFT data set to internally validate the ML model. The accuracy of the ML model is evaluated using a randomly split DFT dataset comprising 9498 samples, each representing a carbon atom, its local atomic environment, and the corresponding induced magnetic moment. 10\,\% of this data have been isolated for the testing process and are used for predictions by the trained ML model. The model performance is assessed via the mean squared error (MSE) between ML prediction and DFT results. The correlation plot of this data is shown in Fig.~\ref{fig:verification}. Using the optimal SOAP parameters (see Sec.~\ref{app:descriptor}), we obtain a MSE of $2\times10^{-9}~\mu_B^2$, which corresponds to an average absolute prediction error of approximately $4.5 \times 10^{-5}~\mu_B$. Compared to typical magnetic moments of $0.4 \times 10^{-3}~\mu_B$, the model achieves an accuracy within $\sim$10\%.
	
	\section{Analytic models}
	\label{app:FuncDep}
	Although several lines of evidence suggest that the magnetic moments arise from a complex hybridization process, we seek to explain the distribution of induced magnetic moments observed in DFT using a simple analytical model. Specifically, we consider the overlap between spherical orbitals centered on the C atoms and those of the CGT-layer atoms.
	Without further specifying these orbitals, we assume that the overlap between the orbitals varies exponentially with the distance between C and CGT atom. Similar to the ML approach, only atoms within a distance of 16\,\AA~are considered. Furthermore, we assign different prefactors $C_j$ and exponential decays $\kappa_j$ for each atom type $j=\{$Cr, Te, Ge$\}$. Correspondingly, we can build an effective interaction as
	$\sum_j \sum_\mathrm{i} C_j \exp(-\kappa_j d_{i}^{(j)})$, 
	where $d_{i}^{(j)}$ denotes the distance from the C atom to each atom $i$ in the CGT layer. By means of the prefactors $C_j$, the interaction between various atom types can be controlled.
	Despite scanning a large parameter space, these efforts did not result in a quantitatively meaningful model. As already indicated by the PDOS-derived magnetization (see Fig.~3), we suspect that the underlying physical interactions are driven by complex hybridization between Te orbitals and states deep in the valence band of graphene -- a process which can not be described by the exponential decay model.
	\begin{figure}[htbp]
		\centering
		\includegraphics[width=.75\linewidth]{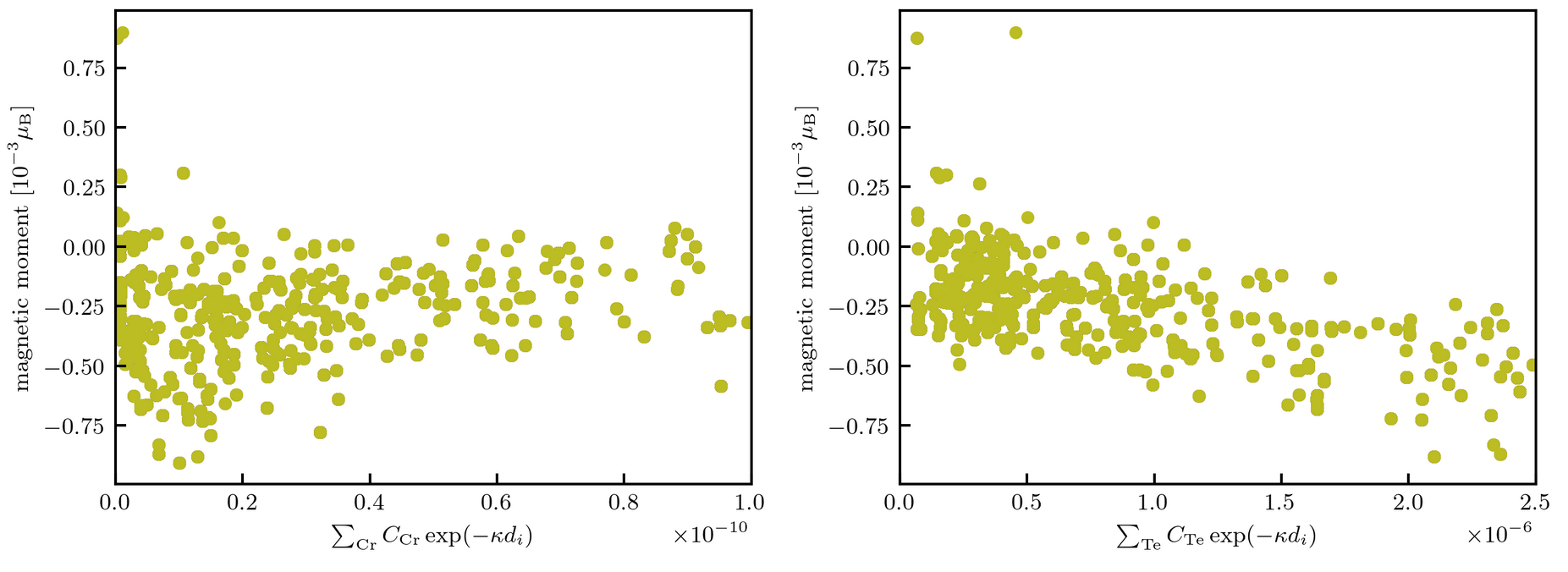}
		\caption{Correlations between DFT magnetic moments and our simple exponential decay model (see text). In the left panel, only Cr atoms are considered and the correlation is virtually non-existent (Pearson correlation coefficient below 0.1). A weak correlation is found when considering only Te atoms as shown in the right panel (Pearson correlation coefficient of 0.3).}
		\label{fig:funcDep}
	\end{figure}
	Interestingly, while at least a weak correlation (Pearson coefficient of 0.3) is found when considering only Te atoms ($C_\mathrm{Cr}=C_\mathrm{Ge}=0$), there is no significant correlation when only Cr atoms are considered as shown in Fig.~\ref{fig:funcDep} (Pearson correlation coefficient below 0.1). This observation is largely independent of $\kappa_j$. The same is true for Ge atoms, and combinations of several atoms (not shown). 
	We conclude that distance vectors to neighboring atoms alone are not suitable to account for the induced magnetic moment.
	However, we take the weak correlation with Te distances as a further indication for our presumption that the proximity-induced magnetization is enabled via hybridization with Te orbitals, the direct neighbors of the graphene layer. 
	
	In addition, we have also tried other functional forms (polynomials) and also to find a correlation by means of symbolic regression~\cite{cranmerDiscovering2020, cranmerInterpretableMachineLearning2023} which is a method to fit to arbitrary functions (build from a pre-defined set of functions containing polynomials, sin, and exp functions in our case). However, also the symbolic regression approach delivers only a simple exponential form which fits the data badly and could not find any other (potentionally more complex) functional dependency.
	The lack of an analytic functional dependency further underlines the need for ML to accurately model the proximity-induced magnetic structure in VdW heterobilayers.
	
	\section{Adaptive $k$-point mesh}
	\label{app:adapt_mesh}
	\begin{figure}
		\centering
		\includegraphics[width=0.4\linewidth]{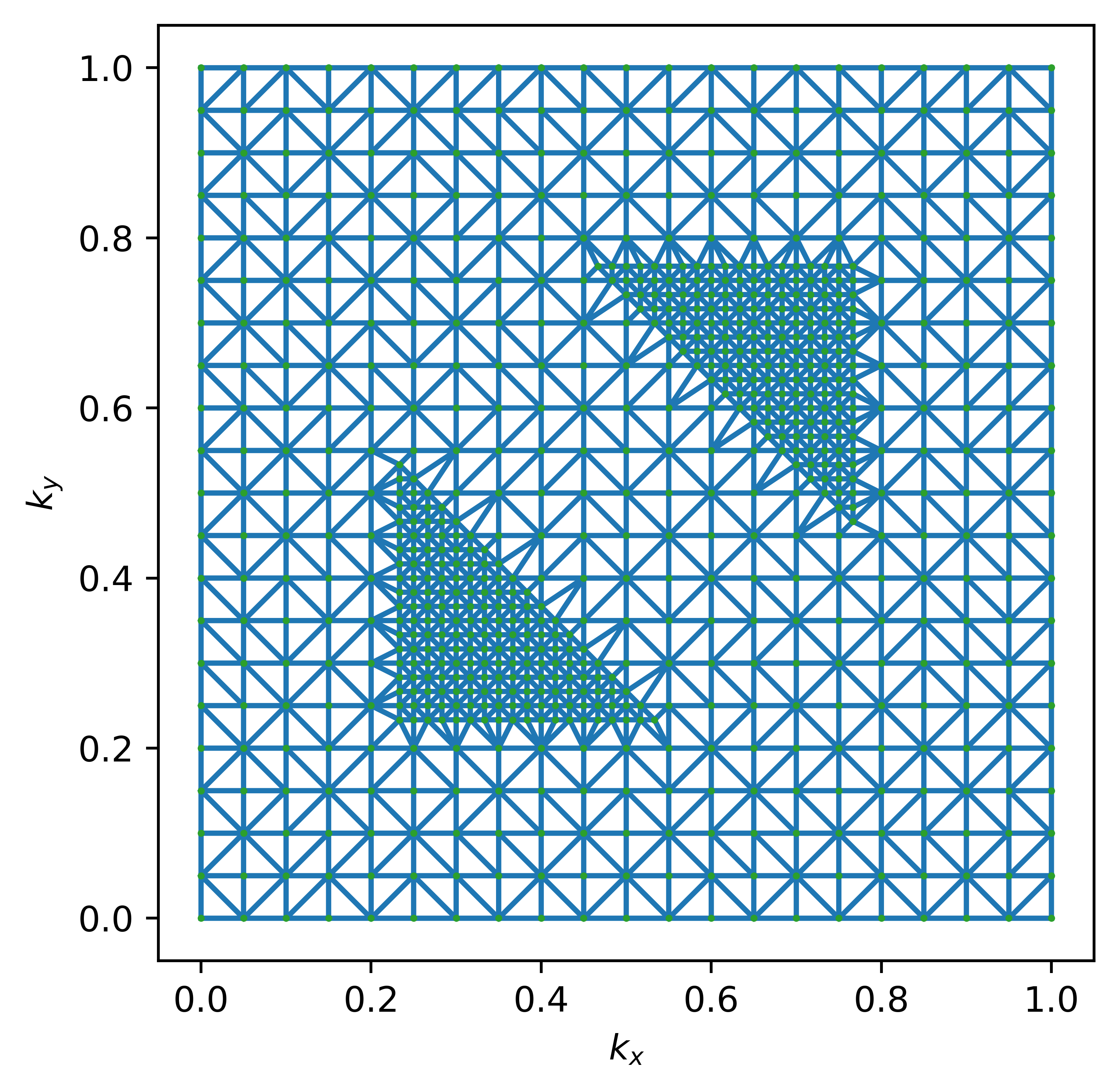}
		\caption{An exemplary adaptive $k$-mesh with 20x20 kpoints across the full BZ. The region around the $K$ point is replaced with a 3 times denser mesh, improving the accuracy of our (P)DOS description around the Dirac point significantly without exceeding practical computational limits.}
		\label{fig:adapt_mesh}
	\end{figure}
	\begin{figure}
		\centering
		\includegraphics[width=0.6\linewidth]{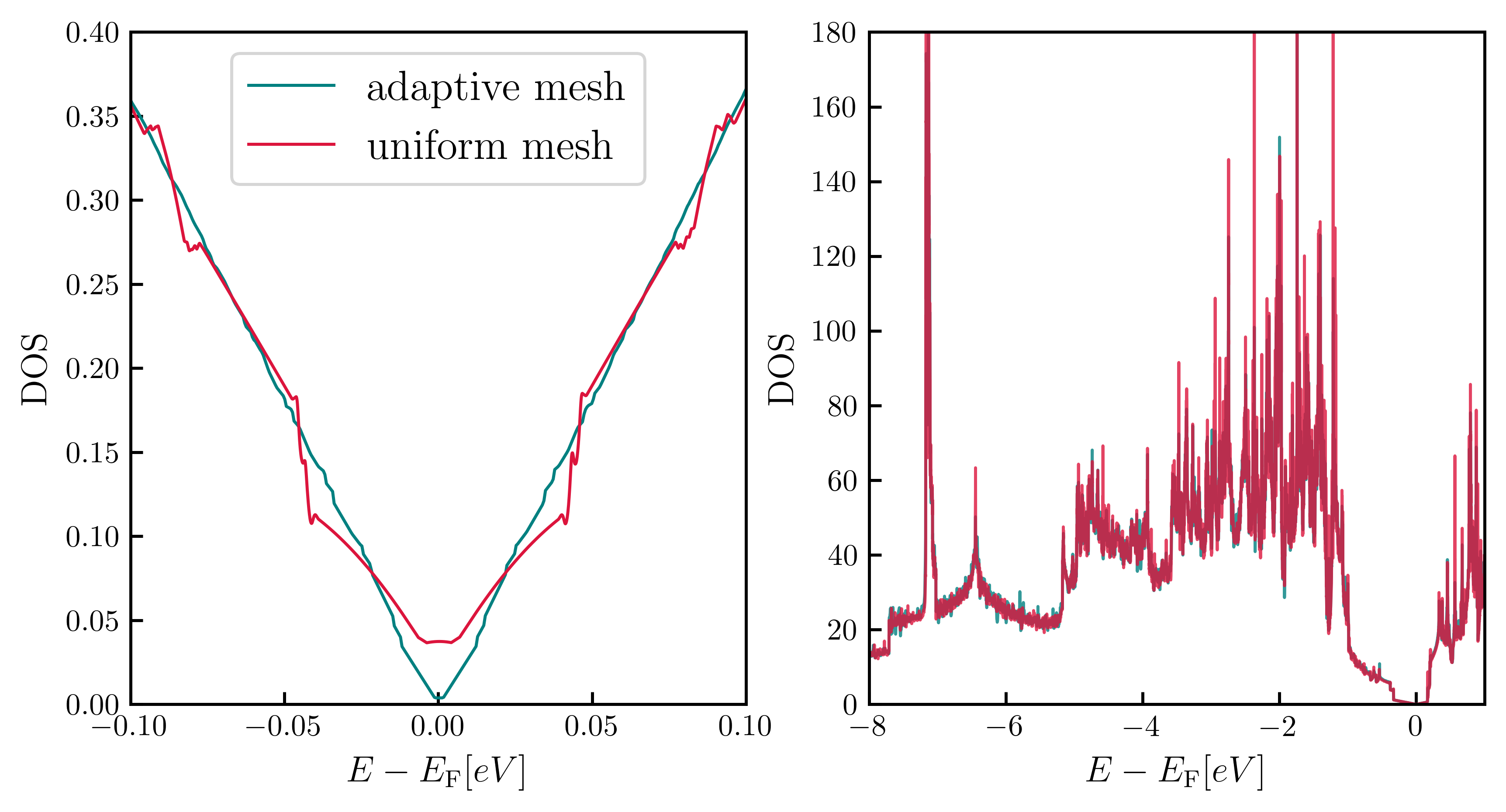}
		\caption{Results for the total DOS as obtained by an irreducible BZ based on a uniform grid with $42\times42$ $k$-points compared to the same calculation using an adaptive $k$-grid using a 5-fold denser mesh around $K$. In this case, the total number of $k$-points is increased by 76\% (compared to 400\% for the same accuracy around $E_\mathrm{F}$ using a uniform mesh). The outcome is more accurate at $K$ (left panel) without distorting the DOS at other energy levels (right panel).}
		\label{fig:qe_vs_adaptive}
	\end{figure}
	
	Obtaining the Dirac cones with exchange splittings on the order of a few meV is a computationally highly challenging task given our large supercells. It requires sampling of the $k$-space with a dense grid around K. Sampling the full BZ with such a dense $k$-mesh is computationally not feasible. Fortunately, other regions in $k$-space can be sampled by a less dense mesh without significant errors. We have therefore implemented a Python code to construct an adaptive mesh for the DFT calculations. On this adaptive mesh, we can then obtain DOS and PDOS by numerically integrating the band energies using a triangular method~\cite{Kurganskii_2D_BZ_integration} as depicted in Fig.~\ref{fig:adapt_mesh}. For the sake of clarity, we show the mesh spanned over the full BZ, while in the DFT calculation, only the points of the irreducible BZ, a much smaller subset depending on the crystal symmetries, are considered.
	
	Adaptive meshes are not implemented in \textit{Quantum Espresso} so this functionality is enabled through the backdoor. Our code first generates irreducible meshes at two specified grid densities, then replaces a predefined region around $K$ of the coarse mesh with the dense one. The resulting list of $k$-points is passed on to \textit{Quantum Espresso}. In Fig.~\ref{fig:qe_vs_adaptive}, a comparison between the DOS obtained via the established non-uniform mesh versus the adaptive mesh, shows a clear improvement, recovering the expected DOS of Dirac cones. The DOS of both spin bands does not go to zero because the exchange-split bands overlap at all energies, also at $E_\mathrm{F}$.

	\section{Hybridization mechanism}
	\label{app:hybridization}
	To investigate possible hybridization processes in the system, we analyze the spin-resolved band structures in the heterostructure projected onto the atomic orbitals of CGT. The band structures are shown separately for spin-up and spin-down channels and distinguish contributions from the various atomic species. In the projection considered here, states with significant weight on the indicated atom type are highlighted, allowing a clear identification of their involvement in any hybridization features. 
	
	\begin{figure}[htbp]
		\centering
		\includegraphics[width=0.7\linewidth]{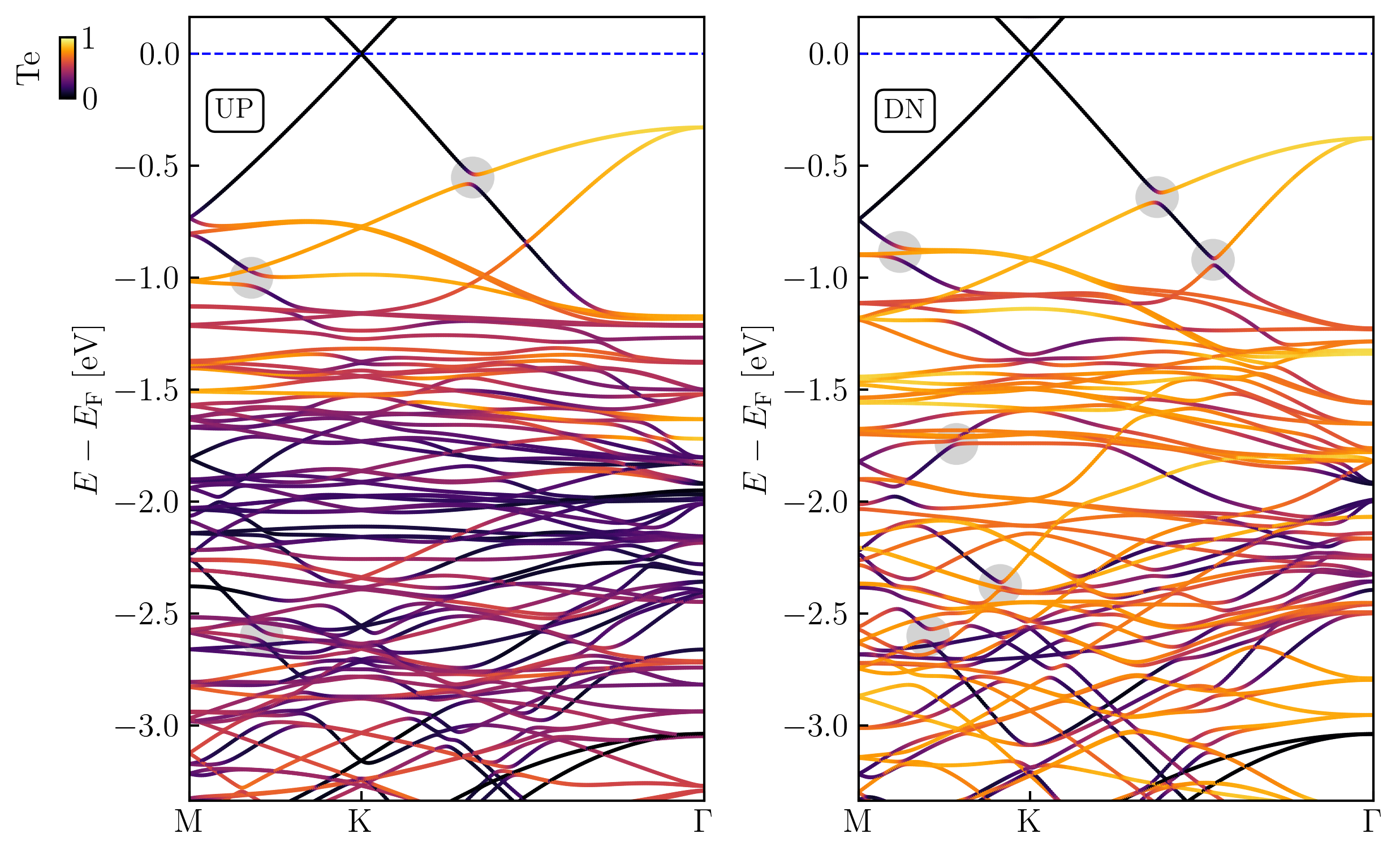}
		\caption{Band structure projected on Te orbitals, where the color scale indicates the contribution of Te for spin up (left) and spin down (right).}
		\label{fig:bands_Te}
	\end{figure}
	
	\begin{figure}[htbp]
		\centering
		\includegraphics[width=0.7\linewidth]{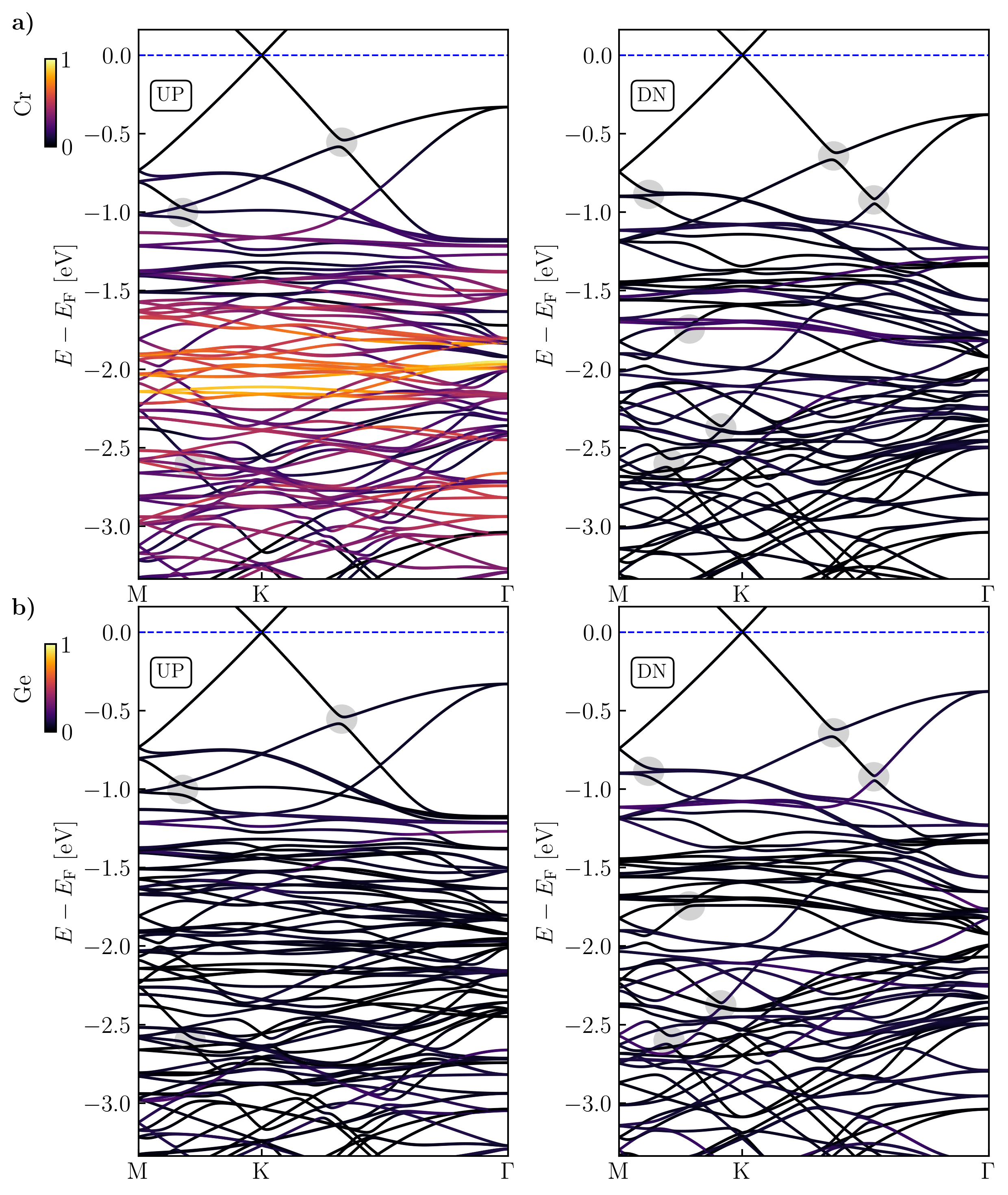}
		\caption{Band structure projected on \textbf{(a)} Cr and \textbf{(b)} Ge orbitals, where the color scale indicates the contribution of the respective atoms for spin up (left) and spin down (right). The gray shaded areas mark anti-crossings which can directly be associated with Te bands, see Fig.~3. Because the herein shown bands are black, hybridization with orbitals from other CGT atoms does not occur.}
		\label{fig:bands_others}
	\end{figure}
	
	Notably, several avoided crossings (anti-crossings, marked by gray areas) with the Graphene orbitals (linear dispersion) are observed within the band structure. However, a closer inspection reveals that these anti-crossings occur exclusively in bands that carry notable spectral weight from Te atoms. This conclusion is supported by complementary plots in Fig.~\ref{fig:bands_Te} and Fig.~\ref{fig:bands_others} using a color code to indicate atomic orbital contributions. Apparent from Fig.~\ref{fig:bands_Te}, the interacting bands predominantly show Te contributions. Furthermore, anti-crossings with large splittings occur mainly in the spin down channel, which is also the predominant magnetization direction of Te atoms, cf.~Sec.~\ref{app:charge_density_plots}. In Fig.~\ref{fig:bands_others}, the absence of color (black lines) signifies a lack of Ge or Cr character in those bands, further confirming that the observed anti-crossings involve states not originating from Ge or Cr orbitals. Therefore, while hybridization effects are present in the system, they do not appear to directly involve the Ge or Cr atoms in the energy window of interest.

\begin{figure}[htbp]
		\centering
		\includegraphics[width=0.7\linewidth]{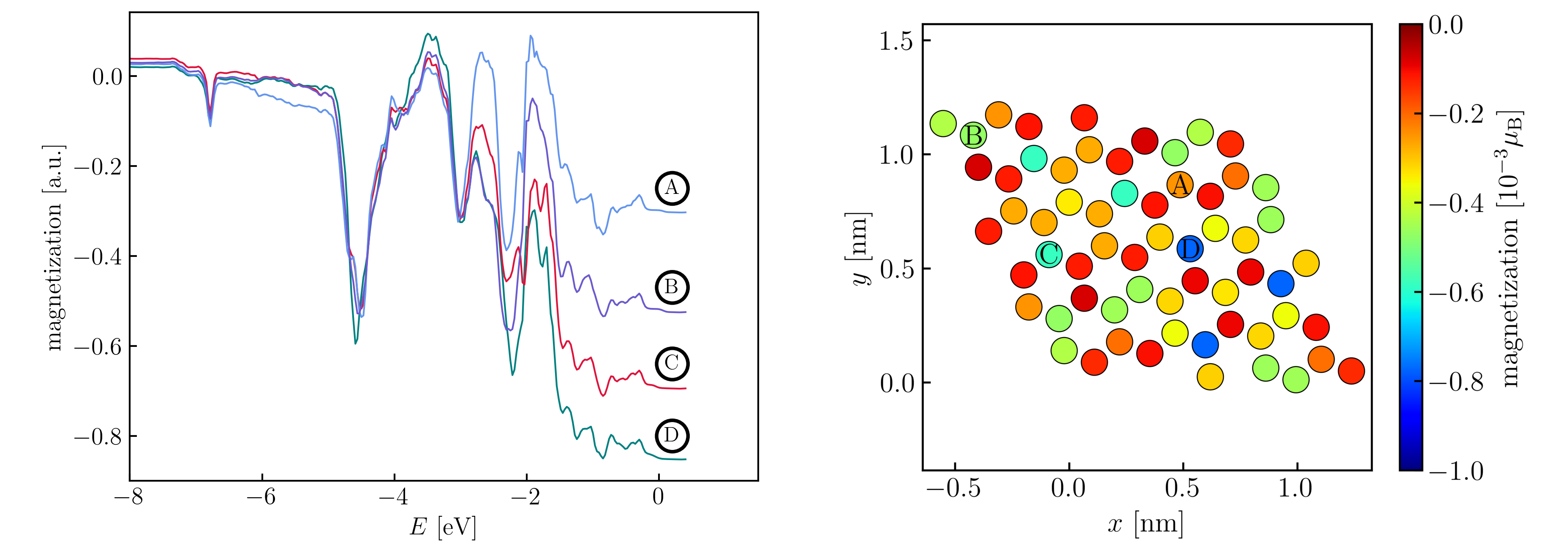}
		\caption{Magnetization $m(E)$ obtained from the DFT-derived spin polarization for some representative C atoms at $\theta=8.948\degree$. Differences are mostly picked up at energies between -3 and -0.5\,eV, the energy range where graphene bands hybridize with the neighboring Te, cf. Fig.~3. Their position in the atomic structure is indicated by the labels.
		}
		\label{fig:magnetization}
	\end{figure}

	Next, we investigate the energy range of the hybridization by means of the PDOS.
	As already pointed out in the main text, we calculate the spin polarization $P(E)= \mathrm{PDOS}_\mathrm{up}(E)-\mathrm{PDOS}_\mathrm{dn}(E)$ along the full energy range which gives the total magnetization of the atom by $m(E)=\int_{-\infty}^{E} P(E') dE'$. As shown in Fig.~\ref{fig:magnetization}, the magnetization is accumulated at lower energies (mainly from -3\,eV to -0.5\,eV). As described in the previous paragraph, this energy range is also where graphene and Te valence states hybridize, compare Fig.~\ref{fig:bands_Te}. The labels refer to the atoms in the atomic structure on the right.

	\section{Charge redistribution in interfaced layers}
	\label{app:charge_density_plots}
\begin{figure}[htbp]
		\centering
		\includegraphics[width=\linewidth]{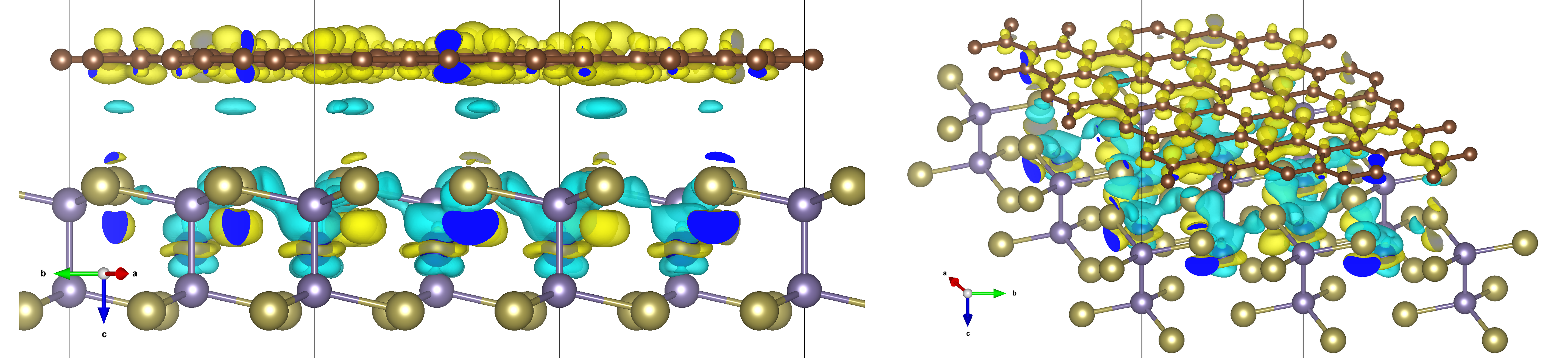}
		\caption{Side (left) and 3D view of the induced spin density in the stacked hetero-bilayer. Light-blue isosurfaces show an increase of spin up density, while yellow isosurfaces depict an increase of spin down density. The dark blue areas are cuts through the bubbles at the periodic supercell border.}
		\label{fig:qe_densities}
	\end{figure}

	In order to further examine the hybridization mechanisms and charge/spin redistributions provoked by the presence of both layers in the heterobilayer structure, we illustrate the induced spin density in Fig.~\ref{fig:qe_densities}. To this end, we first calculate the induced charge density for spin up and down, respectively: $\rho_{ind}^{\uparrow,\downarrow}=\rho_{tot}^{\uparrow,\downarrow}-\rho_\mathrm{CGT}^{\uparrow,\downarrow}-\rho_\mathrm{Gr}^{\uparrow,\downarrow}$, where the charge density of the single layers (CGT/Gr) are subtracted from the complete heterobilayer $\rho_{tot}$. Thus, this quantity contains information about the charges/spins induced by the stacking of the layers. Then, the spin density is readily obtained as $\sigma=\rho_{ind}^{\downarrow}-\rho_{ind}^{\uparrow}$. The depicted spin density is calculated for the energy range from -7 to -2\,eV, covering the energy range where the total magnetic moment in mostly generated, cf.~Fig.~\ref{fig:magnetization} or Fig.~3 in the main text.
	
	In Fig.~\ref{fig:qe_densities}, blue isosurfaces depict an increase of spin up density, while yellow isosurfaces depict an increase of spin down density. A distinctive feature is the emergence of spin up ``channels'' reaching from Cr to Te atoms. Furthermore, the spin up density is increased in the region between the two layers.
	
	Pronounced accumulation of spin down density, on the other hand, is primarily observed at the graphene layer, magnetizing the C atoms in the direction of the CGT layer.
	In addition, an increase of the spin down density directly underneath the Te atoms within the CGT layer is observed. 
	
	
	Comparing the atoms' magnetic moments between single- and bilayer confirms these trends and completes the conceptual understanding. Cr atoms are responsible for the magnetization of CGT itself, and also for the proximitized graphene. Their magnetic moments are slightly decreased when CGT is brought into contact with graphene. The proximitized interaction with the neighboring graphene layer is mediated via Te atoms (showing an anti-parallel magnetization with respect to the Cr atoms). Accompanied by a charge/spin redistribution within the CGT layer, the magnetic moment associated with the Te atoms is increased when the graphene layer is present. The spin density redistributes from the Cr to the Te atoms.
	Magnetic moments induced at the C atoms typically align with the directly adjacent Te moments.

	\section{Proximity-induced charges vs magnetization}
	\label{app:induced_charges}
	Finally, we analyze the induced charges in the graphene layer. Assuming that only hybridization with $p_z$ orbitals and no spin-orbit coupling occurs, the spin polarization of the Dirac cones is sensitive to the position of the Dirac point with respect to the Fermi level.
	Therefore, this naive model suggests correlations between the induced magnetization and the induced charge. 
	
\begin{figure}[!h]
		\centering
		\includegraphics[width=\linewidth]{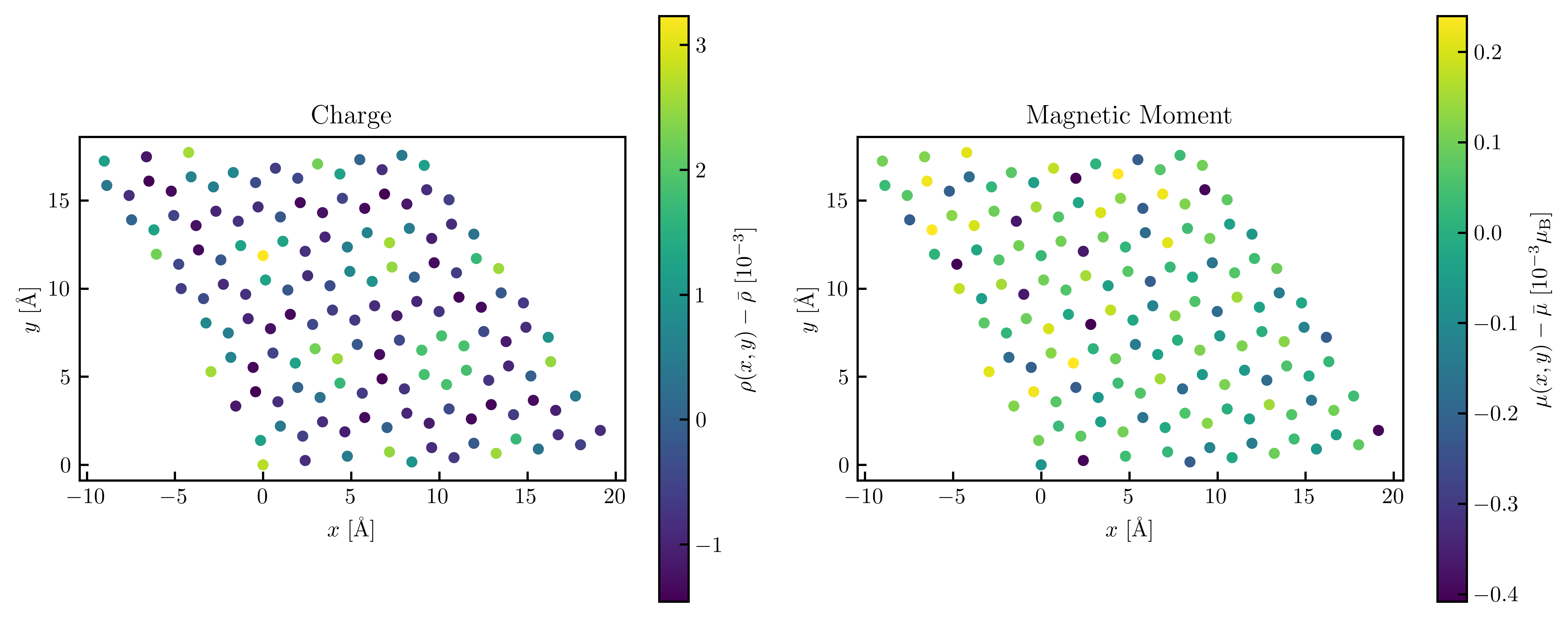}
		\caption{Proximity-induced fluctuations in the charges and magnetic moments of the C atoms in the graphene layer.}
		\label{fig:charges}
	\end{figure}

	However, the (P)DOS of C atoms is notably small close to the Fermi level, and so is the spin polarization. The magnetization, that is the integral of the spin polarization over a large energy range, including energy windows of much larger (P)DOS, is therefore not suspected to change when charges are induced by proximity.
	We have tested the DFT dataset for correlations between fluctuations in induced charge and magnetization. Underpinning the results from Fig.~\ref{fig:magnetization}, such a correlation is not found, as shown for one representative structure in Fig.~\ref{fig:charges}. 
	
	Without touching on the subject, we note here that the induced charge is another quantity which could be effectively predicted by ML.
	
	
\end{document}